\documentclass[a4paper,11pt]{article}
\pdfoutput=1 
\usepackage{verbatim}

\usepackage{jheppub}
\usepackage{amsmath,amsfonts,amssymb,amsthm,graphicx,color}
\usepackage[T1]{fontenc} 
\usepackage{enumerate}
\usepackage{tikz, tikz-3dplot, pgfplots}
\usepackage[utf8]{inputenc}
\usepackage{slashed}
\usepackage{subcaption}

\usepackage{float}
\usepackage{tikz}
\usetikzlibrary{decorations.text}
\usepackage{enumitem}
\setlist{itemsep=0pt}
\usetikzlibrary{math}

\usepackage{tikz}
\usepackage{pgfplots}
\usetikzlibrary{decorations.text,intersections, pgfplots.fillbetween}
\usetikzlibrary{math}
 \usetikzlibrary{snakes,3d,shapes.geometric,shadows.blur}
\usepackage{tikz-3dplot}

\newcommand{\comm}[1]{} 

\def\({\left(}
\def\){\right)}
\def\[{\left[}
\def\]{\right]}

\def\One{{\hbox{ 1\kern-.8mm l}}}

\def\barray{\begin{array}}
\def\earray{\end{array}}
\def\be{\begin{equation}}
\def\ee{\end{equation}}
\def\bea{\begin{eqnarray}}
\def\eea{\end{eqnarray}}
\def\bal{\begin{align}}
\def\eal{\end{align}}

\def\R{R_y} 

\def\-{\,-\,}
\def\={\,=\,}
\def\+{\,+\,}
\def\equi{\,\equiv\,}

\numberwithin{equation}{section} 

\definecolor{cardinal}{rgb}{0.6,0,0}
\definecolor{darkgreen}{rgb}{0,0.4,0}
\definecolor{golden}{rgb}{0.92, 0.7, 0}
\definecolor{midnight}{rgb}{0, 0, 0.5}
\definecolor{darkblue}{rgb}{0, 0, 0.7}
\definecolor{purple}{rgb}{0.5, 0, 0.5}

\def\IR{\mathbb{R}}

\def\cL{{\cal L}}
\def\cM{{\cal M}}

\def\cQ{{\cal Q}}
\def\cR{{\cal R}}

\usetikzlibrary{positioning,calc,fadings,decorations.pathreplacing}

\tikzset{
 diffuse color/.initial = black,                       
}

\tikzset{
 linear opacity/.initial=0.5,                          
 linear stroke/.style = {                              
   preaction={                                         
     draw=\pgfkeysvalueof{/tikz/diffuse color},        
     line width = (2.0-#1)*\pgflinewidth,              
     opacity=\pgfkeysvalueof{/tikz/linear opacity},white}},  
 diffuse gradient/.style={                             
   draw = none,                                        
   linear opacity=#1,                                  
   linear stroke/.list={0.0,#1,...,1.0}},              
 diffuse gradient/.default=1,                          
}

\tikzset{
 non-linear stroke/.style = {                          
   preaction={                                         
     draw=\pgfkeysvalueof{/tikz/diffuse color},        
     line width = (2.0-#1)*\pgflinewidth,              
     opacity=#1,white}},                                     
 diffuse falloff/.style={                              
   draw = none,                                        
   non-linear stroke/.list={0.0,#1,...,1.0}},          
 diffuse falloff/.default=1,                           
}
 
\newcommand\pgfmathsinandcos[3]{%
  \pgfmathsetmacro#1{sin(#3)}%
  \pgfmathsetmacro#2{cos(#3)}%
}
\newcommand\LongitudePlane[3][current plane]{%
  \pgfmathsinandcos\sinEl\cosEl{#2} 
  \pgfmathsinandcos\sint\cost{#3} 
  \tikzset{#1/.style={cm={\cost,\sint*\sinEl,0,\cosEl,(0,0)}}}
}
\newcommand\LatitudePlane[3][current plane]{%
  \pgfmathsinandcos\sinEl\cosEl{#2} 
  \pgfmathsinandcos\sint\cost{#3} 
  \pgfmathsetmacro\yshift{\cosEl*\sint}
  \tikzset{#1/.style={cm={\cost,0,0,\cost*\sinEl,(0,\yshift)}}} %
}

\newcommand\DrawLatitudeCircle[2][1]{
  \LatitudePlane{\angEl}{#2}
  \tikzset{current plane/.prefix style={scale=#1}}
  \pgfmathsetmacro\sinVis{sin(#2)/cos(#2)*sin(\angEl)/cos(\angEl)}
  \pgfmathsetmacro\angVis{asin(min(1,max(\sinVis,-1)))}
  \draw[current plane] (\angVis:1) arc (\angVis:-\angVis-180:1);
  \draw[current plane,dashed] (180-\angVis:1) arc (180-\angVis:\angVis:1);
}

 \newcommand{\bubble}{
    \begin{tikzpicture}[remember picture,trim left=0.1cm]

\def\R{1.1} 
\def\angEl{35} 
\def\angAz{-105} 
\def\angPhi{-40} 
\def\angBeta{19} 

\pgfmathsetmacro\H{\R*cos(\angEl)} 
\tikzset{xyplane/.style={cm={cos(\angAz),sin(\angAz)*sin(\angEl),-sin(\angAz),
                              cos(\angAz)*sin(\angEl),(0,-\H)}}}
\LongitudePlane[xzplane]{\angEl}{\angAz}
\LongitudePlane[pzplane]{\angEl}{\angPhi}
\LatitudePlane[equator]{\angEl}{0}

\fill[ball color=white] (0,0) circle (\R); 
\draw (0,0) circle (\R);

\coordinate[mark coordinate,gray] (O) at (0,0);
\coordinate[mark coordinate] (N) at (0,\H);
\coordinate[mark coordinate] (S) at (0,-\H);
\coordinate (P) at (\H,0);
\path[pzplane] (\R,0) coordinate (PE);
\path[xzplane] (\R,0) coordinate (XE);

\DrawLatitudeCircle[\R]{0} 

\draw[->] (0,-1.3*\H) -- (0,1.6*\R) node[above] {};

\draw (N) +(0.3ex,0.6ex) node[above=0.25, left] {$\mathbf{N}$};
\path (S) +(0.4ex,-0.4ex) node[below=0.27, left] {$\mathbf{S}\,$};
\draw[thin,decorate,decoration={brace,raise=0.5pt,amplitude=1ex}] (\R-0.04,0) -- (0.07,0)
    node[midway, below=0.15] {$r_\text{B}$}; 
 \end{tikzpicture}}

\tikzset{%
  >=latex, 
  inner sep=0pt,%
  outer sep=2pt,%
  mark coordinate/.style={inner sep=0pt,outer sep=0pt,minimum size=3pt,
    fill=black,circle}%
}

\title{\boldmath 	Topological Stars, Black holes and \\ Generalized Charged Weyl Solutions}

\author{Ibrahima Bah and}
\author{Pierre Heidmann} 
\affiliation{Department of Physics and Astronomy, Johns Hopkins University, 3400 North Charles Street, Baltimore, MD 21218, USA}

\emailAdd{iboubah@jhu.edu}
\emailAdd{pheidma1@jhu.edu}

\abstract{We construct smooth static bubble solutions, denoted as topological stars, in five-dimensional Einstein-Maxwell theories which are asymptotic to $\IR^{1,3}\times$S$^1$. The bubbles are supported by allowing electromagnetic fluxes to wrap smooth topological cycles. The solutions live in the same regime as non-extremal static charged black strings, that reduce to black holes in four dimensions. We generalize to multi-body configurations on a line by constructing closed-form generalized charged Weyl solutions in the same theory. Generic solutions consist of topological stars and black strings stacked on a line, that are wrapped by electromagnetic fluxes. We embed the solutions in type IIB String Theory on S$^1\times$T$^4$. In this framework, the charged Weyl solutions provide a novel class in String Theory of multiple charged objects in the non-supersymmetric and non-extremal black hole regime.}

\preprint{}
\begin{document}

\maketitle
\flushbottom

\newpage

%
%
%
%
%
%
%
%
%
%
%
%
%
%
%
%
%
%
%
%
%
%
%
%
%
%
%
%
%
%
%
%
%
%
\section{Introduction}
\label{sec:Intro}

Black holes live at the interface of General Relativity (GR) and Quantum Mechanics.  As such, their theoretical studies have lead to interesting paradoxes that have highlighted and sharpened the fundamental conflicts between the two frameworks.  More importantly, they have provided important windows to the underlying framework that characterizes the basic degrees of freedom of Quantum Gravity. String theory has offered important resolutions to such paradoxes and have provided microscopic descriptions of the basic degrees of freedom of black holes that are counted by the Bekenstein-Hawking entropy \cite{Strominger:1996sh}. These microstates are captured by bound states of strings and branes at weak coupling. When strongly coupled, these microstates can have physical sizes that are large compared to the string and Planck scales, and as large as the horizon of the black hole they correspond to. Such backreaction can resolve the unitarity problem for black hole evaporation as they provide ``quantum hair'' to the black hole \cite{Mathur:2009hf}.  A class of such microstates can be coherent enough to admit classical gravitational descriptions via geometric transition as compact horizonless objects that cap off smoothly at the vicinity of the would-be horizon \cite{Bena:2004de,Bena:2006kb,Bena:2007kg,Bena:2016ypk,Heidmann:2019xrd}.  A classical characterization of such structure must involve \emph{new phases of matter} that arise from fluxes of hidden fields and extra dimensions to prevent from complete collapse \cite{Gibbons:2013tqa}.

In parallel, the upcoming decade will see powerful new observational methods for black holes.  Their close environment can be observed via direct imaging by the Event Horizon Telescope \cite{Akiyama:2019cqa}. The progress of gravitational-wave detection from black hole binaries by the LIGO collaboration \cite{Abbott:2016blz,Abbott:2017oio} and the promise of the eLISA mission \cite{AmaroSeoane:2012km} set also a new incredible playground to directly test ideas in black hole physics by observations. In this new age of astronomy, it is interesting to
wonder whether theoretical results can lead to new observables. In particular, it is natural to ask if classical horizonless microstate candidates or even prototypes can lead to any predictions for beyond-GR black hole physics.

On one side, there are ``bottom-up'' toy models that consist in estimating deviation from black holes in GR by constructing and analyzing exotic compact objects (ECOs) \cite{Cardoso:2017njb,Cardoso:2019rvt}. Those models are usually four-dimensional theories involving exotic matters or mechanisms to construct horizonless ultra-compact objects that resemble a black-hole geometry up to its near-horizon environment. As a non-exhaustive list of such bottom-up models one can refer to boson stars \cite{Schunck:2003kk}, gravastars \cite{Mazur:2001fv} or wormholes \cite{Solodukhin:2005qy} (see \cite{Cardoso:2019rvt} for a review). Such objects are relatively simple to handle which has allowed to derive qualitative departures from black hole in GR through multipole moments, quasi-normal modes, tidal Love number or gravitational wave profile for instance (see \cite{Cardoso:2019rvt} and references thereof). Those computations are important for comparisons with direct observations. However, all the models suffer from significant problems that limit the scope and relevance of their outcomes and predictions. First, they are lacking top-down interpretations as they do not admit a UV origin within the framework of a Quantum Gravity theory. Second, they have fundamental issues that undermine their physical viability. Indeed, because they are mostly four-dimensional models, the no-hair theorem requires to use very exotic matter and unphysical fine-tunings in order to build structure at the scale of the horizon.

On the other side, String theory has provided numerous top-down constructions of horizonless smooth microstate geometries, also seen as classical fuzzballs, that resemble black hole geometries up to Planck scale above the horizon. It naturally realizes the only mechanism to support vast amount of viable microstructures at the vicinity of the horizon \cite{Gibbons:2013tqa}. This mechanism allows to bypass the no-hair theorem via two key ingredients: having \emph{extra compact dimensions} that can degenerate at the vicinity of the horizon and provide ends to spacetime, and turning on \emph{electromagnetic fluxes} to prevent the structure from collapse under its own gravitational attraction. The degeneracy of the extra dimensions at different loci create non-trivial smooth topological cycles, or \emph{bubbles}, supported by fluxes as a replacement for the horizon. However, the construction techniques are rather involved requiring to turn on various degrees of freedom from supergravity theories. The price to pay for those rigorous constructions is that they are complicated to handle involving non-spherically symmetric metrics, numerous gauge fields and scalars. Extracting relevant predictions about new black hole physics as deviations from mutlipole moments \cite{Bena:2020see,Bena:2020uup}, quasi-normal modes \cite{Bena:2020yii} or information recovery \cite{Bena:2019azk} is a challenge (see \cite{Mayerson:2020tpn} for a review). Moreover, almost all the solutions constructed so far live in non-astrophysical regimes. Most of them require supersymmetry, from the first microstate geometries constructed \cite{Bena:2004de} to the large families so far \cite{Bena:2006kb,Bena:2007kg,Bena:2016ypk,Heidmann:2019xrd}. Only few classes of solutions go beyond supersymmetry \cite{Jejjala:2005yu,Goldstein:2008fq,Bena:2009en,DallAgata:2010srl,Bossard:2014ola,Bena:2015drs}, even less are in a valid non-extremal regime of black holes \cite{Bena:2009qv}.

In this paper, we aim to fill the gap between the two philosophies of constructions. We want to settle the simplest framework for the construction of smooth ultra-compact objects that are convenient for phenomenology but keeping the two crucial ingredients of the microstate geometry program in String theory: topology from extra dimensions and fluxes. By doing so, our constructions will be non-supersymmetric and have the benefits of a bottom-up approach while admitting a top-down description from string theory. The minimal framework compatible with our method is \emph{Einstein-Maxwell theories} with one extra compact dimension in addition to the four dimensions. More precisely, we will consider a magnetically sourced one-form gauge field and its electric two-form dual. Those gauge fields may not be considered as the usual gauge fields in Electromagnetism under Kaluza-Klein (KK) reduction, but more as the descendants of ``hidden'' fields from the low-energy and classical limits of Quantum Gravity. For the sake of simplicity, we will focus on \emph{static solutions} only. By creating non-trivial topologies via the degeneracy of the extra dimension and turning on fluxes, we will show that we can construct smooth bubble geometries that we call ``Topological Stars''\footnote{This appellation has been already introduced for similar solutions in \cite{Bena:2013dka}.} in Einstein-Maxwell theories. 

With this approach, we will build single-center two-charge spherically symmetric solutions describing topological stars and black strings in five dimensions first (these were the subject of the short companion paper \cite{Bah:2020ogh}). We will discuss their phase space in four dimensions under Kaluza-Klein reduction and compare to usual GR charged black holes. Then, using the Weyl formalism we will find closed-form two-charge solutions describing multiple topological stars and black strings stacked on a line. We will discuss their top-down origin as D1-D5-KKm objects in type IIB string theory. In this framework our generalized charged Weyl solutions offer brand new multi-center type IIB solutions of static STU black holes and smooth bubbles deep inside the non-supersymmetric and non-extremal regime. Finally, we will discuss generalization to topological stars in $D$ infinite dimensions plus an extra compact dimension. We will have a special attention to $D=5$ which is a common playground for the microstate geometry program. We will compare our topological stars to JMaRT \cite{Jejjala:2005yu} and discuss how they bypass the over-rotating problem of JMaRT.

Our constructions allow for a more direct and qualitative understanding of bubbles as microstate geometries. The basic question about their stability without supersymmetry can be explored. As interestingly, the solutions can be used for phenomenological studies of microstate geometries. This is relevant for black hole astrophysics and gravitational-wave physics. Moreover, our generalized charged Weyl solutions and their embedding in type IIB lay the foundations of non-trivial microstructure constructions replacing black hole horizons by topology without the need for supersymmetry. We hope that the addition of degrees of freedom by degrees of freedom will still allow explicit constructions that will evolve towards more generic solutions far within the astrophysical regime.

The structure of the paper is as follows. We start with a summary of results in section \ref{sec:Summary}. In section \ref{sec:bubble&BH5d}, we construct and study single-center spherically symmetric topological stars and two-charge black strings in five-dimensional Einstein-Maxwell theory. In section \ref{sec:Multi}, we derive the generalized two-charge Weyl solutions in the same framework and construct the axisymmetric two-charge solutions describing multiple topological stars and black strings on a line. In section \ref{sec:ExtensionDdim}, we discuss the embedding of those solutions in type IIB String Theory and the generalization to arbitrary dimensions. We conclude in section \ref{sec:concl} and discuss future directions. The interested reader can find complementary details about the construction of the charged Weyl solutions in Appendix \ref{app:WeylSol} and the properties of topological stars and black strings in $D+1$ dimensions in Appendix \ref{app:D+1dimext}.


\section{Summary of Results}\label{sec:Summary}

\subsection*{Single-center Topological Stars and two-charge Black Strings}

Our discussion starts by considering spherically symmetric solutions to characterize the main features of topological stars and the black strings they correspond to. We generalize the results of \cite{Bah:2020ogh} by adding an electric two-form gauge field as well as the magnetic one-form gauge field. Therefore, we consider static two-charge solutions, sourced by a magnetic monopole and a line electric charge along the extra dimension. The solutions superpose a Schwarzschild string, that is a S$^1$ fibration over a four-dimensional Schwarzschild black hole, and a static bubble of nothing \cite{Witten:1981gj}, that is an Euclidean Schwarzschild solution with a time direction. The former has a horizon where the timelike Killing vector $\partial_t$ shrinks while the latter has a smooth bubble, or \emph{bolt}, where the orbits of a spacelike Killing vector $\partial_y$ shrink. We defined $y$ as the coordinate of the extra compact dimension. The two solutions are related by \emph{double Wick rotation} $(t,y) \rightarrow (iy,it)$. We will show that superposing those solutions into one solution indeed requires to turn on electromagnetic fluxes wrapping the bubble. The final class of two-charge solutions will be invariant under the double Wick rotation. Depending on which of the horizon or the bubble locus comes first, the solution is either a two-charge black string or a two-charge smooth bubble solution, a topological star. Under KK reduction, the black strings correspond to charged non-rotating black holes while the topological stars give naked singularity due to the degeneracy of the extra dimension. By studying the phase space according to the four-dimensional mass and charges, we will show that topological stars live in the same regime as black strings

\begin{figure}
\begin{subfigure}[h]{0.5\linewidth}
\includegraphics[width=\linewidth]{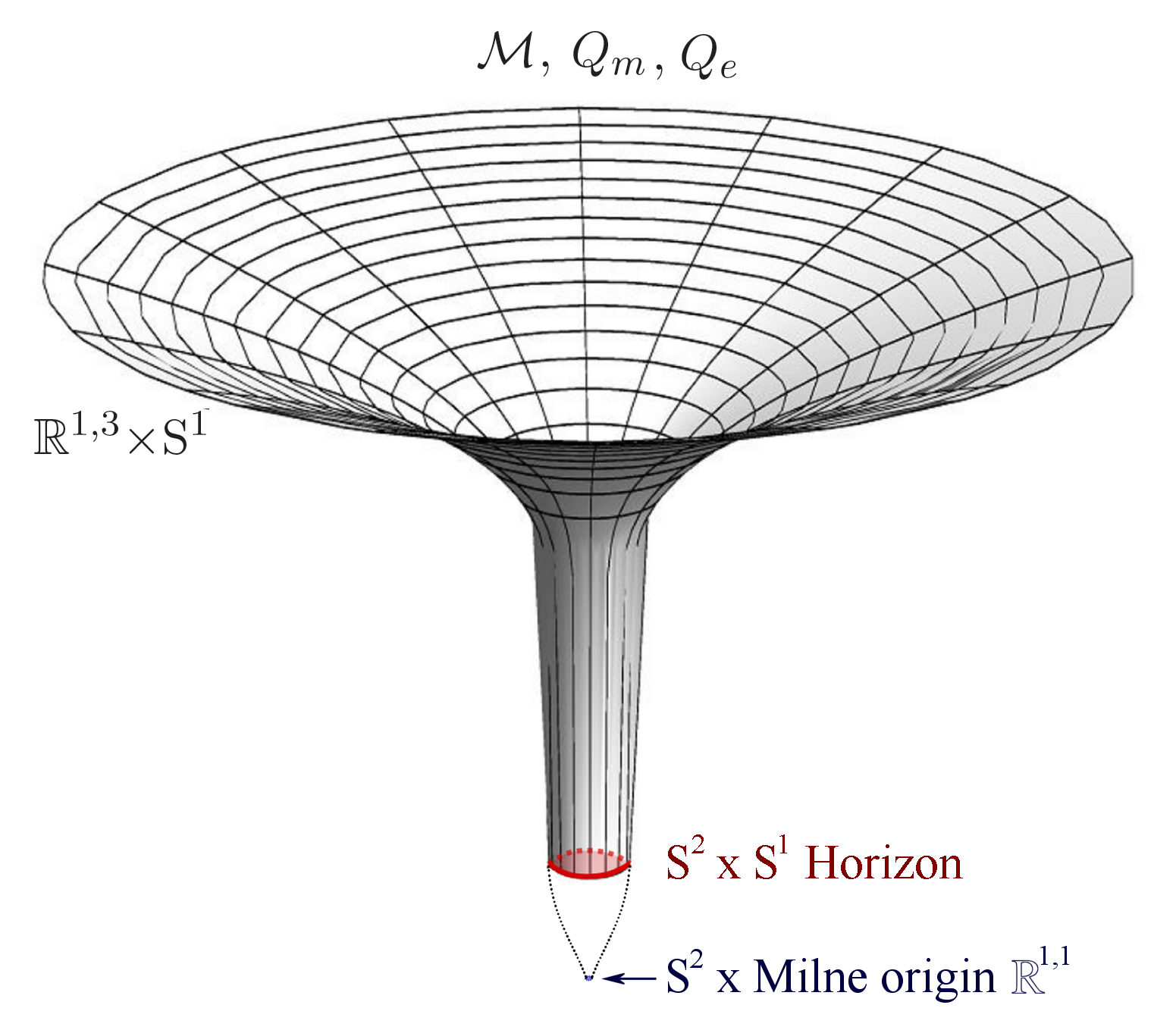}
\caption{Black string.}
\end{subfigure}
\hspace{0.5cm}
\begin{subfigure}[h]{0.5\linewidth}
\includegraphics[width=\linewidth]{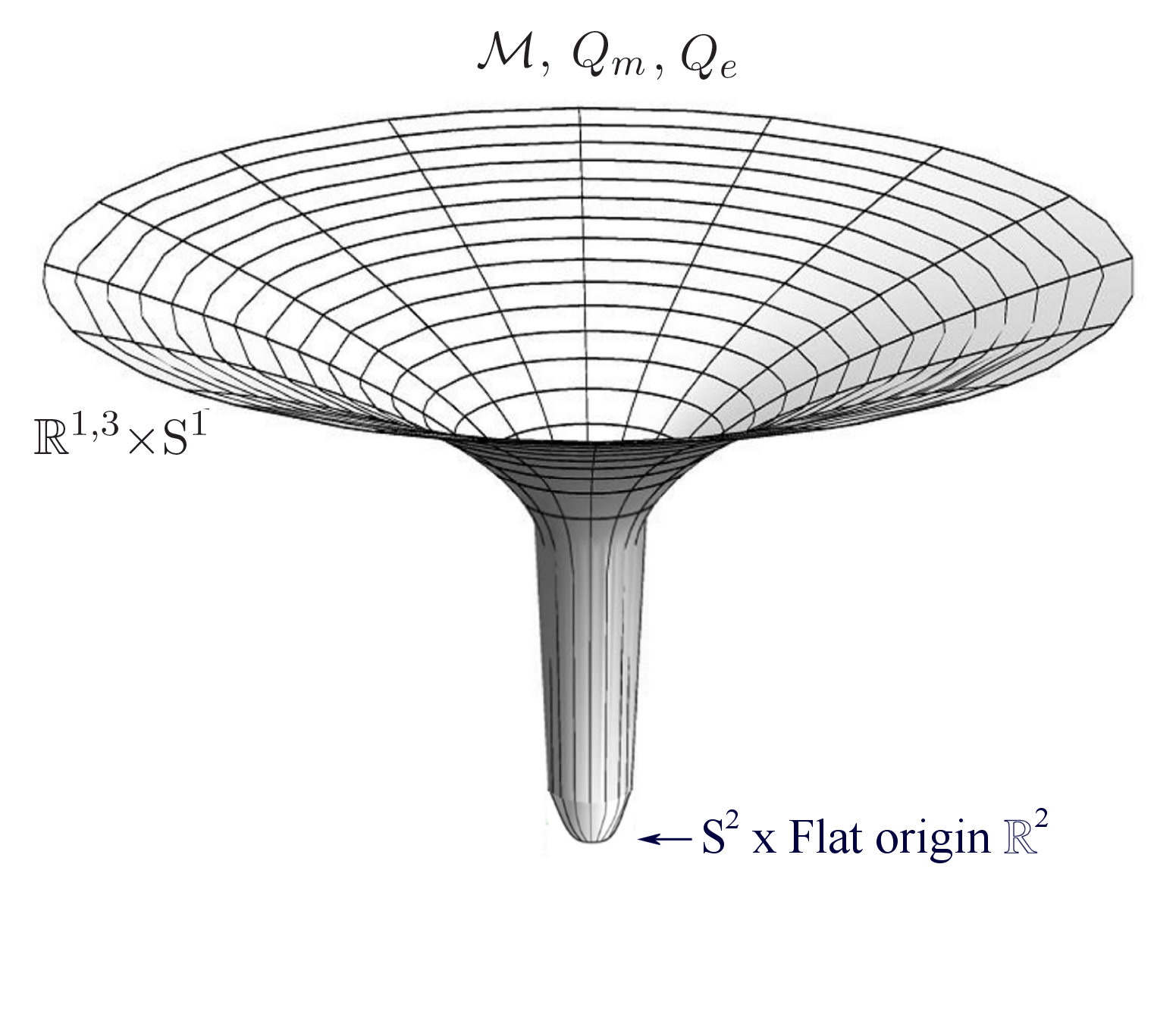}
\caption{Topological star.}
\end{subfigure}%
\caption{The two types of spherically symmetric static solutions of our five-dimensional Einstein-Maxwell theory. The solutions have a four-dimensional mass $\cM$ and are supported by electromagnetic fluxes with conserved magnetic and electric charges $(\cQ_m, \cQ_e)$.}
\label{fig:SolPict}
\end{figure}

Both objects have interesting properties (see Fig.\ref{fig:SolPict} for a schematic description). First, the black string (Fig.\ref{fig:SolPict}(a)) has a bubble in its interior that hides its curvature singularity. The topology at the bubble  corresponds to a round S$^2$ sitting at the origin of a smooth Milne space \cite{MilneBook,Horowitz:1990ap}. Such a space has singular properties under certain perturbations but is well studied in cosmology as a transition from a Big Crunch to a Big Bang \cite{Khoury:2001bz} which might suggest some traversability properties for our black string solutions. As for topological stars (Fig.\ref{fig:SolPict}(b)), the solutions resemble the black strings but cap off smoothly as a bolt, described by a round S$^2$ at the origin of a flat $\IR^2$. The regularity at the bolt constrains their overall size depending on the size of the extra dimension. The only way to get around this problem is to add a conical defect to the bolt that allows to consider topological stars of astrophysical size. A conical defect is not considered as a singularity in String Theory as long as it acts as a discrete smooth quotient on the local geometry. Moreover, we will argue that the conical defect can be classically resolved by blowing up Gibbons-Hawking bubbles at the poles of the bolt. This classical resolution brings to light a richer microstructure that our spherical symmetry hypothesis has swept under the rug.

\subsection*{Generalized two-charge Weyl solutions}

To resolve the conical defect and to build more generic topological stars, one needs to construct multi-bubble solutions in a non-perturbative manner and to derive the backreaction of charged bubbles on the spacetime by solving non-linear Einstein-Maxwell equations. This is a highly non-trivial task without the rescue of supersymmetry. However, generic topological stars must have an angular momentum, potential NUT charges along the extra dimension and multiple bubbles supported by fluxes on the three-dimensional base space. 

As a first step towards this goal, we construct the axisymmetric generalization of our solutions by allowing multiple charged bubbles and black strings on a line using the \emph{Weyl formalism} \cite{Weyl:book}. This has been successfully applied to classify axisymmetric vacuum static solutions of Einstein theory in four dimensions with an extra compact dimension \cite{Costa:2000kf,Emparan:2001wk,Elvang:2002br,Emparan:2008eg,Charmousis:2003wm}. The solutions are entirely determined by two functions that solve a Laplace equation which can be sourced by rods or point particles on a line. A generic solution consists of Schwarzschild black strings and bubbles of nothing stacked on a line and separated by \emph{struts}, or string with negative tension, to prevent the structure from collapse. The addition of electromagnetic gauge fields leads to non-linear differential equations. We have found closed-form solutions that exhibit a linear structure in terms of two functions that solve the Laplace equation. We introduce five different branches of possible gauge-field backreactions. Those branches highlight strong non-linear non-perturbative phenomenon. We show that a generic solution for one of this branch consists of two-charge black strings and topological stars stacked on a line. In our constructions so far, regularity requires that the fluxes on each object have the same orientation. Therefore, the solutions are again supported by struts instead. It is possible that another branch can solve those struts or that one needs to add other degrees of freedom as NUT charges or angular momenta.  The role of the struts in general are to account for binding energy of the system. 

\subsection*{Embedding in Type IIB and arbitrary dimensions}

Our constructions are obtained from a bottom-up approach, however they can be directly derived from String Theory and compared to known class of solutions. We will show that our five-dimensional solutions can be embedded in type IIB String Theory on S$^1\times$T$^4$ by turning only the two-form Ramond-Ramond field, $C_2$. The electric two-form gauge field in five dimensions arises directly from $C_2$ while the magnetic one-form is actually made of two indistinguishable gauge fields with equal charges but with very different UV origins: one arises from $C_2$ while the other arises from a non-trivial connection along the new S$^1$. More concretely, our solutions are D1-D5-KKm non-BPS solutions in type IIB with equal charges and with $\IR^{1,3}\times$S$^1\times$S$^1\times$T$^4$ asymptotics. In that framework, our black strings solutions are a three-charge non-rotating subclass of the generic four-charge STU black hole \cite{Cvetic:1995kv,Cvetic:2011dn}, while our topological stars are brand new smooth D1-D5-KKm solutions obtained from an analytic continuation of the parameter space of the STU black hole. Moreover, the generalized charged Weyl solutions we have constructed in five dimensions can be similarly embedded in type IIB. They consist in D1-D5-KKm black holes and D1-D5-KKm smooth bubbles stacked on a line and separated by struts. They give the first non-trivial examples of multiple D1-D5-KKm objects deep inside the non-BPS and non-extremal regime!

Finally, the discussion is not necessarily restricted to solutions that live with four infinite dimensions. We will construct spherically symmetric topological stars and black strings in $(D+1)$-dimensional Maxwell-Einstein theories with an electric two-form gauge field and a magnetic $(D-3)$-form gauge field. The construction works similarly by superposing a S$^1$-fibered Schwarzschild-Tangherlini solution \cite{Tangherlini:1963bw} and a $(D+1)$-dimensional bubble of nothing using electromagnetic fluxes. We will pay a special attention to $D=5$. We will show that the solutions can be embedded in type IIB on T$^4$ as D1-D5 non-BPS solutions with equal charges. In this framework, the black strings correspond to a non-rotating two-charge subclass of Cvetic-Youm black holes \cite{Cvetic:1996xz,Cvetic:1997uw,Giusto:2004id} while the topological stars are obtained from an analytic continuation of the parameter space. We will compare our smooth solutions to the known JMaRT solutions \cite{Jejjala:2005yu}. These solutions are non-BPS smooth solutions in type IIB with a similar topology as the topological stars. However, their regularity requires more angular momenta that the Cvetic and Youm black hole can have, and therefore, unlike topological stars, they do not live in the same regime as non-extremal black holes. We will discuss how our solutions have bypassed this issue.

\section{Topological stars and black strings in five dimensions}
\label{sec:bubble&BH5d}

We consider an Einstein-Maxwell theory in five dimensions defined by the action\footnote{We define the norm of a $p$-form $\mathcal{F}$ as
\begin{equation}
\left|\mathcal{F}\right|^2 \equiv \frac{1}{p!} \,g^{\alpha_1 \beta_1}\ldots g^{\alpha_p \beta_p} \,\mathcal{F}_{\alpha_1\ldots \alpha_p}\,\mathcal{F}^{\beta_1\ldots \beta_p}\,,\qquad \text{where}\quad \mathcal{F} \equiv \frac{1}{p!} \,\mathcal{F}_{\alpha_1\ldots \alpha_p} \,dx^{\alpha_1} \wedge \ldots \wedge dx^{\alpha_p}\,.
\end{equation}}

\begin{equation}
S_5=\int \mathrm{d}^5 x \sqrt{-\det g}\left(\frac{1}{2 \kappa_{5}^{2}} R-\frac{1}{2} \left|F^{(m)}\right|^2-\frac{1}{2} \left|F^{(e)}\right|^2\right)\,,
\label{eq:ActionGen}
\end{equation}
where $\kappa_5 $ is the five-dimensional Einstein gravitational constant, $F^{(m)}$ and $F^{(e)}$ are the magnetic two-form and electric three-form field strengths respectively, $g$ is the five-dimensional metric, $R$ is the Ricci scalar. The equations of motion are 
\begin{equation}
d\star F^{(m)} = 0 \,,\qquad d\star  F^{(e)} = 0 \,,\qquad R_{\mu \nu} = \kappa_5 ^2\left( T_{\mu \nu} - \frac{1}{3} \,g_{\mu \nu} \, {T_\alpha}^\alpha\right)\,,
\label{eq:EOMgen}
\end{equation}
where $\star$ is the Hodge star operator in five dimensions, $R_{\mu \nu}$ is the Ricci curvature tensor and $T_{\mu \nu}$ is the stress tensor of the gauge fields
\begin{equation}
\begin{split}
T_{\mu \nu} ~=~ & {F^{(m)}}_{\mu \alpha}{{F^{(m)}}_{\nu}}^{ \alpha} - \frac{1}{4}\,g_{\mu \nu} {F^{(m)}}_{\alpha \beta}{F^{(m)}}^{\alpha \beta} +  \frac{1}{2} \, \left[{F^{(e)}}_{\mu \alpha \beta }{{F^{(e)}}_{\nu}}^{ \alpha\beta} - \frac{1}{6}\,g_{\mu \nu} {F^{(e)}}_{\alpha \beta \gamma}{F^{(e)}}^{\alpha \beta \gamma}\right]\,.
\end{split} \nonumber
\end{equation}
We aim to construct spherically symmetric solutions that are asymptotic to a S$^1$ fibration over a four-dimensional Minkowski, $\IR^{1,3}\times$S$^1$. 
We use the spherical coordinates $(r,\theta,\phi)$ and the time coordinate $t$ to parametrize the four-dimensional spacetime while the extra dimension is denoted by $y$ with radius $R_y$.

\subsection{The class of two-charge solutions}
\label{sec:class4d}
We consider an ansatz for the spacetime metric as:
\begin{equation}
ds^2_5= - f_\text{S}(r)\, dt^2 + f_\text{B}(r)\, dy^2 + \frac{dr^2}{h(r) } + r^2\left( d\theta^2 + \sin^2 \theta \,d\phi^2\right), 
\label{eq:metric5d}
\end{equation} 
We want to translate the shrinking of the $y$-circle as a construction constraint. For that purpose, we exhibit a \emph{double Wick rotation symmetry} for our constructions.  The solutions will be symmetric under Wick exchange of $(t,y)$,
\begin{equation}
\left(t\,,\,y\,,\,f_\text{S}(r)\,,\,f_\text{B}(r) \right) \quad \longrightarrow \quad \left(i\,t\,,\,i\,y\,,\,f_\text{B}(r)\,,\,f_\text{S}(r) \right)\,.
\end{equation}
Thus, if we initially prepare a solution with a horizon where the timelike Killing vector $\partial_t$ shrinks at some loci, its symmetric counterpart solution has a spacelike Killing vector $\partial_y$ that shrinks. The spherical symmetry and the double Wick rotation symmetry drastically constrain the field strengths to be
\begin{equation}
F^{(e)} = \frac{Q}{r^2}\, dr \wedge dt \wedge dy\,,\qquad F^{(m)} = P\,\sin\theta\,d\theta \wedge d\phi\,.
\label{eq:MaxField5d}
\end{equation}
Thus, $F^{(e)}$ is sourced by a line charge $Q$ along the $y$ circle and $F^{(m)}$ is sourced by a magnetic monopole $P$.  In the vacuum limit $(P=Q=0)$ we have
\begin{equation}
\begin{split}
\text{Bubble of Nothing:}& \quad f_\text{B}(r) = h(r) = 1 - \frac {r_\text{B}}{r}, \quad f_\text{S}(r) =1 \,,\qquad F^{(e)} = F^{(m)} = 0\,,\\
\text{SW Black hole $\times$ S$^1$:}&  \quad f_\text{S}(r) = h(r) = 1 - \frac {r_\text{S}}{r}, \quad f_\text{B}(r)=1\,,\qquad F^{(e)} = F^{(m)} = 0\,.  \label{BH&BuoN}
\end{split}
\end{equation}
We consider a superposition of the two vacuum solutions and consider\footnote{Bubbles and black strings of these solutions were also studied in \cite{Stotyn:2011tv}.}
\begin{equation}
f_\text{B}(r) \= 1 - \frac {r_\text{B}}{r}\,,\qquad f_\text{S}(r) \= 1 - \frac {r_\text{S}}{r}\,,\qquad h(r) \= f_\text{B}(r)\,f_\text{S}(r)\,. \label{eq:Metric5dGen}
\end{equation}
The equations of motion \eqref{eq:EOMgen} are solved by turning on the fluxes \eqref{eq:MaxField5d} as
\begin{equation}
 P^2 +Q^2 = \frac{3\, r_\text{S} r_\text{B}}{2 \kappa_5^2}\,.
\label{eq:Field5dGen}
\end{equation}
Thus, the superposition is prevented from collapse by electromagnetic fluxes with fixed total charge. The solutions have a curvature singularity at $r=0$ and two coordinate singularities at $r=r_\text{B}$ and $r=r_\text{S}$. The first corresponds to a bolt coordinate singularity where the $y$-circle degenerates while the second corresponds to a horizon coordinate singularity where the timelike Killing vector, $\partial_t$, shrinks. Depending on the order, $r_\text{B} \lessgtr r_\text{S}$, the topology of the solution is very different. Before describing each type of solutions in the class, we first discuss their main characteristics from a four-dimensional perspective after compactification along $y$.

\subsection{Reduction to four dimensions}
\label{eq:reduction4d}

We keep the minimal number of ingredients for the Kaluza-Klein reduction along $y$ by turning off several degrees of freedom that are trivial for our class of solutions. Therefore, we will not consider the gauge field that arises from the connection in five dimensions, we will consider that $F^{(m)}$ has no leg on $dy$ while $F^{(e)}$ has only leg on $dy$,
\begin{equation}
F^{(e)} \= F^{(e)}_y \,\wedge dy\,, 
\label{eq:Fey}
\end{equation}
where $F^{(e)}_y$ is the field strength of a common electric U(1) gauge field. The five-dimensional action \eqref{eq:ActionGen} reduces to a Einstein-Maxwell-Dilaton action given by
\begin{equation}
\begin{split}
S_{4}= \int \mathrm{d}^{4} x \sqrt{-\det g_{4}}& \left(\frac{1}{2 \kappa_4^{2}} R_{4} - \frac{3}{\kappa_4^{2}}\, \partial_a \Phi\, \partial^a \Phi -\frac{e^{-2\Phi}}{2e^2}\left|F^{(m)}\right|^2- \frac{e^{2\Phi}}{2e^2} \left|F^{(e)}_y\right|^2\right)\,,
\end{split}
\label{eq:Action4d}
\end{equation}
where the gravitational and the electric couplings are
\begin{equation}
\kappa_4^2 \equi \frac{\kappa_5^2}{2\pi R_y}\,,\qquad e^2 \equi \frac{1}{2\pi R_y}\,.
\end{equation}
In this framework, the solutions are given by 
\begin{equation}
\begin{split}
ds_{4}^2 & =  \left(1-\frac{r_\text{B}}{r} \right)^{\frac{1}{2}} \left[- \left(1-\frac{r_\text{S}}{r} \right) dt^2 + \frac{r^{2} dr^2}{(r-r_\text{S})(r-r_\text{B})} + r^2 \left( d\theta^2 + \sin^2 \theta \,d\phi^2\right) \right]\,,\\
e^{2 \Phi} &= \left(1-\frac{r_\text{B}}{r} \right)^{-\frac{1}{2}}\,,\\
\end{split}
\label{eq:metric4d}
\end{equation}
and the electric and magnetic U(1) gauge fields have the following field strengths
\begin{equation}
F^{(e)}_y = \frac{Q}{r^2}\, dr \wedge dt \,,\qquad F^{(m)} = P\,\sin\theta\,d\theta \wedge d\phi\,,\qquad P^2 +Q^2 = \frac{3 e^2\, r_\text{S} r_\text{B}}{2 \kappa_4^2}\,.
\label{eq:MaxField4d}
\end{equation}
From a four-dimensional perspective, the solutions have an electric charge and a magnetic charge. The conserved quantities in four dimensions, as the ADM mass, $\cM$, the electric and the magnetic charges, $Q_e$ and $Q_m$, are given, following the conventions of \cite{Myers:1986un}, by
\begin{equation}
\cM \= \frac{2 \pi}{\kappa_4^2}\,\left( 2 r_\text{S} +r_\text{B}\right)\,,\quad Q_m \= \frac{P}{e}\,,\quad Q_e \= \frac{Q}{e}\,,\quad \mathcal{Q}^2 \equiv Q_m^2 + Q_e^2 \= \frac{3 r_\text{S} r_\text{B}}{2 \kappa_4^2}\,.
\label{eq:ADMmass&charges4d}
\end{equation}
We have then constructed a three-parameter family of two-charge solutions. It is worth for what will follow to express the initial parameters according to the asymptotic quantities and we obtain two pairs $(r_\text{S},r_\text{B})$ for given $(\cM,\cQ)$:
\begin{equation}
\begin{split}
r_\text{S}^{(1)} \= \frac{\kappa_4^2}{8\pi}  \left[ \cM  - \cM_{\Delta}  \right] \,,\quad r_\text{B}^{(1)} \= \frac{\kappa_4^2}{4\pi}  \left[ \cM  + \cM_{\Delta}  \right]\,,\\
r_\text{S}^{(2)} \=  \frac{\kappa_4^2}{8\pi}  \left[ \cM  + \cM_{\Delta}  \right]\,,\quad r_\text{B}^{(2)} \=  \frac{\kappa_4^2}{4\pi}  \left[ \cM  - \cM_{\Delta}  \right]\,,
\end{split} \qquad \cM_\Delta^2 = \cM^2 - \left(\frac{8\pi \cQ}{ \sqrt{3} \kappa_4}\right)^2.
\label{eq:rs&rbgivenM&Q}
\end{equation}
Therefore, we have physical solutions only in the regime where
\begin{equation}
\sqrt{3}\kappa_4 \,\cM \geq 8 \pi \,\cQ \=8 \pi \, \sqrt{Q_m^2+Q_e^2}\,.
\label{eq:eq:RangeOur}
\end{equation}
As an illustration, one can compare to the two-charge Reissner-Norstr\"om solution. It would be only illustrative since it is not a solution of \eqref{eq:Action4d} due to the dilaton equation of motion. The metric and fields are given by
\begin{equation}
\begin{split}
ds_{RN}^2 &\= -\left(1 -  \frac{\kappa_4^2 \,\cM}{4 \pi \,r}+ \frac{\kappa_4^2 \,\cQ^2}{2 \,r^2}\right) \,dt^2 +\left(1 -  \frac{\kappa_4^2 \,\cM}{4 \pi \,r}+ \frac{\kappa_4^2 \,\cQ^2}{2 \,r^2}\right) ^{-1}dr^2 + r^2 d\Omega_2^2\,,\\
F^{(e)}_y &\= \frac{Q}{r^2}\, dr \wedge dt \,,\qquad F^{(m)} \= P\,\sin\theta\,d\theta \wedge d\phi\,,\qquad P^2 +Q^2 \= e^2 \,\cQ^2\,.
\end{split}
\label{eq:RN4d}
\end{equation}
The two horizons are then at
\begin{equation}
r^{RN}_\pm = \frac{\kappa_4^2\,\cM \pm \sqrt{\kappa_4^4 \cM^2-32 \pi^2 \cQ^2}}{8\pi}\,, \qquad \kappa_4 \cM \geq 4\sqrt{2} \pi \,\cQ \,.
\end{equation}
The range of our class of solutions is larger than the Reissner-Norstr\"om as implied by the cosmic censorship bound.

\subsection{Topological star}
\label{sec:TS4d/5d}

We now describe in details the different types of solutions contained in the class \eqref{eq:metric5d} and \eqref{eq:MaxField5d} with \eqref{eq:Metric5dGen} and \eqref{eq:Field5dGen}. We first assume that $r_\text{B} > r_\text{S}$. Thus, the outermost coordinate singularity corresponds to $r=r_\text{B}$ where the $y$-circle shrinks to zero size forming an end to spacetime. The horizon and the singularity are not part of the spacetime. The solutions are smooth geometries provided that the metric is regular at $r=r_B$ where the $y$-circle shrinks\footnote{The regularity outside the coordinate singularity, as the absence of closed timelike curves or the degeneracy of the $\phi$-circle at $\theta=0$ and $\pi$, is fairly straightforward from the form of the metric and gauge fields.}. The region near $r=r_\text{B}$ is best described by the local radial direction
\begin{equation}
\rho^2 \equi \frac{4 \,(r-r_\text{B})}{r_\text{B}-r_\text{S}}\,,
\end{equation}
and taking the limit $\rho \rightarrow 0$. The five-dimensional metric \eqref{eq:metric5d} with \eqref{eq:Metric5dGen} converges to
\begin{equation}
\begin{split}
ds^2_5 & \sim  - \frac{r_\text{B} - r_\text{S}}{r_\text{B}}\,dt^2 \+ r_\text{B}^2 \,\left[d\rho^2 + \frac{r_\text{B}-r_\text{S}}{4 \,r_\text{B}^3}\,\rho^2 \,dy^2+d\theta^2+\sin^2\theta\,d\phi^2\right]  \,.
\end{split}
\label{eq:bubblemetricBP}
\end{equation}
The $(\theta,\phi)$-subspace describes a round S$^2$ of radius $r_\text{B}$ while the $(\rho,y)$-subspace corresponds to a smooth origin of $\IR^2$ if 
\begin{equation}
R_y^2 \= \frac{4 r_\text{B}^3}{r_\text{B} - r_\text{S}}\,.
\label{eq:RyconstraintNocon}
\end{equation}
With this condition, the topology at the coordinate singularity corresponds to a \emph{bolt}, a smooth S$^2$ bubble sitting at an origin of a $\IR^2$. One needs to also check the regularity of the gauge fields at this locus. The regularity is satisfied if the components along the shrinking circle, $dy$, vanishes as $\rho \rightarrow 0$. Therefore, the magnetic field is straightforwardly regular and the electric gauge field goes to a constant value that can be gauged away as $\rho \rightarrow 0$. Both gauge fields are then regular. 

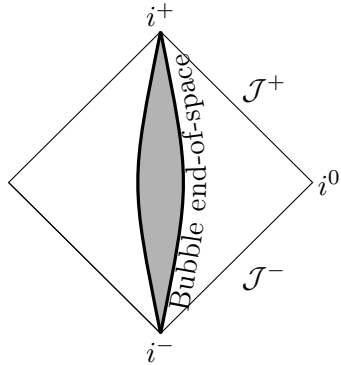
\begin{figure}[ht]
\setlength\belowcaptionskip{-0.5cm}
\begin{center}
\begin{tikzpicture}
\tikzmath{\scale=1;} 
\node (I)    at ( 2*\scale,0)   {};
\node (II)   at (-2*\scale,0) {};
\node (III)  at (0, 1.25*\scale) {};
\node (IV)   at (0,-1.25*\scale) {};

\path 
   (I) +(90:2*\scale)  coordinate[label=90:$i^+$] (Itop)
       +(-90:2*\scale) coordinate[label=-90:$i^-$] (Ibot)
       +(180:2*\scale) coordinate[label=180:$ $]  (Ileft)
       +(0:2*\scale)   coordinate[label=0:$i^0$] (Iright)
       ;
       \draw 
      (Ileft) --
      (Ibot) ;
\draw  (Ileft) --    (Itop) --node[midway, above right]    {$\cal{J}^+$}          (Iright) --    node[midway, below right]    {$\cal{J}^-$}    (Ibot) --     (Ileft) -- cycle;

\draw[very thick,black,postaction={decorate, decoration={text align={center},raise={-3mm},text along path, text={|\color{black}|Bubble end-of-space}}},name path=A] (Ibot) .. controls (2.4*\scale,0) .. (Itop) ;
\draw[very thick,black,name path=B] (Ibot) .. controls (1.6*\scale,0) .. (Itop) ;
\tikzfillbetween[of=A and B]{black, opacity=0.3};
\end{tikzpicture}
\caption{Penrose diagram of a topological star.}
\label{fig:PenroseBubble}
\end{center}
\end{figure}

Thus, we have constructed a solution that caps off smoothly before the horizon as a bolt supported by electromagnetic fluxes. The geometry is depicted by the Penrose diagram Fig.\ref{fig:PenroseBubble}. It has the same structure as a S$^1$ fibration over a four-dimensional Minkowski spacetime but the time slices end smoothly at $r=r_\text{B}$ as a $\IR^2\times$S$^2$. 

In four dimensions the mass, the electric and magnetic charges are given by \eqref{eq:ADMmass&charges4d}. The four-dimensional metric \eqref{eq:metric4d} has a curvature singularity at $r=r_\text{B}$ due to the dilaton blowing there.  In this region, one must use the five-dimensional theory to describe the solutions.  

The solutions can be characterized by the asymptotic data $(R_y,\cM,\cQ)$.  This will be useful when we study the phase space of objects for fixed mass and charges $(\cM,\cQ)$.  For the topological star, the regularity condition for the smooth shrinking of the $y$-circle, \eqref{eq:RyconstraintNocon}, can be recast as
\begin{equation}
r_\text{S} \= r_\text{B} \,\left(1 \- \frac{4\,r_\text{B}^2}{R_y^2} \right)\,.
\end{equation}
Since $r_\text{S}$ and $r_\text{B}$ must have the same sign in order to have physical charges \eqref{eq:Field5dGen}, this gives a significant bound to the size of the bubble
\begin{equation}
r_\text{B}^2 \,\leq\, \frac{R_y^2}{4}\,.
\end{equation}
Therefore, the topological star is at best the size of the extra dimension. To have an astrophysical bubble, one needs to have a large extra dimension which is incompatible with the real world physics.  Our ``topological star'' appellation is in that sense too optimistic and should be replaced by ``topological particle''. However, if we assume that the local metric around the bubble has a conical defect and has the topology of a smooth $\mathbb{Z}_k$ quotient over $\IR^2 \times$S$^2$, the constraint \eqref{eq:RyconstraintNocon} transforms to
\begin{equation}
R_y^2 \= \frac{4 r_\text{B}^3}{k^2\left( r_\text{B} - r_\text{S}\right)}\qquad \Leftrightarrow \qquad r_\text{S} \= r_\text{B} \,\left(1 \- \frac{4\,r_\text{B}^2}{k^2\,R_y^2} \right)\,.
\label{eq:Ryconical}
\end{equation}
Taking $k$ to be large allows for the construction of astrophysical bubbles with a size much larger than the extra dimensions.  

The orbifold parameter has important implication onto the classical degrees of freedom that can make topological stars.  Indeed, within string theory, spacetimes with conical singularities can be smoothed, and often describe localized objects.  An interesting question is if we can make sense of this singularity with the context of Einstein-Maxwell theory, and as interestingly if we can provide physical interpretation for their presences.  We will make the observation that the conical defect is induced by KK monopole charges at the poles of the bolt and discuss the possibility of blowing up those monopoles into smooth small bubbles.

\subsubsection{Conical defect and geometric transition}
\label{sec:ResolutionConical}

In this section, we show that the conical defect arises as two Kaluza-Klein monopoles of charge $k$ and $-k$ at the north and south poles of the bolt, $\theta=0$ and $\pi$, and that each one can be replaced by $k-1$ smooth bubbles without conical defects. More precisely, the spherical coordinates around the North and South poles are given by
\begin{equation}
\begin{split}
\rho_N& \equi \frac{R_y}{2} \left(2 r  -r_\text{B}-r_\text{S}- (r_\text{B}-r_\text{S})\,\cos \theta \right)\,,\quad \cos \vartheta_N \equi R_y\,\frac{r_\text{S}-r_\text{B} - (r_\text{B}+r_\text{S}-2r)\,\cos \theta}{2\rho_N}\,,\\
\rho_S &\equi \frac{R_y}{2} \left(2 r  -r_\text{B}-r_\text{S}+ (r_\text{B}-r_\text{S})\,\cos \theta \right)\,,\quad \cos \vartheta_S \equi R_y\,\frac{r_\text{B}-r_\text{S} - (r_\text{B}+r_\text{S}-2r)\,\cos \theta}{2\rho_S}\,.
\end{split} \nonumber
\end{equation} The time slices of the metric at the vicinity of the poles, $\rho_{N/S} \rightarrow 0$, are
\begin{equation}
\begin{split}
\ell \,ds_5^2& \underset{\rho_N\rightarrow 0}{\sim} \frac{k }{\rho_N} \,\left(d\rho_N^2 + \rho_N^2  \left( d\vartheta_N^2 + \sin^2\vartheta_N d\chi^2 \right)\right)+\frac{\rho_N}{k}\, \left(2d\varphi + k (\cos \vartheta_N-1) \,d\chi \right)^2, \\
\ell \,ds_5^2& \underset{\rho_S\rightarrow 0}{\sim} \frac{k}{\rho_S} \,\left(d\rho_S^2 + \rho_S^2  \left( d\vartheta_S^2 + \sin^2\vartheta_S d\chi^2 \right)\right)+\frac{\rho_S}{k}\, \left(2d\varphi - k(\cos \vartheta_N+1) \,d\chi \right)^2,
\end{split}
\label{eq:metricsatpoles}
\end{equation}
where we have also defined
\begin{equation}
\varphi \equi \frac{y}{R_y}\,,\qquad \chi \equi \phi + \frac{\varphi}{k}\, \qquad  \ell \equiv  2\sqrt{\frac{r_\text{B}-r_\text{S}}{r_\text{B}}}\,.
\label{eq:chiDef}
\end{equation}
The local metrics are then in the class of Gibbons-Hawking spaces \cite{GIBBONS1978430}\footnote{This transformation was motivated by a similar that was used to understand geometric descriptions of punctures in class $\mathcal{S}$ field theory constructions \cite{Bah:2018jrv,Bah:2019jts}.}. They are Hyper-K\"ahler spaces described as a S$^1$ fibration over a flat three-dimensional base with lattice of periodicities $\psi \to \psi +4\pi$ and $(\phi,\psi) \to (\psi,\phi) +(2\pi,2\pi)$, given by
\begin{equation}
ds_{GH}^2 \= V\,\left[d\rho^2 + \rho^2 (d\vartheta^2+\sin^2 \vartheta d\phi^2)\right] \+ V^{-1} \left( d\psi + A\right)^2\,, \quad \star_3 d_3 A \= \epsilon\,d_3 V\,, \quad \epsilon=\pm 1\,,  \nonumber
\end{equation}
where $\epsilon=1$ for the north pole and $\epsilon=-1$ for the south pole. This difference of sign corresponds to the sign of the KK charge with respect to the orientation of the base. For both local metrics \eqref{eq:metricsatpoles}, the Gibbons-Hawking space is sourced by a single Gibbons-Hawking center of charge $\epsilon k$ giving $\IR^4/\mathbb{Z}_k$.  It is well-know that such space can have a geometric transition to $k$ centers of charge $\epsilon$ as follows
\begin{equation}
V \= \frac{k}{\rho} \quad \longrightarrow \quad V \= \sum_{i=1}^k \frac{1}{\rho_i}\,,
\end{equation}
where $\rho_i$ defines the distance to the $i^\text{th}$ center. The new metric looks like $\IR^4$ at each center, $\IR^3\times$S$^1$ in between and the space is now free from conical defect. The new geometry defines $k-1$ new bubbles. We assume that these new bubbles exist in a limit where their characteristic size, $r_{\mu\text{B}}$, is much smaller than the original bubble of size $r_\text{B}$. The space far away from these bubbles but very close to the poles of the bolt will be still given by $V \sim \frac{k}{\rho}$ and the original bubble will look like a bolt with conical defect $k$. However, as soon as we get very close to the poles of the bolt the structure of the small bubbles will be manifest and will resolve the conical defect.

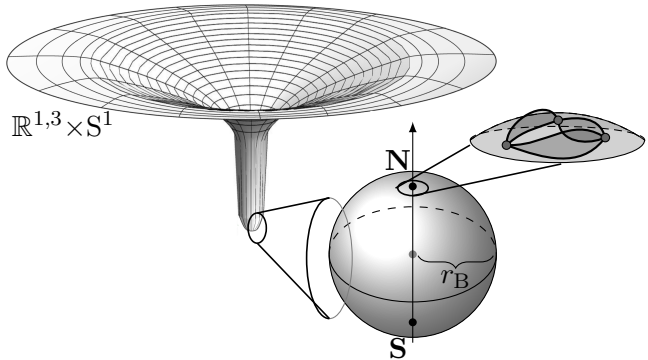
\begin{figure}[h]
\setlength\belowcaptionskip{-0.5cm}
\begin{center}
\begin{tikzpicture}
\tikzmath{\xxpic = -4; \ycap=-1.3;\sizefirstlens=0.2;\xleftfirstlens=0.05;\xrightfirstlens=1.;\yrightfirstlens=-0.4;\sizerightfirstlens=0.8;\xbubble=2.7;\ybubble=-0.1;\sizeseclens=0.23;\xleftseclens=\xbubble-0.6;\yleftseclens=0.53;\yrightseclens=1.15;\sizerightseclens=1;\xrightseclens=\xbubble+1.3;} 

\node[inner sep=0pt] (bubblefar) at (\xxpic,0)
    {\includegraphics[width=0.45\textwidth]{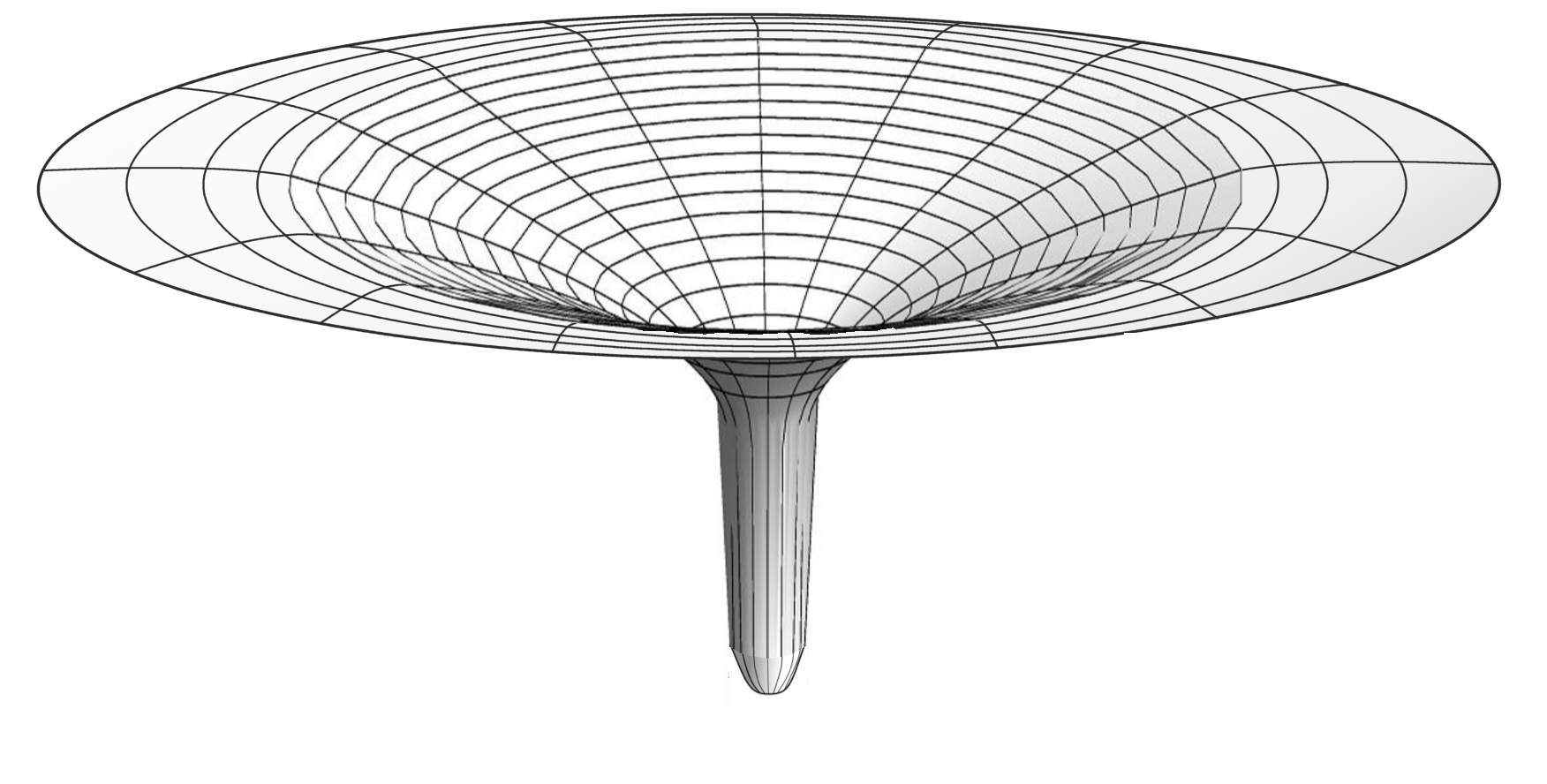}};
\node at (\xxpic-2.5,0.07)  {$\mathbb{R}^{1,3}\times$S$^1$} ;
    
\node (bubble) at (\xxpic+\xbubble,\ycap+\ybubble)
    {\bubble};

	\draw[color=gray] (\xxpic+\xrightfirstlens,\ycap+\yrightfirstlens-\sizerightfirstlens) arc (-90:90:0.3 and \sizerightfirstlens);
	\draw[semithick] (\xxpic+\xrightfirstlens,\ycap+\yrightfirstlens-\sizerightfirstlens) -- (\xxpic+\xleftfirstlens,\ycap-\sizefirstlens);
	\draw[semithick] (\xxpic+\xrightfirstlens,\ycap+\yrightfirstlens+\sizerightfirstlens) -- (\xxpic+\xleftfirstlens,\ycap+\sizefirstlens);
	\draw[semithick] (\xxpic+\xrightfirstlens,\ycap+\yrightfirstlens-\sizerightfirstlens) arc (270:90:0.3 and \sizerightfirstlens);
	\draw[semithick] (\xxpic+\xleftfirstlens,\ycap) ellipse (0.12 and 0.2);

	\draw[semithick,] (\xxpic+\xrightseclens-\sizerightseclens-0.15,\ycap+\yrightseclens-0.05) -- (\xxpic+\xleftseclens-\sizeseclens,\ycap+\yleftseclens);

		\draw[semithick] (\xxpic+\xrightseclens+0.05,\ycap+\yrightseclens-0.30) -- (\xxpic+\xleftseclens,\ycap+\yleftseclens-0.1);
	\draw[semithick] (\xxpic+\xleftseclens,\ycap+\yleftseclens) ellipse (0.2 and 0.1);


  \draw[rotate=90,fill=gray,fill opacity=0.35]
      (\ycap+\yleftseclens-1,\xxpic+\xrightseclens) ++(35:2cm)          
      arc (90:-90:-0.3 and {2*sin(35)}) 
      arc (-35:35:2cm);                 
 \draw[rotate=90,densely dashed,fill=gray,fill opacity=0.25]
   (\ycap+\yleftseclens-1,\xxpic+\xrightseclens) ++ (35:2cm)              
     arc (90:-90:0.18 and {2*sin(35)})  
    arc (-35:35:2cm);   
\coordinate[mark coordinate] (N) at  (\ycap+\yleftseclens+0.78,\xxpic+\xrightseclens+0.13);
\coordinate[mark coordinate] (B1) at  (\ycap+\yleftseclens+1.4,\xxpic+\xrightseclens-0.1);
\coordinate[mark coordinate] (B2) at  (\ycap+\yleftseclens+0.1,\xxpic+\xrightseclens-0.2);


\filldraw [fill={rgb:black,1.5;white,1}] (B1) circle (1.6pt);
\filldraw [fill={rgb:black,1.5;white,1}](B2)circle (1.6pt);
\filldraw [fill={rgb:black,1.5;white,1}](N)circle (1.6pt);

\draw[opacity=0] (N) +(0.3ex,0.6ex) node[above=0.25, left,opacity=1] {};
\draw[thick,name path= B1N1]    (B1) to[out=200,in=-80] (N);
\draw[thick,name path= B1N2]    (B1) to[out=110,in=10] (N);
\draw[thick,name path= B1B21]    (B1) to[out=230,in=-20] (B2);
\draw[thick,name path= B1B22]    (B1) to[out=150,in=00] (B2);
\draw[thick,name path= B2N1]    (B2) to[out=85,in=120] (N);
\draw[thick,name path= B2BN2]    (B2) to[out=40,in=200] (N);
\tikzfillbetween[of=B1N1 and B1N2]{black, opacity=0.17};
\tikzfillbetween[of=B1B21 and B1B22]{black, opacity=0.2};
\tikzfillbetween[of=B2N1 and B2BN2]{black, opacity=0.23};

\end{tikzpicture}
\caption{Schematic description of a generic topological star.}
\label{fig:Schematic}
\end{center}
\end{figure}

From that aspect, a generic topological-star geometry is made of two different scales as depicted in Fig.\ref{fig:Schematic}. We have a large scale $r_\text{B}$ corresponding to the size of the large S$^2$ bubble of order $k R_y$ and a small scale $r_{\mu\text{B}} \ll r_\text{B}$ corresponding to the scale for which the small bubbles at the poles make the geometry entirely smooth.

\vspace{0.3cm}

Note that we were a bit too fast in our argumentation since the local metrics at the poles \eqref{eq:metricsatpoles} are not strictly speaking in the class of Gibbons-Hawking space. The issue is coming from the periodicity of $\chi$ \eqref{eq:chiDef} which is not $2\pi$ but $2\pi/k$. This has to be understood as an artifact of taking a spherically symmetric probe limit for the muti-bubble system. First, in order to grow the additional Gibbons-Hawking bubbles we will need to turn on NUT charge along the $y$-circle which will break the spherical symmetry and change the periodicity constraints of the various circles.  Such monopoles will also add additional asymptotic charges that can provide further macroscopic data to characterize topological stars.  The analysis of the spherically symmetric system then suggests a larger phase space of smooth classical solutions.  To move towards a complete derivation of these solutions, we
should consider more general ansatz without spherical symmetry and allowing NUT charges along $y$. In section \ref{sec:Multi}, we will take the first step and consider axially symmetric systems of multi-bubble geometries.

\subsection{Black string}
\label{sec:BS4d/5d}

When $r_\text{S} \geq r_\text{B}$, the locus where the timelike Killing vector $\partial_t$ vanishes is now part of the spacetime. This degeneracy highlights an event horizon and the geometries correspond to black objects. We will see that for $r_\text{S} > r_\text{B}$, the solutions are non-extremal black strings and that for $r_\text{S}=r_\text{B}$, they correspond to extremal black strings.

\subsubsection{The non-extremal two-charge black string}
\label{sec:nonextremalBS4d/5d}

If $r_\text{S} > r_\text{B}$, the first coordinate singularity is a horizon at $r=r_\text{S}$. The topology of the horizon can be made manifest by considering the local metric with the radial direction
\begin{equation}
\rho^2 \equi \frac{4 \,(r-r_\text{S})}{r_\text{S}-r_\text{B}}\,r_\text{S}^2\,,
\end{equation}
and taking $\rho \rightarrow 0$. The five-dimensional metric \eqref{eq:metric5d} with \eqref{eq:Metric5dGen} leads to
\begin{equation}
ds_5^2 \,\sim \, - \frac{r_\text{S} -r_\text{B}}{4 r_\text{S}^3}\,\rho^2 \,dt^2 \+ d\rho^2 \+ r_\text{S}^2 \,\left( d\theta^2+ \sin^2 \theta \,d\phi^2 \right) + \frac{r_\text{S}-r_\text{B}}{r_\text{S}} \,dy^2\,.
\end{equation}
The horizon has a S$^2\times$S$^1$ topology and the radii of the S$^2$ and S$^1$ are $r_\text{S}$ and $\sqrt{\frac{r_\text{S}-r_\text{B}}{r_\text{S}}}R_y$ respectively. The Bekenstein-Hawking entropy gives
\begin{equation}
S \= \frac{8\pi^2}{\kappa_4^2}\,\sqrt{r_\text{S}^{3} \,(r_\text{S}-r_\text{B})}\,.
\label{eq:entropy4d}
\end{equation}
We have then defined a two-charge non-extremal black string that reduces to a two-charge non-extremal black hole in four dimensions given by \eqref{eq:metric4d} and \eqref{eq:MaxField4d}, with mass and charges  \eqref{eq:ADMmass&charges4d}. For more details on the thermodynamic properties, we refer to the exhaustive analysis \cite{Cvetic:2011dn} where a class of four-dimensional black holes has been studied in which our black string solutions are contained.

Unlike the topological star, the second coordinate singularity is part of the full spacetime and is in the interior of the black string. Rigorously, one should consider the Kruskal coordinates to extend the metric in the whole interior. However, the local metric at $r=r_\text{B}$ will look the same as if we directly consider the spherical coordinate $(r,\theta,\phi)$ which we will do. We therefore consider the coordinate
\begin{equation}
\rho^2 \equi \frac{4 (r-r_\text{B})}{r_\text{S} -r_\text{B}}\,,
\end{equation}
and the metric \eqref{eq:metric5d} behaves when $\rho \rightarrow 0$ as
\begin{equation}
ds_5^2 \,\sim\, \frac{r_\text{S}-r_\text{B}}{r_\text{B}}\,dt^2 \+ r_\text{B}^2 \left[-d\rho^2 + \frac{r_\text{S}-r_\text{B}}{4 r_\text{B}^3} \,\rho^2 dy^2 + d\Omega_2^2  \right]\,.
\end{equation}
The spacelike Killing vector $\partial_y$ shrinks, thereby defining a S$^2$ bubble behind the horizon. However, because $r$ is the timelike direction beyond the horizon, the $(\rho,y)$-subspace does not correspond to a $\IR^2$ as for the topological star but describes a quotient of $\IR^{1,1}$ by a boost. We consider again the $2\pi$-periodic angle $\varphi$ as $y=\varphi \, R_y$ and perform the transformation
\begin{equation}
T \equi -\rho\, \cosh \left(\gamma \,\varphi \right)\,,\qquad R \equi -\rho\, \sinh \left(\gamma \,\varphi \right)\,,\qquad \gamma^2 \equi \frac{r_\text{S}-r_\text{B}}{4 r_\text{B}^3}\,R_y^2\,.
\end{equation} 
The two-dimensional metric transforms to
\begin{equation}
-d\rho^2 + \frac{r_\text{S}-r_\text{B}}{4 r_\text{B}^3} \,\rho^2 dy^2 \= - dT^2 \+ dR^2\,.
\end{equation}
From the expressions of $T$ and $R$, the $(\rho,\varphi)$-subspace describes only a lower cone in the two-dimensional $(T,R)$ Minkowski space. The apex of the cone corresponds to the coordinate singularity where the $y$-circle shrinks. The extension to negative values of $\rho$ defines the upper cone connected at the apex. The two cones form the Milne region of a Misner space defined as the quotient of $\IR^{1,1}$ by the boost $\gamma$ \cite{MilneBook,MisnerBook}. Milne space are smooth and free from closed timelike curves. It is well studied in cosmology as a smooth transition from a Big Crunch to a Big Bang \cite{Khoury:2001bz}. It is not surprising to find such a topology for the black string solutions. In the interior, $r$ is the time line and $r=r_\text{B}$ can be then compared to a Big Crunch inside the black string due to the degeneracy of the $y$-circle. The fact that the Milne space can be extended to the other part of the cone as a Big Bang could describes a new class of possible wormholes, and deserves further study. However, it should be noted that geodesics that have momentum along $y$ or string probe along $y$ are singular at this location, which could make the traversability analysis subtle \cite{Horowitz:1990ap,Berkooz:2004re}.  The wormholes would be at best stable for probes with energies bellow the KK scales of the external spacetime.

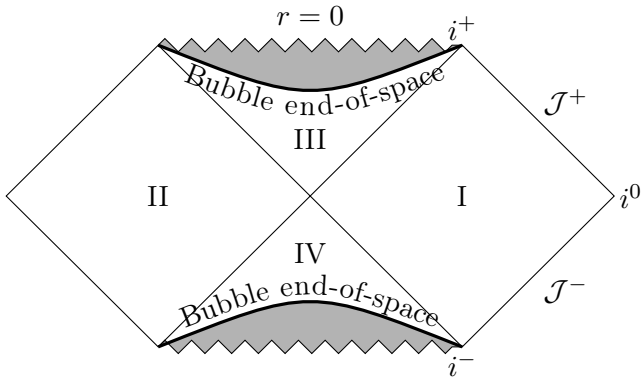
\begin{figure}[ht]
\setlength\belowcaptionskip{-0.7cm}
\begin{center}
\begin{tikzpicture}
\tikzmath{\scale=1;} 
\node (I)    at ( 2*\scale,0)   {I};
\node (II)   at (-2*\scale,0)   {II};
\node (III)  at (0, 0.75*\scale) {III};
\node (IV)   at (0,-0.75*\scale) {IV};

\path  
  (II) +(90:2*\scale)  coordinate  (IItop)
       +(-90:2*\scale) coordinate(IIbot)
       +(0:2*\scale)   coordinate                  (IIright)
       +(180:2*\scale) coordinate (IIleft)
       ;

\path 
   (I) +(90:2*\scale)  coordinate[label=90:$i^+$] (Itop)
       +(-90:2*\scale) coordinate[label=-90:$i^-$] (Ibot)
       +(180:2*\scale) coordinate (Ileft)
       +(0:2*\scale)   coordinate[label=0:$i^0$] (Iright)
       ;
       \draw (IIleft) -- (IItop) --   (IIright) --  (IIbot) -- (IIleft) -- cycle;
\draw  (Ileft) -- (Itop) --node[midway, above right]    {$\cal{J}^+$}  (Iright) --    node[midway, below right]    {$\cal{J}^-$}  (Ibot) -- (Ileft) -- cycle;

\draw[decorate,decoration=zigzag,name path=A] (IItop) -- (Itop)
      node[midway, above, inner sep=2mm] {$r=0$};

\draw[decorate,decoration=zigzag,name path=C] (IIbot) -- (Ibot)
      node[midway, below, inner sep=2mm] {};

\draw[very thick,black,postaction={decorate, decoration={text align={center},raise={-3.25mm},text along path, text={|\color{black}| Bubble end-of-space}}},name path=B] (IItop) .. controls (0,1.2*\scale) .. (Itop) ;
\tikzfillbetween[of=A and B]{black, opacity=0.3};
\draw[very thick,black,postaction={decorate, decoration={text align={center},raise={1mm},text along path, text={|\color{black}|Bubble end-of-space}}},name path=D] (IIbot) .. controls (0,-1.2*\scale) .. (Ibot) ;
\tikzfillbetween[of=C and D]{black, opacity=0.3};
\end{tikzpicture}
\caption{Penrose diagram of the black string.}
\label{fig:PenroseBH}
\end{center}
\end{figure}

The causal structure of the spacetime is depicted by the Penrose diagram Fig.\ref{fig:PenroseBH}. Even if it is not clear how the different regions maybe connected when crossing the bubble, the black-hole singularity has been hidden by a S$^2$ bubble at the origin of a Milne space without curvature singularity.

\subsubsection{The extremal two-charge black string}
\label{sec:extremalBS4d/5d}

We now study the last type of solutions obtained when $r_\text{S}=r_\text{B}=m$. The five-dimensional solution \eqref{eq:metric5d} with \eqref{eq:Metric5dGen} and \eqref{eq:Field5dGen} is now given by 
\begin{align}
ds^2_5 &\=  \left(1+ \frac{m}{\rho} \right)^{-1}\,\left(-dt^2 + dy^2 \right)+ \left(1+ \frac{m}{\rho} \right)^2\left[d\rho^2 + \rho^2 \,d\Omega_2^2 \right]\,,\\
F^{(e)} &\= \frac{Q}{m} d\left(\left(1 + \frac{m}{\rho} \right)^{-1}\right) \wedge dt \wedge dy\,,\qquad F^{(m)} = P\,\sin\theta\,d\theta \wedge d\phi\,,  \qquad P^2 +Q^2 = \frac{3 m^2}{2 \kappa_5^2}\,, \nonumber
\end{align} 
where we have defined the isotropic coordinate $\rho = r-m$. We recognize a two-charge extremal black string. At $\rho=0$, both $\partial_y$ and $\partial_t$ Killing vectors degenerate defining an AdS$_3\times$S$^2$ near-horizon geometry. Considering $\rho =4 m^3 \bar{\rho}^2$ and taking the limit $\bar{\rho}\rightarrow 0$, we have
\begin{equation}
ds^2_5 \sim 4 m^2 \left[\frac{d\bar{\rho}^2}{\bar{\rho}^2} + \bar{\rho}^2 \left( -dt^2+dy^2\right)\right]  \+ m^2\,d\Omega_2^2 \,.
\end{equation}
The radius of AdS$_3$ and S$^2$ are $2m$ and $m$ respectively. Due to the degeneracy of the $y$ circle, the horizon area vanishes and the black string does not have a macroscopic horizon. Those solutions are well-understood as D1-D5-KKm extremal black holes when embedded in type IIB string theory as we will discuss in section \ref{sec:ExtensionDdim}.

\subsection{Phase space}
\label{sec:phasespace}

The class of spherically symmetric solutions contains two types of solutions: the smooth topological stars of section \ref{sec:TS4d/5d}, the black strings of section \ref{sec:BS4d/5d}. Upon Kaluza-Klein reduction, they correspond to four-dimensional solutions as discussed in section \ref{eq:reduction4d} with mass $\cM$ and electric and magnetic charges $(Q_e,Q_m)$ given by \eqref{eq:ADMmass&charges4d}. In this section, we aim to describe the phase space of solutions for given asymptotic quantities. In addition to the four-dimensional mass and charges, one must also consider $R_y$ as a fixed quantity. If $R_y$ is free for the black string, it constrains the topological stars as in \eqref{eq:RyconstraintNocon} or \eqref{eq:Ryconical}. Moreover, the electromagnetic duality induces a degree of freedom between the magnetic and electric charge, and only the ``total'' charge, $\cQ = \sqrt{Q_e^2+Q_m^2}$ \eqref{eq:ADMmass&charges4d}, is fixed.

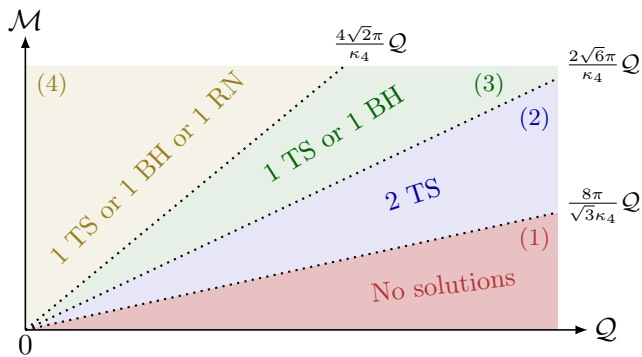
\begin{figure}[ht]
\setlength\belowcaptionskip{-0.7cm}
\begin{center}
 \usetikzlibrary{snakes}
 \definecolor{myred}{rgb}{0.7,0.2,0.2}
 \definecolor{mygreen}{rgb}{0.1,0.6,0.3}
\definecolor{myblue}{rgb}{0.2,0.2,0.7}
\definecolor{myyellow}{rgb}{0.6,0.5,0.1}
\begin{tikzpicture}
\tikzmath{\xxm = 0; \xxM =3.5; \yym=0; \yyM = 7;\xxf=0.2*3/4;\xxs=6.55*3/4;\dashinter=2;\step=0.7*3/4;\npos=1*3/4;\Qpos=7.5*3/4;\sl=4.5;\slb=2.1;\slc=1.2;} 
\path[fill=myred!30] (\yym,\xxm ) rectangle (\yyM,\xxM);
\path[fill=darkblue!10] (\yym,\yyM/\sl) rectangle (\yyM,\xxM);
\fill [darkblue!10, domain=\yyM/\sl:\yym, variable=\x]
      (0,\xxM) -- plot ({\sl*(\x)}, {\x}) --  (0,0) -- cycle;
\path[fill=darkgreen!10] (\yym,\yyM/\slb) rectangle (\yyM,\xxM);
\fill [darkgreen!10, domain=\yyM/\slb:\yym, variable=\x]
      (0,\xxM) -- plot ({\slb*(\x)}, {\x}) --  (0,0) -- cycle;
\path[fill=darkgreen!10] (\yym,\yyM/\slb) rectangle (\yyM,\xxM);
\fill [darkgreen!10, domain=\yyM/\slb:\yym, variable=\x]
      (0,\xxM) -- plot ({\slb*(\x)}, {\x}) --  (0,0) -- cycle;
\fill [myyellow!10, domain=\xxM:\yym, variable=\x]
      (0,\xxM) -- plot ({\slc*(\x)}, {\x}) --  (0,0) -- cycle;

\draw[scale=1,domain=\yym:\yyM/\sl,smooth,variable=\x,dotted,thick=2] plot ({\sl*(\x)},{\x});
\draw[scale=1,domain=\yym:\yyM/\slb,smooth,variable=\x,dotted,thick=2] plot ({\slb*(\x)},{\x});
\draw[scale=1,domain=\yym:\xxM,smooth,variable=\x,dotted,thick=2] plot ({\slc*(\x)},{\x});
\draw[black]	(\yyM+0.6,\yyM/\sl+0.4/\sl) node{{\footnotesize $\frac{8 \pi}{\sqrt{3}\kappa_4} \cQ$}};
\draw[black]	(\yyM+0.6,\yyM/\slb+0.4/\slb) node{{\footnotesize $\frac{2\sqrt{6} \pi}{\kappa_4} \cQ$}};
\draw[black]	(\xxM*\slc+0.4/\slc,\xxM+0.3) node{{\footnotesize $\frac{4\sqrt{2} \pi}{\kappa_4} \cQ$}};

\draw[myred]	(\yyM-1.5,0.6) node[align=center,rotate=7]{{\small No solutions}};
\draw[darkblue]	(\yyM-1.9,1.8) node[align=center,rotate=18]{{\small 2 TS}};
\draw[darkgreen]	(\yyM-2.95,2.6) node[align=center,rotate=32]{{\small 1 TS or 1 BH}};
\draw[myyellow]	(1.6,\xxM-1.35) node[align=center,rotate=45]{{\small 1 TS or 1 BH or 1 RN}};
\draw[myred]	(\yyM-0.3,1.2) node[align=center]{{\footnotesize (1)}};
\draw[darkblue]	(\yyM-0.3,2.8) node[align=center]{{\footnotesize (2)}};
\draw[darkgreen]	(\yyM-0.9,3.25) node[align=center]{{\footnotesize (3)}};
\draw[myyellow]	(0.35,3.25) node[align=center]{{\footnotesize (4)}};

\draw[->,semithick] (\yym,\xxm) -- (\yym,\xxM+0.4) node[anchor=south] {\textbf{$\cM$}};
\draw[->,semithick] (\yym,\xxm) -- (\yyM+0.4,\xxm) node[anchor=west] {\textbf{$\cQ$  }};
\draw (0,\xxm) node[anchor=north]{$0$};
\end{tikzpicture}
\caption{Phase space of spherically symmetric solutions. ``TS'', ``BH'' or ``RN'' stand for the topological star of section \ref{sec:TS4d/5d}, the black string of section \ref{sec:BS4d/5d} and the magnetic Reissner-Nordstr\"om \eqref{eq:RN4d} respectively . The graph should be read as ``1 TS or 1 BH''$=$ one topological star solution and one black string solution for the same $(\cM,\cQ)$.}
\label{fig:phasespace}
\end{center}
\end{figure}

Let us first consider the different regimes for given $(\cM,\cQ)$. This phase space is depicted in Fig.\ref{fig:phasespace}. By inverting the expressions of $\cM$ and $\cQ$ \eqref{eq:ADMmass&charges4d}, we obtain two solutions of $(r_\text{S},r_\text{B})$ given by \eqref{eq:rs&rbgivenM&Q}. 
\begin{itemize}
\item[•] \underline{For $\sqrt{3}\,\kappa_4\,\cM < 8 \pi\,\cQ$: }

In this regime, the solutions $(r_\text{S}^{(i)},r_\text{B}^{(i)})_{i=1,2}$ \eqref{eq:rs&rbgivenM&Q} are not real. Therefore, the corresponding solutions are unphysical and no solutions in our class exist. This corresponds to the regime (1) in Fig.\ref{fig:phasespace}.

\item[•] \underline{For $ \frac{8 \pi}{\sqrt{3}}\,\cQ\leq \kappa_4\,\cM < 2 \sqrt{6} \pi \,\cQ$: }

In this regime, both solutions $(r_\text{S}^{(i)},r_\text{B}^{(i)})_{i=1,2}$ are real and $r_\text{S}^{(i)} < r_\text{B}^{(i)}$. Therefore, they correspond to topological stars. However, if we consider $R_y$ fixed and if we allow for a conical defect \eqref{eq:Ryconical}, we have an extra quantization constraint for each solution:
\begin{align}
k^{(1)} \= \frac{\kappa_4}{3 \sqrt{6} \pi R_y} \, \frac{\left(3 \kappa_4 \cM + \sqrt{(3\kappa_4 \cM)^2 -192 \pi^2 \cQ^2} \right)^{\frac{3}{2}}}{\left(\kappa_4 \cM + \sqrt{(3\kappa_4 \cM)^2 -192 \pi^2 \cQ^2} \right)^{\frac{1}{2}}} \,\in \, \mathbb{Z}_+\,,\label{eq:quantization1stSol}\\
k^{(2)} \= \frac{\kappa_4}{3 \sqrt{6} \pi R_y} \, \frac{\left(3 \kappa_4 \cM - \sqrt{(3\kappa_4 \cM)^2 -192 \pi^2 \cQ^2} \right)^{\frac{3}{2}}}{\left(\kappa_4 \cM - \sqrt{(3\kappa_4 \cM)^2 -192 \pi^2 \cQ^2} \right)^{\frac{1}{2}}} \,\in \, \mathbb{Z}_+\,.
\end{align}
Therefore, this regime of mass and charges, depicted by the region (2) in Fig.\ref{fig:phasespace}, should not be considered as a continuum of bubble solutions but as two discrete lattices of solutions for which each node corresponds to a topological star with a specific orbifold parameter.

\item[•] \underline{For $ 2 \sqrt{6} \pi \,\cQ \leq \kappa_4\,\cM$: }

When approaching the line $2 \sqrt{6} \pi \,\cQ = \kappa_4\,\cM$ from the region (2), the second topological star has $r_\text{S}^{(2)} \rightarrow r_\text{B}^{(2)}$. On the line, the solution then becomes the extremal black string of section \ref{sec:extremalBS4d/5d}. Therefore, in the regime $ 2 \sqrt{6} \pi \,\cQ \leq \kappa_4\,\cM$, depicted by the regions (3) and (4) in Fig. \ref{fig:phasespace}, the first solution corresponds to a topological star while the second corresponds to a black string. Once again, the topological stars exist on a lattice given by the quantization \ref{eq:quantization1stSol}. On this lattice, both topological stars and black strings exist for the same mass and charges. Moreover, from the expressions \eqref{eq:rs&rbgivenM&Q}, we have
\begin{equation}
r_\text{B}^{(1)} \= 2 \,r_\text{S}^{(2)}\,,
\end{equation}
and therefore the spacetime caps off for the topological star at a distance twice larger than where the horizon of its corresponding black string is. The topological star has a S$^2$ topology while the black string has a S$^2\times$S$^1$ topology in five dimensions which renders the comparison of size subtle. However, one can still compare the size of the S$^2$. In that sense, the size of the topological star is also twice bigger than the size of the black string.

The region (4) in Fig.\ref{fig:phasespace} corresponds to the domain of validity of the two-charge Reissner-N\"ordstrom in four dimensions \eqref{eq:RN4d}. We remind that such a solution is not a solution of our theory \eqref{eq:Action4d} and should be considered as an illustrative comparison with usual GR objects. In this regime, topological stars, black strings and Reissner-N\"ordstrom exist for the same mass and charges. From a four-dimensional perspective, one can compare the size of the black string solution with the Reissner-Nordstr\"om, which means to compare the radius of the S$^2$ at the horizons. We essentially find that the size matches when $\cM \gg \cQ$ and the difference is maximal when $\kappa_4 \cM \sim 4 \sqrt{2}\pi \cQ$ where the size of the Reissner-N\"ordstrom is twice smaller than the black string.

\end{itemize}

The take-away message in the context of black hole microstates is that, even if we restrict to spherically symmetric solutions, we have smooth five-dimensional bubble solutions that have the same charge and mass as the non-extremal four-dimensional black holes. It might be appropriate to interpret this solution as a microstate of the thermal ensemble given by the Bekenstein-Hawking entropy. It is already surprising that such a state can be built with spherical symmetry. Moreover, because the topological star is twice as large as the size of its corresponding black hole, it is a rather atypical state. This is a common story for microstate geometries. Having solutions that scale very close to the horizon requires to consider multi-bubble solutions, that is, to break the spherical symmetry.

\section{Generalized two-charge Weyl solutions in five dimensions}
\label{sec:Multi}

In the previous section, smooth solutions that follow the spirit of microstate geometries have been constructed with the minimum complexity, that is within a class of spherically symmetric solutions of Einstein-Maxwell theory in five dimensions. We aim to extend to more typical bubbling geometries in the same theory where the solutions can have multiple sources in the three-dimensional base space, angular momentum, NUT charges along $\phi$ and momentum along $y$. Such solutions will follow the most general ansatz
\begin{equation}
\begin{split}
ds_5^2 & \= - f_\text{S} \,\left( dt + \omega \,d\phi \right)^2 + f_\text{B} \,\left( dy+ A_t \,dt +A_\phi \,d\phi \right)^2 + h_r \,dr^2 + h_\theta \,d\theta^2 + h_\phi \,d\phi^2\,,\\
F^{(m)} &\= dH \wedge d\phi\,,\qquad F^{(e)} \= d\left[Z \, \left( dt + \omega \,d\phi \right) \wedge \left( dy+ A_t \,dt +A_\phi \,d\phi \right)\right]\,.
\end{split}
\end{equation}
It will be also very interesting to add Chern-Simons terms to the Maxwell-Einstein theory which will give non-trivial contribution with this present ansatz and which is known to be important for constructing microstructure at the vicinity of the would-be horizon \cite{Gibbons:2013tqa}.

The price to pay for this ansatz will be to have highly non-linear equations of motion for which closed-form solutions might be very complicated to find. Instead of jumping directly into the fire, we will allow one degree of complication at a time and try to see if closed-form solutions can be derived. 

In section \ref{sec:ResolutionConical}, we have seen the importance of blowing up Gibbons-Hawking bubbles at the poles of the spherically symmetric topological stars in order to give them a macroscopic size and to classically resolve their conical defect. This requires to consider at least axisymmetric bubble configurations and additional NUT charges. As a first step, we will apply the well-known Weyl formalism \cite{Weyl:book} which will allow to find axisymmetric solutions that are asymptotic to four-dimensional flat spacetime times an extra compact dimension. 

The Weyl formalism has been very successful to test the uniqueness theorem in four-dimensional Einstein theory (see \cite{Bonnor} for a short review). We can cite the Israel-Khan solutions \cite{Israel1964,Gibbons:1979nf} which consist in a chain of Schwarzschild black holes that are either supported from collapse by \emph{struts}, that are strings of negative tension, or by the fact that the chain is made of a succession of positive-mass and negative-mass objects. The rotating generalization has been initiated by Kramer and Neugebauer \cite{KRAMER1980259} who found the solutions corresponding to a superposition of two Kerr black holes supported by struts.  However, they can be balanced by the spin-spin interaction when their horizons touch. Generalization to multi-parameter solutions of two Kerr-NUTs can be found in \cite{Chen:2015vva} and references thereof. Ernst's extensions to four-dimensional solutions with a U(1) gauge field are in \cite{Ernst:1967wx,Ernst:1967by}.

The extension to vacuum multi-body solutions in five dimensions for which one dimension is an internal compact circle has been done in \cite{Costa:2000kf,Emparan:2001wk,Elvang:2002br,Emparan:2008eg,Charmousis:2003wm}. The much richer nature of gravity with extra dimensions has highlighted a larger diversity of solutions. Depending if we source along the time direction or along the $y$ direction we have either black string centers or bubble-of-nothing centers respectively. As in four dimensions, all solutions require struts between the objects to prevent from its collapse.

We can consider generalizations of our construction for multi-body solutions using the Weyl formalism. The difference with vacuum solutions \cite{Costa:2000kf,Emparan:2001wk,Elvang:2002br,Emparan:2008eg,Charmousis:2003wm} is that the addition of electric and magnetic gauge fields makes the equations of motion non-linear. Nonetheless we can solve them and generalize the single-body solutions of the previous section into a class of multi-body solutions made of two-charge black strings and topological stars on an axis. This class of charged solutions is novel and has a very large phase space due to the non-trivial possibilities from the backreactions of the gauge fields. 

As in vacuum, the solutions will be given according to two functions that solve a Laplace equation, but the interaction with the gauge fields makes the functions appear in the metric in non-perturbative and highly non-trivial ways. The functions can be sourced on the $z$-axis by rod sources, point sources etc. We will treat only the solutions of rod sources with a specific type of gauge-field backreaction. For these specific examples, the solutions admit struts between the centers as for vacuum solutions, but it is possible that different type of sources and gauge-field configurations might resolve this issue. Moreover, we also expect that the addition of angular momentum or NUT charges allowing Gibbons-Hawking type of sources will also decrease the needs of struts to sustain the geometry. This will be the subject of further study.

\subsection{Ansatz and equations}
\label{sec:EOMaxisym}

We consider axisymmetric solutions of the five-dimensional Einstein-Maxwell theory \eqref{eq:ActionGen}.

\noindent The ansatz of metric and field strengths in the Weyl form are
\begin{equation}
\begin{split}
ds^2 &\= -f_\text{S}(\rho,z) \,dt^2 + f_\text{B}(\rho,z) \, dy^2 +h(\rho,z)  \left[ e^{2\nu(\rho,z)} \left(d\rho^2 + dz^2 \right) + \rho^2 d\phi^2 \right]\,,\\
F^{(m)} &\= d H(\rho,z) \wedge d\phi\,,\qquad F^{(e)} \= dZ(\rho,z) \wedge dt\wedge dy\,,
\end{split}
\end{equation}
where $(\rho,z,\phi)$ are the cylindrical coordinates of the three-dimensional base space and $(f_\text{S},f_\text{B},h,\nu)$ are scalars that are functions of $\rho$ and $z$ and that must solve the Maxwell and Einstein equations \eqref{eq:EOMgen}. We define the Laplace operator on the three-dimensional base as
\begin{equation}
\cL \equi \frac{1}{\rho} \,\partial_\rho \left( \rho \partial_\rho \right) \+ \partial_z^2\,.
\label{eq:LaplaceOperator}
\end{equation}

The derivations of the equations of motion are given in the Appendix \ref{app:WeylSol}.  The equations of motion simplify greatly with the field redefinition
\begin{align}
V(\rho,z) \equi \rho\,\sqrt{f_\text{S}\,f_\text{B}\, h}\,,\quad W_0(\rho,z) \equi \sqrt{\frac{f_\text{S}}{f_\text{B}}}\,,\quad W_\text{GF}(\rho,z) \= \frac{1}{\sqrt{f_\text{S}\,f_\text{B}}}\,, \quad \nu \equi \nu_0 + \nu_\text{GF}\,. \nonumber
\end{align}
The solutions are then given by
\begin{align}
ds^2 &\= W_\text{GF}^{-1} \left[ -W_0 \,dt^2 +W_0^{-1} \, dy^2 \right]+ W_\text{GF}^2 \left(\frac{V}{\rho} \right)^2 \left[ e^{2(\nu_0 + \nu_\text{GF})} \left(d\rho^2 + dz^2 \right) + \rho^2 d\phi^2 \right]\,,\nonumber \\
F^{(m)} &\= d H(\rho,z) \wedge d\phi\,,\qquad F^{(e)} \= dZ(\rho,z) \wedge dt\wedge dy\,.
\label{eq:metricAxisym}
\end{align}
This redefinition is appropriate to pair the scalars into physically meaningful quantities. As we will see, the pair $(W_0,\nu_0)$ will be completely independent of the gauge fields, they correspond to the purely ``massive'' warp factors and will be governed by the same equations as in vacuum. The pair $(W_\text{GF},\nu_\text{GF})$ corresponds to the gauge-field contribution, that is why we use the index ``GF'' as gauge field. They will be non-trivially sourced by and coupled with $H$ and $Z$. Note that the dependence on the gauge fields appears as an overall warp factor along the $(t,y)$ direction. Moreover, the redefinition makes the double Wick rotation symmetry manifest since the class of solutions is symmetric under 
\begin{equation}
(t,y,W_0) \,\rightarrow\, (i\,y,i\,t, W_0^{-1})\,.
\end{equation}

Before listing the equations, we use electromagnetic duality to fix $Z$ according to $H$ by imposing $\star_5 F^{(e)} \propto F^{(m)}$ to obtain
\begin{equation}
dZ \= \frac{q}{V W_\text{GF}^2} \,\star_2 dH\,.
\end{equation}
The parameter $q$ represents a charge ratio between the total electric and magnetic charges, and $\star_2$ corresponds to the Hodge star operator on the flat $(\rho,z)$-subspace.

We decompose the equations of motion into four distinct layers:
\begin{itemize}
\item[•] \underline{The zeroth layer:}

The zeroth layer fixes the potential $V$ as
\begin{equation}
\partial_\rho^2 V \+ \partial_z^2 V \= 0\,.
\label{eq:zerothlayer}
\end{equation}

\item[•] \underline{The mass layer:}

The ``mass'' layer corresponds to the equation that governs the purely massive warp factor,
\begin{equation}
\partial_\rho \left( V \,\partial_\rho \log W_0 \right) \+ \partial_z \left( V \,\partial_z \log W_0 \right) \= 0\,.
\label{eq:masslayer}
\end{equation}

\item[•] \underline{The Maxwell layer:}

The ``Maxwell'' layer corresponds to the equations of motion for the gauge fields and their backreaction on the spacetime
\begin{align}
&\partial_\rho \left(\frac{1}{V\,W_\text{GF}^2} \partial_\rho H \right) \+ \partial_z \left(\frac{1}{V\,W_\text{GF}^2} \partial_z H \right) \= 0\,, \label{eq:Maxwelllayer} \\
&V\,W_\text{GF}^2 \,\left[\partial_\rho \left( V \,\partial_\rho \log W_\text{GF} \right) \+ \partial_z \left( V \,\partial_z \log W_\text{GF} \right)  \right] \= - \frac{2(1+q^2)\kappa_5^2}{3} \,\left[ \left(\partial_\rho H\right)^2 +\left(\partial_z  H\right)^2 \right]\,. \nonumber
\end{align}
\item[•] \underline{The  base layer:}

The base layer corresponds to simple integral equations for the last scalars $(\nu_0,\nu_\text{GF})$. They are non-trivially sourced by the other fields and fix the nature of the three-dimensional base,
\begin{equation}
\begin{split}
\partial_\rho \log V ~\partial_{z} \nu_{0} \+ \partial_z \log V ~\partial_\rho \nu_0&\=\mathcal{S}_{z}^{(0)}\left(V,W_{0}\right), \\
\partial_\rho \log V ~\partial_{\rho} \nu_{0} \- \partial_z \log V ~\partial_z \nu_0 &\=\mathcal{S}_{\rho}^{(0)}\left(V,W_{0}\right)\,,\\
\partial_\rho \log V ~\partial_{z} \nu_{\text{GF}}\+ \partial_z \log V ~\partial_\rho \nu_{\text{GF}}&\=\mathcal{S}_{z}^{(\text{GF})}\left(V,W_{\text{GF}}\right), \\
 \partial_\rho \log V ~\partial_{\rho} \nu_{\text{GF}} \- \partial_z \log V~ \partial_z \nu_{\text{GF}} &\=\mathcal{S}_{\rho}^{(\text{GF})}\left(V,W_{\text{GF}}\right)\,,
\end{split}
\label{eq:baselayer}
\end{equation}
where the explicit forms of the source functions are given in \eqref{eq:EqfornuApp}.
\end{itemize}

We see that one can solve  the equations almost linearly layer by layer. Except for the non-linear coupled Maxwell layer, the other equations can be solved one by one as linear partial differential equations.

\subsection{The class of axisymmetric solutions}

The zeroth layer of equation is an irrelevant constraint that can be fixed by coordinate transformation, usually referred as ``Weyl's canonical coordinates'' (see \cite{Emparan:2001wk} for all the details of this transformation). In other words, one can consider without restriction that
\begin{equation}
V \= \rho\,.
\end{equation}

The Weyl's canonical coordinates have the benefit to transform the equations for the warp factors to Laplace equations in the three-dimensional base. The mass layer can be solved with purely mass sources,
\begin{equation}
\cL \left(\log W_0 \right) \= 0\,,
\label{eq:EOMW0}
\end{equation}
where $\cL$ is the Laplace operator \eqref{eq:LaplaceOperator}. This equation has well-known solutions for rod sources or point sources \cite{Israel1964,Emparan:2001wk,Bonnor}.

The complications arise when solving the Maxwell layer since the presence of gauge field makes the Maxwell equation and the Laplace equation for $\log W_\text{GF}$ to be two coupled non-linear partial differential equations. To find closed-form solutions we split the Maxwell equation in two pieces that will be taken to be zero,
\begin{equation}
\begin{split}
 \rho\, \partial_\rho \left(\dfrac{1}{\rho} \partial_\rho H \right) \+ \partial_z^2 H\= 0 \,, \\
 \partial_\rho W_\text{GF}^{-2}\,\partial_\rho H \+  \partial_z W_\text{GF}^{-2}\,\partial_z H \=0\,,
\end{split} \, \Rightarrow \,\,\, \partial_\rho \left(\frac{1}{\rho\,W_\text{GF}^2} \partial_\rho H \right) \+ \partial_z \left(\frac{1}{\rho\,W_\text{GF}^2} \partial_z H \right) \= 0.
\label{eq:Maxwellrestriction}
\end{equation}
This is the only assumption we make to solve the system of equations.  This is motivated by how the spherically symmetric solutions  in section \ref{sec:class4d} solve the system.  

The solutions of the Maxwell layer with fluxes turned on are given by an arbitrary function $K(\rho,z)$ that solves the following Laplace equation
\begin{equation}
\cL \left(\frac{1}{\rho} \,\partial_\rho K \right) \= 0\,.
\label{eq:Keq}
\end{equation}
See the Appendix \ref{app:closedformSol} for details.  The fields are given as\footnote{At first glance, we have an integration freedom $K \to K + f(z)$ that does not change the nature of $\rho^{-1} \partial_\rho K$ but changes the gauge field given by $H$. This integration freedom $f(z)$ must be actually fixed such that $\rho \partial_\rho \left( \frac{1}{\rho} \partial_\rho K \right) + \partial_z^2 K =0$, which can always be found from solutions satisfying \eqref{eq:Keq}. See Appendix \ref{app:closedformSol} for more details.}
\begin{equation}
W^2_{\text{GF}} \=- \frac{\cosh^2 \left(\frac{a}{\rho}\, \partial_\rho K +b\right)}{a^2}\,,\qquad H \= \sqrt{\frac{3}{2(1+q^2)\kappa_5^2}}\,\partial_z K\,,\qquad (a,b)\in \mathbb{C}\,.
\label{eq:SolWGF&H}
\end{equation}
Vacuum solutions with the fluxes turned off are also given in terms of a function $K$ that satisfy \eqref{eq:Keq}, but the fields would take the much simpler form
\begin{equation}
W^2_{\text{GF}} \=\exp \left[ \frac{1}{\rho} \,\partial_\rho K \right]\,,\qquad H \=0\,.
\end{equation}
We observe that the addition of the gauge fields keeps the same structure of sources as with vacuum solutions, but they change the metric in a non-perturbative way. The free parameters $(a,b)$ in \eqref{eq:SolWGF&H} are complex. We then find five possible branches of backreactions with real coefficients that fix  $W_\text{GF}$,
\begin{align}
F_1 (x) &\= \left(\frac{\sinh(a x+b)}{a} \right)^2\,,\qquad F_2(x) \= -\left(\frac{\cosh(a x+b)}{a} \right)^2\,,\qquad F_5 (x) \= (x +b)^2\,,\nonumber \\
F_3 (x) &\= \left(\frac{\sin(a x+b)}{a} \right)^2\,,\qquad F_4(x) \= \left(\frac{\cos(a x+b)}{a} \right)^2\,,\qquad (a,b) \in \mathbb{R}\,,
\label{eq:DefFi}
\end{align}
and we can take $$W^2_{\text{GF}} \= F_I(\rho^{-1} \partial_\rho K)\,, $$for arbitrary $I=1,...,5$. In this paper, we will only analyze the branch of solutions given by $F_1$ since it is the direct generalization of the single-center solutions in section \ref{sec:class4d}. However, the other branch might also give interesting class of solutions. For instance, the branch given by $F_2$ has the benefit to have no potential zeroes to avoid for $W_\text{GF}$, but it gives solutions with $(-,-,-,-,-)$ signature since it requires $(W_\text{GF},W_0)\in i\IR$. The branch given by $F_5$ is also interesting since $W_\text{GF}$ is a linear function of the solution of Laplace equation. It corresponds to a kind of ``extremal'' branch, and as we will see the base layer is not affected by the gauge fields for this branch.

The base layer \eqref{eq:baselayer} also drastically simplifies considering that $V \= \rho$. The equations for $\nu_0$ are the usual equations for vacuum solutions and we know how to integrate them when $\log W_0$ solves Laplace equation with rod sources or point sources,
\begin{equation}
\partial_z \nu_0=  \frac{\rho}{2}\, \partial_\rho \log W_0 \partial_z \log W_0 \,,\qquad \partial_\rho \nu_0= \frac{\rho}{4}\left[ \left(\partial_\rho \log W_0 \right)^2 - \left(\partial_z \log W_0 \right)^2 \right]\,.
\label{eq:Eqfornu1}
\end{equation}
The equations for $\nu_\text{GF}$ has the same form after replacing $\log W_0$ by $\rho^{-1} \partial_\rho  K$, but the coefficients differ depending on which $F_I$ is taken,\footnote{If we take the generic solutions \eqref{eq:SolWGF&H} with complex parameters $(a,b)$, one would obtain $$\partial_z \nu_{\text{GF}}  = \frac{3 a^2 \,\rho}{2} \,\partial_\rho \left(\frac{1}{\rho} \partial_\rho K \right)\,\partial_z \left(\frac{1}{\rho} \partial_\rho K \right) \,,\qquad \partial_\rho \nu_{\text{GF}} =\frac{3 a^2 \,\rho}{4} \,\left[\left(\partial_\rho \left(\frac{1}{\rho} \partial_\rho K \right)\right)^2-\left(\partial_z \left(\frac{1}{\rho} \partial_\rho K \right)\right)^2 \right]   \,.$$}
\begin{equation}
\begin{split}
\partial_z \nu_{\text{GF}} &\= \begin{cases} 
\frac{3 a^2 \,\rho}{2} \,\partial_\rho \left(\frac{1}{\rho} \partial_\rho K \right)\,\partial_z \left(\frac{1}{\rho} \partial_\rho K \right) \quad \text{if} \quad I=1,2 \,,\\
-\frac{3 a^2 \,\rho}{2} \,\partial_\rho \left(\frac{1}{\rho} \partial_\rho K \right)\,\partial_z \left(\frac{1}{\rho} \partial_\rho K \right) \quad \text{if} \quad I=3,4 \,, \\
~~ 0 \qquad \text{if} \quad I=5\,,
\end{cases}\,,
\end{split}
\end{equation}
\begin{equation}
\begin{split}
 \partial_\rho \nu_{\text{GF}} &\= \begin{cases} 
\frac{3 a^2 \,\rho}{4} \,\left[\left(\partial_\rho \left(\frac{1}{\rho} \partial_\rho K \right)\right)^2-\left(\partial_z \left(\frac{1}{\rho} \partial_\rho K \right)\right)^2 \right]\quad \text{if} \quad I=1,2 \,,\\
-\frac{3 a^2 \,\rho}{4} \,\left[\left(\partial_\rho \left(\frac{1}{\rho} \partial_\rho K \right)\right)^2-\left(\partial_z \left(\frac{1}{\rho} \partial_\rho K \right)\right)^2 \right] \quad \text{if} \quad I=3,4 \,, \\
~~0 \qquad \text{if} \quad I=5\,,
\end{cases}\,.
\end{split}
\label{eq:EqfornuGF}
\end{equation}
The base layer is therefore simple to integrate and depends to what type of sources have been considered for $(\log W_0, \rho^{-1} \partial_\rho K)$.

\subsubsection*{Summary}

We have defined a class of axisymmetric two-charge solutions of the five-dimensional Einstein-Maxwell theory \eqref{eq:ActionGen} in a closed form. The metric and gauge fields are given by 
\begin{equation}
\begin{split}
ds^2 &\= W_\text{GF}^{-1} \left[ -W_0 \,dt^2 +W_0^{-1} \, dy^2 \right]+ W_\text{GF}^2 \left[ e^{2(\nu_0 + \nu_\text{GF})} \left(d\rho^2 + dz^2 \right) + \rho^2 d\phi^2 \right]\,,\\
F^{(m)} &\= d H \wedge d\phi\,,\qquad F^{(e)} \= \frac{q}{\rho\,W_\text{GF}^2} \,\star_2 dH\,\wedge dt\wedge dy\,.
\end{split}
\label{eq:WeylSolfin}
\end{equation}
The solutions are determined by two arbitrary functions that solve a Laplace equation on the three-dimensional base
\begin{equation}
\cL \left(\log W_0 \right) \=0 \,,\qquad \cL \left(\frac{1}{\rho} \,\partial_\rho K \right) \= 0\,, \qquad \text{with} \quad \cL \equi \frac{1}{\rho} \,\partial_\rho \left( \rho \partial_\rho \right) \+ \partial_z^2\,.
\end{equation}
The scalars $(W_\text{GF}, H)$ are given by
\begin{equation}
W_\text{GF}^2 \= F_I\left(\frac{1}{\rho} \partial_\rho K \right) \,,\qquad H \= \sqrt{\frac{3}{2(1+q^2)\kappa_5^2}}\,\partial_z K\,,
\label{eq:WGF&H}
\end{equation}
where $F_I$ is one of the five generating functions of one variable given by two real parameters \eqref{eq:DefFi}. The base scalars $(\nu_0,\nu_\text{GF})$ are obtained by integrating \eqref{eq:Eqfornu1} and \eqref{eq:EqfornuGF}. These integrals must be treated in a case-by-case manner depending on the type of sources chosen for $\log W_0$ and $\rho^{-1} \partial_\rho K$. In the next section, we will study solutions obtained from rod sources using $F_1$.

\subsection{Multiple topological stars and black strings on a line}
\label{sec:Rodsources}

We consider sources for $\log W_0$ and $\frac{1}{\rho} \partial_\rho K$ that are given by $n$ distinct rods of length $M_i$ along the $z$-axis centered around $z=a_i$. Without lost generality we can order them as $a_i < a_j$ for $i<j$ (see Fig.\ref{fig:profile} below). The coordinates of the endpoints of the rods on the $z$-axis are given by
\begin{equation}
z^\pm_i \equi a_i \pm \frac{M_i}{2}\,.
\label{eq:coordinatesEndpoints}
\end{equation}
We define the distances to the endpoints $r_\pm^{(i)}$ and the distances $R_\pm^{(i)}$ as
\begin{equation}
r_\pm^{(i)} \equi \sqrt{\rho^2 + \left(z-z^\pm_i\right)^2}\,, \qquad R_\pm^{(i)} \equi r_+^{(i)}+r_-^{(i)}\pm M_i\,\,.
\end{equation}
The harmonic functions associated to such sources are 
\begin{equation}
\log W_0 = \sum_{i=1}^n G_i \,\log  \frac{R_+^{(i)}}{R_-^{(i)}}\,,\qquad \frac{1}{\rho} \partial_\rho K = \sum_{i=1}^n P_i \,\log  \frac{R_+^{(i)}}{R_-^{(i)}}\,,
\end{equation}
and we take the branch of solutions \eqref{eq:WGF&H} with $I=1$. The metric warp factors and the gauge fields \eqref{eq:WeylSolfin} are then given by
\begin{equation}
\begin{split}
H &= \frac{\sqrt{3}}{\sqrt{2(1+q^2)}\kappa_5} \sum_{i=1}^n P_i \left(r_+^{(i)}-r_-^{(i)} \right)\,,\qquad W_0 = \prod_{i=1}^n \left(  \frac{R_+^{(i)}}{R_-^{(i)}}\right)^{G_i} \,,\\
W_\text{GF} &= \frac{1}{2a} \left[e^b \prod_{i=1}^n \left(  \frac{R_+^{(i)}}{R_-^{(i)}}\right)^{a P_i}-e^{-b} \prod_{i=1}^n \left(  \frac{R_-^{(i)}}{R_+^{(i)}}\right)^{a P_i} \right]\,.
\end{split}
\label{eq:WarpfactorsRods}
\end{equation}
One can check from the form of $H$ that the magnetic field strength $F^{(m)}=dH\wedge d\phi$ is indeed sourced by magnetic monopoles on each rod and then that its electric dual $F^{(e)}$ is sourced by electric charges along the $y$-circle. It is manifest using the spherical coordinates $(r_i,\theta_i,\phi)$ around the $i^\text{th}$ rod as follows
\begin{equation}
2\,r_i \equi r_+^{(i)}+r_-^{(i)}-M_i \,,\qquad M_i\,\cos \theta_i \equi r_+^{(i)}-r_-^{(i)}\,,
\end{equation}
and $F^{(m)}$ has the usual form for a magnetic monopole around the  $i^\text{th}$ rod
\begin{equation}
F^{(m)} \sim \frac{\sqrt{3}}{\sqrt{2}\,\kappa_5} \left(- \frac{P_i}{M_i} \,\sin \theta_i \+ \text{cst} \right) d\theta_i \wedge d\phi\,.
\end{equation}
Therefore, our multi-rod solutions are indeed the multi-body generalization of the single-center two-charge solutions described in section \ref{sec:class4d}.

\noindent Moreover, the solutions are invariant under the following transformations 
\begin{equation}
\begin{split}
(M_i,P_i,G_i) \rightarrow (-M_i,-P_i,-G_i)\, ~~\forall i\quad \mbox{and} \quad (t,y,a) \rightarrow (iy,it,-a)\,.
\end{split}
\end{equation}
We can then fix without loss of generality that $a>0$ and $M_i>0$ $\forall i$. We now need to integrate the base layer \eqref{eq:Eqfornu1} and \eqref{eq:EqfornuGF} to get $(\nu_0,\nu_\text{GF})$. We define for that purpose
\begin{equation} 
E_{\pm \pm}^{(i,j)} \equi r_\pm^{(i)} r_\pm^{(j)} + \left(z-z_i^\pm \right)\left(z-z_j^\pm \right) +\rho^2\,,
\end{equation}
and the generating functions $\nu_{ij}$,
\begin{equation}
\nu_{ij} \equi \log \frac{E_{+-}^{(i,j)}E_{-+}^{(i,j)}}{E_{++}^{(i,j)}E_{--}^{(i,j)}}\,.
\label{eq:GeneratingNu}
\end{equation}
The base layer \eqref{eq:Eqfornu1} and \eqref{eq:EqfornuGF} gives
\begin{equation}
\nu_\text{GF} \= \frac{3 a^2}{4} \,\sum_{i,j=1}^n P_i P_j \,\nu_{ij} \,,\qquad \nu_0 \= \frac{1}{4} \, \sum_{i,j=1}^n G_i G_j \,\nu_{ij}\,.
\label{eq:nufinal}
\end{equation}
We have constructed a family of solutions given by $4n+2$ parameters $(M_i,G_i,P_i,a_i,a,b)$. We now have to study the regularity of the solutions that constrains the parameter space. The potential constraints arise from coordinate singularities on the $z$-axis, regularity of the spacetime elsewhere and from conditions on the asymptotics. We discuss in greater details the regularity analysis in the Appendix \ref{app:MultiRod}. As a summary, we found that

\begin{itemize}\setlength{\itemsep}{0pt}\setlength{\parskip}{0pt}
\item[•] The solutions must be asymptotic to $\IR^{1,3}\times$S$^1$ at large $\rho$ or/and $z$ and regular everywhere out of the $z$-axis. This requires
\begin{equation}
a \= \sinh b \,,\qquad P_i > 0\,.
\end{equation}
Note that this implies that all the charges have the same sign. It comes from the requirement that $W_\text{GF}$ \eqref{eq:WarpfactorsRods} does not change sign. Having taken $W_\text{GF}$ to be a $\sinh$ of $\rho^{-1} \partial_\rho K$ induces many zeroes around two rods of different sign of charges. If we would have considered $W_\text{GF}$ as a $\cosh$ using $F_2$ \eqref{eq:DefFi}, we would have been able to consider rods with different sign of charges but it would have led to issue with the signature of the solutions. Because all the charges have the same sign, we cannot use the gauge fields to make the rods to repulse each other or to have a neutral system from far away.
\item[•] At each rod where $G_i < 0$, the timelike Killing vector $\partial_t$ shrinks and the rod corresponds to a regular S$^2\times$S$^1$ horizon of the black string discussed in section \ref{sec:BS4d/5d} if
\begin{equation}
G_i \= -\frac{1}{2} \,,\qquad \,P_i \= \frac{1}{2 \sinh b} \,.
\end{equation}
Its four-dimensional ADM mass, $\cM^{(i)}$, electric charge, $Q_e^{(i)}$, and magnetic charge, $P^{(i)}$, are given by
\begin{equation}
\cM^{(i)} \= \frac{\pi\,M_i}{\kappa_4^2} \left(3 \coth b+1\right)\,,\quad {Q_e^{(i)}}^2 \= q^2\,{Q_m^{(i)}}^2 \= \frac{3\,q^2}{8(1+q^2)\,\kappa_4^2} \,\frac{M_i^2}{\sinh^2 b}\,.
\end{equation}
The presence of a black-string rod induces a temperature to the whole solution which can be derived by regularity of the Euclidean version of the metric. We find that
\begin{equation}
T^{-2} \= \frac{2 \pi^2 \,M_i^2 \,e^{3 b}}{\sinh^3 b}\,d_i^2\,  \prod_{j\neq i} \left(\frac{z_j^+ - z_i^-}{z_j^- -z_i^-} \right)^{\text{sign}(j-i) \,\frac{3 - 2 G_j }{2}} \,,
\end{equation}
where the $z$-coordinate of the rod endpoints, $z^\pm_j$, is given in \eqref{eq:coordinatesEndpoints} and $d_i$ corresponds to the following product of aspect ratios
\begin{equation}
d_1 \equi 1\,,\qquad d_i \equi  \prod_{j=1}^{i-1} \prod_{k=i}^n \left(\dfrac{(z_k^- - z_j^+)(z_k^+ - z_j^-)}{(z_k^+ - z_j^+)(z_k^- - z_j^-)}  \right)^{ \frac{3 + 4 G_j G_k}{4}}\quad \text{when } i=2,\ldots n\,.
\label{eq:di}
\end{equation}
\item[•] At each rod where $G_i > 0$, the spacelike Killing vector $\partial_y$ shrinks and the rod corresponds to a degeneracy of the $y$-circle if
\begin{equation}
G_i \= \frac{1}{2} \,,\qquad \,P_i \= \frac{1}{2 \sinh b} \,.
\end{equation}
As for the single-center solutions of section \ref{sec:TS4d/5d}, this degeneracy corresponds to $\IR^2 / \mathbb{Z}_{k_i} \times S^2$ with $k_i \in \mathbb{Z}_+$ if
\begin{equation}
R_y^2 \=  \frac{M_i^2 \,e^{3 b}}{2\,k_i^2\,\sinh^3 b}\,d_i^2\,  \prod_{j\neq i} \left(\frac{z_j^+ - z_i^-}{z_j^- -z_i^-} \right)^{\text{sign}(j-i) \,\frac{3 + 2 G_j }{2}}\,.
\end{equation}
The rod corresponds to a bolt, that is a S$^2$ bubble similar to the single-center topological star depicted in section \ref{sec:TS4d/5d}. However, the S$^2$ is now warped due to the presence of the other sources as detailed in the Appendix \ref{app:regularity}. Its four-dimensional ADM mass, $\cM^{(i)}$, electric charge, $Q_e^{(i)}$, and magnetic charge, $P^{(i)}$, are given by
\begin{equation}
\cM^{(i)} \= \frac{\pi\,M_i}{\kappa_4^2} \left(3 \coth b-1\right)\,,\quad {Q_e^{(i)}}^2 \= q^2\,{Q_m^{(i)}}^2 \= \frac{3\,q^2}{8(1+q^2)\,\kappa_4^2} \,\frac{M_i^2}{\sinh^2 b}\,.
\end{equation}
\item[•] On the $z$-axis in between the rods, the $\phi$-circle shrinks as the usual cylindrical coordinate degeneracy. To be smooth at those loci, the time slices of the spacetime should be locally $\IR^3\times$S$^1$. We showed in the Appendix \ref{app:regularity} that this condition is guaranteed if $d_i \= 1$ for all $i=1,\ldots,n$. However, the arguments in the product of $d_i$ \eqref{eq:di} are necessarly smaller than one and the powers, $\frac{3+4 G_j G_k}{4}$, are necessarly positive\footnote{It should be noted that the powers are initially assigned by $3 a^2 P_j P_k+ G_j G_k$. If we were allowed to have different sign of charges, these powers could be negative and therefore $d_i$ could be set to $1$.}. Therefore, the solutions are forced to have conical excesses on the segments between the rods given by $n-1$ rational numbers $0<d_i<1$ for $i=2,...,n$. Those conical excesses correspond to struts or strings with negative tension that are necessary to prevent the rods from collapse.
\end{itemize}

To conclude, the parameters are strongly constrained by regularity such that
\begin{equation}
a \= \sinh b \,,\qquad P_i \= \frac{1}{2\sinh b}\,,\qquad G_i \= \frac{\epsilon_i}{2}\,,
\end{equation}
where $\epsilon_i \= \pm 1$ is a sign lattice determining the nature of the $i^\text{th}$ rod: for $\epsilon_i \= 1$ the rod corresponds to a topological star while for $\epsilon_i \= -1$ the rod corresponds to a two-charge black string. In the $n-1$ segments in between the rods, there is a conical excess given by the parameter $0<d_i<1$:
\begin{equation}
d_i \equi  \prod_{j=1}^{i-1} \prod_{k=i}^n \left(\dfrac{(z_k^- - z_j^+)(z_k^+ - z_j^-)}{(z_k^+ - z_j^+)(z_k^- - z_j^-)}  \right)^{ \frac{3 + \epsilon_j \epsilon_k}{4}}\quad \text{when } i=2,\ldots n\,.
\end{equation}
In the presence of black strings and bubbles the temperature and the radius of the $y$-circle are respectively 
\begin{equation}
\begin{split}
T^{-2} &\= \frac{2 \pi^2 \,M_i^2 \,e^{3 b}}{\sinh^3 b}\,d_i^2\,  \prod_{j\neq i} \left(\frac{z_j^+ - z_i^-}{z_j^- -z_i^-} \right)^{\text{sign}(j-i) \,\frac{3 - \epsilon_j }{2}} \,,\qquad \forall i\text{ s.t. } \epsilon_i \= -1\,,\\
R_y^2& \=  \frac{M_i^2 \,e^{3 b}}{2\,k_i^2\,\sinh^3 b}\,d_i^2\,  \prod_{j\neq i} \left(\frac{z_j^+ - z_i^-}{z_j^- -z_i^-} \right)^{\text{sign}(j-i) \,\frac{3 + \epsilon_j }{2}}\,,\qquad \forall i\text{ s.t. } \epsilon_i \= 1\,.
\end{split}
\label{eq:Ry&Trods}
\end{equation}

From far away, the solutions are asymptotic to $\IR^{1,3}\times$S$^1$ and have the following conserved charges in four dimensions
\begin{equation}
\cM \= \frac{\pi}{\kappa_4^2}\sum_{i=1}^n M_i \left(3 \coth b-\epsilon_i\right)\,,\qquad {Q_e} \= q\,{Q_m} \= \frac{\sqrt{3}\,q}{2\sqrt{2(1+q^2)}\,\kappa_4} \,\frac{\sum_{i=1}^n M_i}{\sinh b}\,.
\end{equation}

\begin{figure}[h]
\setlength\belowcaptionskip{-0.7cm}
\begin{center}
 \usetikzlibrary{snakes,3d,shapes.geometric,shadows.blur}
\tikzoption{canvas is xy plane at z}[]{%
    \def\tikz@plane@origin{\pgfpointxyz{0}{0}{#1}}%
    \def\tikz@plane@x{\pgfpointxyz{1}{0}{#1}}%
    \def\tikz@plane@y{\pgfpointxyz{0}{1}{#1}}%
    \tikz@canvas@is@plane}
 \definecolor{myred}{rgb}{0.7,0.2,0.2}
 \definecolor{mygreen}{rgb}{0.1,0.6,0.3}
\definecolor{myblue}{rgb}{0.2,0.2,0.7}
\definecolor{myyellow}{rgb}{0.6,0.5,0.1}
\begin{tikzpicture}
\tikzmath{\xxm = -2; \xxM =5; \yym=-4.4; \yyM = 4;\aa=\yyM/2;\rodsize=1;\rodsizeb=1.55;\xlabelM=2.5;\xlabelm=0.5;} 
\draw[->,semithick] (0,\yym) -- (0,\yyM+1/3) node[anchor=south east] {\textbf{$z$}};
\draw[line width=0.65mm, |-|] (0,\aa-\rodsize) -- (0,\aa+\rodsize);
\draw[line width=0.65mm, |-|] (0,-\aa-\rodsizeb) -- (0,-\aa+\rodsizeb);
\draw (0,-\aa)node[circle,fill,inner sep=2.0pt]{};
\draw (0,\aa)node[circle,fill,inner sep=2.0pt]{};
\draw (0,\aa) node[anchor=east]{$a_2\hspace{0.4cm}$};
\draw (0,\aa+\rodsize) node[anchor=east]{$a_2 + \frac{M_2}{2}\hspace{0.6cm}$};
\draw (0,\aa-\rodsize) node[anchor=east]{$a_2 - \frac{M_2}{2}\hspace{0.6cm}$};
\draw (0,-\aa) node[anchor=east]{$a_1\hspace{0.5cm}$};
\draw (0,-\aa+\rodsizeb) node[anchor=east]{$a_1 + \frac{M_1}{2}\hspace{0.6cm}$};
\draw (0,-\aa-\rodsizeb) node[anchor=east]{$a_1 - \frac{M_1}{2}\hspace{0.6cm}$};
\draw (0.7,\yyM) node[anchor=west,myred]{$y$ circle};
\draw (0.5,-\aa-\rodsizeb+0.4) node[anchor=west,myblue]{$\phi$ circle};

   \begin{scope} 
	\clip(\xxm,\yym)rectangle(\xxM,\yyM);     
    \draw [fill=myred!40, myred!40,opacity=0.2]  (0,\yyM) ellipse [x radius=17pt, y radius=28pt];
    \draw [myred]  (0,\yyM) ellipse [x radius=17pt, y radius=28pt];
    \draw [myred] (0,\yyM) ellipse [x radius=5pt, y radius=28pt,];
    \draw [myred] (0,\yyM) ellipse [x radius=17pt, y radius=5pt,];
   \end{scope}  
   
 \begin{scope}
\clip(\xxm,\yym)rectangle(\xxM,\yyM);     
 \draw [fill=myred!40, myred!40,opacity=0.2]   (-0.5,-\aa+\rodsizeb+0.2) .. controls (-0.4,\aa-\rodsize-0.3)  .. (0,\aa-\rodsize) .. controls (0.4,\aa-\rodsize-0.3)  .. (0.5,-\aa+\rodsizeb+0.2);
     \draw [myred]   (-0.5,-\aa+\rodsizeb+0.2) .. controls (-0.4,\aa-\rodsize-0.3)  .. (0,\aa-\rodsize) .. controls (0.4,\aa-\rodsize-0.3)  .. (0.5,-\aa+\rodsizeb+0.2);
\draw [myred]   (-0.5,-\aa+\rodsizeb+0.2) .. controls (-0.4,\aa-\rodsize-0.3)  .. (0,\aa-\rodsize) .. controls (0.4,\aa-\rodsize-0.3)  .. (0.5,-\aa+\rodsizeb+0.2);
	\draw [myred]   (-0.17,-\aa+\rodsizeb+0.2) .. controls (-0.11,\aa-\rodsize-0.3)  .. (0,\aa-\rodsize) .. controls (0.11,\aa-\rodsize-0.3)  .. (0.17,-\aa+\rodsizeb+0.2);
	    \draw [myred] (0,-\aa+\rodsizeb+0.7) ellipse [x radius=12pt, y radius=4pt,];
	    	\draw [dashed,thick=1,myred] (0.5,-\aa+\rodsizeb+0.2) -- (0.51,-\aa+\rodsizeb-0.5);
	   	\draw [dashed,thick=1,myred] (-0.5,-\aa+\rodsizeb+0.2) -- (-0.51,-\aa+\rodsizeb-0.5);
	   	\draw [dashed,thick=1,myred] (0.5,-\aa-\rodsizeb+0.45) -- (0.51,-\aa-\rodsizeb-0.5);
	   	\draw [dashed,thick=1,myred] (-0.5,-\aa-\rodsizeb+0.45) -- (-0.51,-\aa-\rodsizeb-0.5);
	   	\filldraw [fill=myred!40, myred!40,opacity=0.25,path fading = south]   (0.5,-\aa+\rodsizeb+0.2) -- (0.51,-\aa+\rodsizeb-0.5) --  (-0.51,-\aa+\rodsizeb-0.5) -- (-0.5,-\aa+\rodsizeb+0.2);
   \end{scope}  
   
   \draw[decorate,thick=1,myred] (0,\aa-\rodsize) -- (0,\aa+\rodsize);
          \draw[decorate,thick,myred] (0,-\aa-\rodsizeb) -- (0,-\aa+\rodsizeb);
\draw[<-,semithick] (\xlabelm,\aa+\rodsize) -- (\xlabelM,\aa+\rodsize) node[anchor=west] {$\IR^4$};
\draw[<-,semithick] (\xlabelm,\aa-\rodsize) -- (\xlabelM,\aa-\rodsize) node[anchor=west] {$\IR^4$ with conical excess};
\draw [decorate,decoration={brace,amplitude=10pt,mirror,raise=4pt},yshift=0pt]
(\xlabelM-0.5,\aa-\rodsize+0.15) -- (\xlabelM-0.5,\aa+\rodsize-0.15) node [black,midway,xshift=3.65cm] {\textbf{Two-charge bubble} as $\IR^2/\mathbb{Z}_{k_2}\times$S$^2$};
\draw [decorate,decoration={brace,amplitude=10pt,mirror,raise=4pt},yshift=0pt]
(\xlabelM-0.5,-\aa-\rodsizeb+0.15) -- (\xlabelM-0.5,-\aa+\rodsizeb-0.15) node [black,midway,xshift=2.9cm] {\textbf{Two-charge black string}};
\draw [decorate,decoration={brace,amplitude=10pt,mirror,raise=4pt},yshift=0pt]
(\xlabelM-0.5,-\aa+\rodsizeb+0.15) -- (\xlabelM-0.5,\aa-\rodsize-0.15) node [black,midway,xshift=3.65cm] {$\IR^3\times$S$^1$ with {\bf a strut} (conical excess)};
   \begin{scope}
	\clip(\xxm,\yym)rectangle(\xxM,\yyM);  
	\draw [decorate,decoration={brace,amplitude=10pt,mirror,raise=4pt},yshift=0pt]
(\xlabelM-0.5,\aa+\rodsize+0.15) -- (\xlabelM-0.5,\yyM+2) node [black,midway,xshift=1.2cm] {$\IR^3$S$^1$};
   \end{scope}  
   \draw	(\xlabelM+0.6,\yyM) node[align=left]{$\IR^3\times$S$^1$};
   \begin{scope}
	\clip(\xxm,\yym)rectangle(\xxM,\yyM);  
	\draw [decorate,decoration={brace,amplitude=10pt,mirror,raise=4pt},yshift=0pt]
(\xlabelM-0.5,\yym-2) -- (\xlabelM-0.5,-\aa-\rodsizeb-0.15) node [black,midway,xshift=1.2cm] {$\IR^3$S$^1$};
   \end{scope}  
   \draw	(\xlabelM+0.6,\yym) node[align=left]{$\IR^3\times$S$^1$};
   \begin{scope}
	\clip(\xxm,\yym)rectangle(\xxM,\yyM);     
    \draw [fill=myblue!40, myblue!40,opacity=0.2]  (0,-\aa) ellipse [x radius=17pt, y radius=42pt];
    \draw [myblue]  (0,-\aa) ellipse [x radius=17pt, y radius=42pt];
    \draw [myblue] (0,-\aa) ellipse [x radius=7pt, y radius=42pt,];
    \draw [myblue] (0,-\aa) ellipse [x radius=17pt, y radius=7pt,];
   \end{scope}

      \begin{scope}
	\clip(\xxm,\yym)rectangle(\xxM,\yyM);     
    \draw [fill=myblue!40, myblue!40,opacity=0.2]  (0,\aa) ellipse [x radius=8pt, y radius=26pt];
    \draw [myblue]  (0,\aa) ellipse [x radius=8pt, y radius=26pt];
    \draw [myblue] (0,\aa) ellipse [x radius=3pt, y radius=26pt,];
    \draw [myblue] (0,\aa) ellipse [x radius=8pt, y radius=3pt,];
   \end{scope}  
   
   \draw[decorate, draw=darkgreen,
        decoration={zigzag,amplitude=1.5pt, segment length=3pt},thick=1,darkgreen] (0,-\aa+\rodsizeb) -- (0,\aa-\rodsize);
          \draw[decorate,thick,myblue] (0,\aa+\rodsize) -- (0,\yyM);
          \draw[decorate,thick,myblue] (0,-\aa-\rodsizeb) -- (0,\yym);

	\draw [fill=black!80, black!80,opacity=0.4]  (-0.12,-\aa+\rodsizeb) --(0.12,-\aa+\rodsizeb) -- (0.12,-\aa-\rodsizeb)--(-0.12,-\aa-\rodsizeb)--(-0.12,-\aa+\rodsizeb); 
        
\end{tikzpicture}
\caption{Illustration of the topology on the $z$-axis for a solution with one black string and one topological star. The behavior of the $y$-circle is depicted in red: it shrinks at the bubble rod and have a finite size otherwise. The $\phi$ circle is in blue: it shrinks out of the rods and has a finite size at the rods. The magnetic and electric fluxes are wrapped on the blue bubbles at the rods.}
\label{fig:profile}
\end{center}
\end{figure}
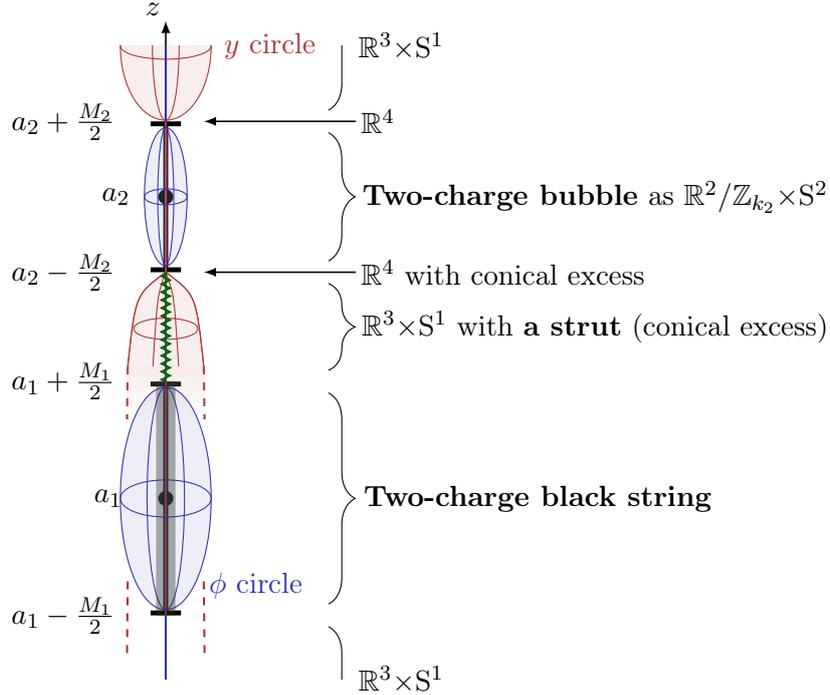

In Fig.\ref{fig:profile}, we have depicted the typical topology of multi-rod solutions by considering an illustrative example of a two-body configuration with a topological star and a black string. In addition to the rod profile, we have depicted the behavior of the $y$-circle (red) and the $\phi$-circle (blue). The $y$-circle shrinks to zero size on the bubble rod and has finite size elsewhere while the $\phi$-circle shrinks on the $z$-axis except on the rods.

\subsubsection{A simple example}

\begin{itemize}
\item[•] \underline{One-rod configurations:}
\end{itemize}

We can retrieve the class of single-center solutions discussed in section \ref{sec:class4d} by considering only one rod, $n=1$, changing the parameters to
\begin{equation}
a_1\= 0\,,\qquad M_1 \= \left| r_\text{S} - r_\text{B} \right|\,,\quad\epsilon_1 \= \text{sign}(r_\text{B}-r_\text{B}) \,,\quad b \= \frac{\text{sign}(r_\text{B}-r_\text{S})}{2}\,\log \frac{r_\text{B}}{r_\text{S}}\,,
\label{eq:maptosinglecenter}
\end{equation}
and changing the coordinates to the spherical $(r,\theta,\phi)$ as
\begin{equation}
\rho \= \sqrt{r(r+M_1)}\,\sin \theta\,,\qquad z \= \left(r+\frac{M_1}{2} \right) \,\cos \theta\,.
\end{equation}

\begin{itemize}
\item[•] \underline{Two-bubble-rod configurations:}
\end{itemize}
We now consider two distinct bubble rods, $n=2$ and $\epsilon_1 = \epsilon_2 = 1$, 
We place the origin of the $z$-axis such that $- a_1 = a_2 = u >0 $. Moreover, we assume by symmetry that the rods are ordered such that $0< M_2 \leq M_1$. The condition to have two distinct rods translates to
\begin{equation}
0 < M_1 +M_2 < 4\, u\,.
\label{eq:distinctrodcond}
\end{equation}
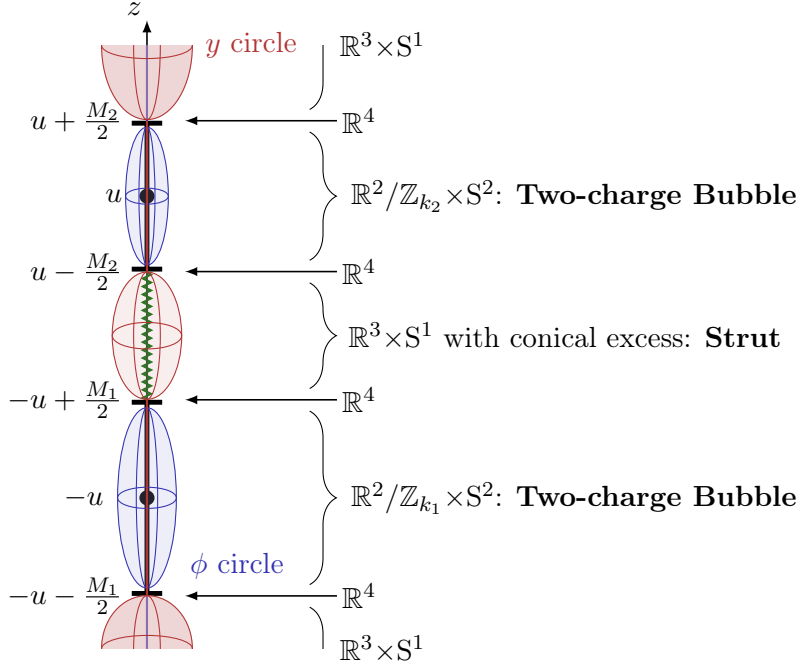
\begin{figure}[ht]
\setlength\belowcaptionskip{-0.7cm}
\begin{center}
 \usetikzlibrary{snakes,3d,shapes.geometric,shadows.blur}
\tikzoption{canvas is xy plane at z}[]{%
    \def\tikz@plane@origin{\pgfpointxyz{0}{0}{#1}}%
    \def\tikz@plane@x{\pgfpointxyz{1}{0}{#1}}%
    \def\tikz@plane@y{\pgfpointxyz{0}{1}{#1}}%
    \tikz@canvas@is@plane}
 \definecolor{myred}{rgb}{0.7,0.2,0.2}
 \definecolor{mygreen}{rgb}{0.1,0.6,0.3}
\definecolor{myblue}{rgb}{0.2,0.2,0.7}
\definecolor{myyellow}{rgb}{0.6,0.5,0.1}
\begin{tikzpicture}
\tikzmath{\xxm = -2; \xxM =5; \yym=-4; \yyM = 4;\aa=\yyM/2;\rodsize=1;\rodsizeb=1.3;\xlabelM=2.5;\xlabelm=0.5;} 
\draw[->,semithick] (0,\yym) -- (0,\yyM+1/3) node[anchor=south east] {\textbf{$z$}};
\draw[line width=0.65mm, |-|] (0,\aa-\rodsize) -- (0,\aa+\rodsize);
\draw[line width=0.65mm, |-|] (0,-\aa-\rodsizeb) -- (0,-\aa+\rodsizeb);
\draw (0,-\aa)node[circle,fill,inner sep=2.0pt]{};
\draw (0,\aa)node[circle,fill,inner sep=2.0pt]{};
\draw (0,\aa) node[anchor=east]{$u\,\,\,\,$};
\draw (0,\aa+\rodsize) node[anchor=east]{$u + \frac{M_2}{2}\,\,\,\,$};
\draw (0,\aa-\rodsize) node[anchor=east]{$u - \frac{M_2}{2}\,\,\,\,$};
\draw (0,-\aa) node[anchor=east]{$-u\hspace{0.5cm}$};
\draw (0,-\aa+\rodsizeb) node[anchor=east]{$-u + \frac{M_1}{2}\,\,\,\,$};
\draw (0,-\aa-\rodsizeb) node[anchor=east]{$-u - \frac{M_1}{2}\,\,\,\,$};
\draw (0.7,\yyM) node[anchor=west,myred]{$y$ circle};
\draw (0.5,-\aa-\rodsizeb+0.4) node[anchor=west,myblue]{$\phi$ circle};

\draw[<-,semithick] (\xlabelm,\aa+\rodsize) -- (\xlabelM,\aa+\rodsize) node[anchor=west] {\textbf{$\IR^4$}};
\draw[<-,semithick] (\xlabelm,\aa-\rodsize) -- (\xlabelM,\aa-\rodsize) node[anchor=west] {\textbf{$\IR^4$}};
\draw[<-,semithick] (\xlabelm,-\aa+\rodsizeb) -- (\xlabelM,-\aa+\rodsizeb) node[anchor=west] {\textbf{$\IR^4$}};
\draw[<-,semithick] (\xlabelm,-\aa-\rodsizeb) -- (\xlabelM,-\aa-\rodsizeb) node[anchor=west] {\textbf{$\IR^4$}};
\draw [decorate,decoration={brace,amplitude=10pt,mirror,raise=4pt},yshift=0pt]
(\xlabelM-0.5,\aa-\rodsize+0.15) -- (\xlabelM-0.5,\aa+\rodsize-0.15) node [black,midway,xshift=3.62cm] {$\IR^2/\mathbb{Z}_{k_2}\times$S$^2$: {\bf Two-charge Bubble}};
\draw [decorate,decoration={brace,amplitude=10pt,mirror,raise=4pt},yshift=0pt]
(\xlabelM-0.5,-\aa-\rodsizeb+0.15) -- (\xlabelM-0.5,-\aa+\rodsizeb-0.15) node [black,midway,xshift=3.62cm] {$\IR^2/\mathbb{Z}_{k_1}\times$S$^2$: {\bf Two-charge Bubble}};
\draw [decorate,decoration={brace,amplitude=10pt,mirror,raise=4pt},yshift=0pt]
(\xlabelM-0.5,-\aa+\rodsizeb+0.15) -- (\xlabelM-0.5,\aa-\rodsize-0.15) node [black,midway,xshift=3.5cm] {$\IR^3\times$S$^1$ with conical excess: {\bf Strut}};
   \begin{scope}
	\clip(\xxm,\yym)rectangle(\xxM,\yyM);  
	\draw [decorate,decoration={brace,amplitude=10pt,mirror,raise=4pt},yshift=0pt]
(\xlabelM-0.5,\aa+\rodsize+0.15) -- (\xlabelM-0.5,\yyM+2) node [black,midway,xshift=1.2cm] {$\IR^3$S$^1$};
   \end{scope}  
   \draw	(\xlabelM+0.6,\yyM) node[align=left]{$\IR^3\times$S$^1$};
   \begin{scope}
	\clip(\xxm,\yym)rectangle(\xxM,\yyM);  
	\draw [decorate,decoration={brace,amplitude=10pt,mirror,raise=4pt},yshift=0pt]
(\xlabelM-0.5,\yym-2) -- (\xlabelM-0.5,-\aa-\rodsizeb-0.15) node [black,midway,xshift=1.2cm] {$\IR^3$S$^1$};
   \end{scope}  
   \draw	(\xlabelM+0.6,\yym) node[align=left]{$\IR^3\times$S$^1$};
   \begin{scope}
	\clip(\xxm,\yym)rectangle(\xxM,\yyM);     
    \draw [fill=myblue!40, myblue!40,opacity=0.2]  (0,-\aa) ellipse [x radius=11pt, y radius=34pt];
    \draw [myblue]  (0,-\aa) ellipse [x radius=11pt, y radius=34pt];
    \draw [myblue] (0,-\aa) ellipse [x radius=4pt, y radius=34pt,];
    \draw [myblue] (0,-\aa) ellipse [x radius=11pt, y radius=4pt,];
   \end{scope}

      \begin{scope}
	\clip(\xxm,\yym)rectangle(\xxM,\yyM);     
    \draw [fill=myblue!40, myblue!40,opacity=0.2]  (0,\aa) ellipse [x radius=8pt, y radius=26pt];
    \draw [myblue]  (0,\aa) ellipse [x radius=8pt, y radius=26pt];
    \draw [myblue] (0,\aa) ellipse [x radius=3pt, y radius=26pt,];
    \draw [myblue] (0,\aa) ellipse [x radius=8pt, y radius=3pt,];
   \end{scope}  
   
   \draw[decorate, draw=darkgreen,
        decoration={zigzag,amplitude=1.5pt, segment length=3pt},thick=1,darkgreen] (0,-\aa+\rodsizeb) -- (0,\aa-\rodsize);
   \draw[decorate, draw=darkgreen,thick=1,myblue] (0,\aa+\rodsize) -- (0,\yyM);
   \draw[decorate, draw=darkgreen,thick=1,myblue] (0,-\aa-\rodsizeb) -- (0,-\yyM);
   \begin{scope}
	\clip(\xxm,\yym)rectangle(\xxM,\yyM);     
    \draw [fill=myred!40, myred!40,opacity=0.2]  (0,\rodsizeb/2-\rodsize/2) ellipse [x radius=13pt, y radius=24pt];
    \draw [myred]  (0,\rodsizeb/2-\rodsize/2) ellipse [x radius=13pt, y radius=24pt];
    \draw [myred] (0,\rodsizeb/2-\rodsize/2) ellipse [x radius=5pt, y radius=24pt,];
    \draw [myred] (0,\rodsizeb/2-\rodsize/2) ellipse [x radius=13pt, y radius=5pt,];
   \end{scope}  
   
   \begin{scope}
	\clip(\xxm,\yym)rectangle(\xxM,\yyM);     
    \draw [fill=myred!40, myred!40,opacity=0.5]  (0,\yyM) ellipse [x radius=17pt, y radius=28pt];
    \draw [myred]  (0,\yyM) ellipse [x radius=17pt, y radius=28pt];
    \draw [myred] (0,\yyM) ellipse [x radius=5pt, y radius=28pt,];
    \draw [myred] (0,\yyM) ellipse [x radius=17pt, y radius=5pt,];
   \end{scope}  
   
      \begin{scope}
	\clip(\xxm,\yym)rectangle(\xxM,\yyM);     
    \draw [fill=myred!40, myred!40,opacity=0.5]  (0,\yym) ellipse [x radius=17pt, y radius=20pt];
    \draw [myred]  (0,\yym) ellipse [x radius=17pt, y radius=20pt];
    \draw [myred] (0,\yym) ellipse [x radius=5pt, y radius=20pt,];
    \draw [myred] (0,\yym) ellipse [x radius=17pt, y radius=5pt,];
   \end{scope}  
   
   \draw[decorate, draw=darkgreen,thick=1,myred] (0,\aa-\rodsize) -- (0,\aa+\rodsize);		   \draw[decorate, draw=darkgreen,thick=1,myred] (0,-\aa-\rodsizeb) -- (0,-\aa+\rodsizeb);
          
\end{tikzpicture}
\caption{The profile of the solutions along the $z$-axis. We have depicted the behavior of the $y$-circle in red which shrinks on the rods and have a finite size otherwise and of the $\phi$-circle in blue which has a finite size on the rods and shrinks otherwise. The electric and magnetic fluxes are wrapped on the blue bubbles on the rods.}
\label{fig:profile2}
\end{center}
\end{figure}
The profile of the solutions is depicted in Fig.\ref{fig:profile2}. The metric components and the gauge fields are given by the functions in \eqref{eq:WarpfactorsRods} and \eqref{eq:nufinal}:
\begin{align}
H &= \frac{\sqrt{3}}{2\sqrt{2}\,\kappa_5\, \sinh b}\, \sum_{i=1}^2 \left(r_+^{(i)}-r_-^{(i)} \right)\,,\qquad W_\text{GF} = \frac{1}{2\sinh b} \left[e^b \prod_{i=1}^2 \sqrt{ \frac{R_+^{(i)}}{R_-^{(i)}}} -e^{-b} \prod_{i=1}^2 \sqrt{  \frac{R_-^{(i)}}{R_+^{(i)}}} \right]\,, \nonumber\\
 W_0 &= \prod_{i=1}^2 \sqrt{ \frac{R_+^{(i)}}{R_-^{(i)}}}\,,\hspace{3cm} e^{2(\nu_0 + \nu_\text{GF})} = \prod_{i,j=1}^2  \,\sqrt{ \frac{E_{+-}^{(i,j)}E_{-+}^{(i,j)}}{E_{++}^{(i,j)}E_{--}^{(i,j)}} }\,.
\end{align}
The regularity conditions introduce three conical-defect parameters, $d_2\in \mathbb{Q}$ with $0<d_2<1$ \eqref{eq:di} and $(k_1,k_2)\in \mathbb{Z}_+$ \eqref{eq:Ry&Trods}. In between the two rods on the $z$-axis, the $\phi$-circle shrinks and the time slices of the solution correspond to the cylindrical degeneracy of $\IR^3\times$S$^1$ with a conical excess
\begin{equation}
d_2\= \frac{16 \,u^2 - \left(M_1 +M_2 \right)^2}{16 \,u^2 - \left(M_1 -M_2 \right)^2}  \,.
\end{equation}
Note that we have indeed $0<d_2<1$ for $M_1+M_2<4 u$. Moreover, bringing the two bubbles closer to each other $u\to  \frac{1}{4}(M_1+M_2)$ implies that the tension on the strut becomes greater and greater $d_2 \to 0$, which is intuitive since the gravitational attraction between the two bubbles diverges.

\noindent The regularity on the rods gives
\begin{equation}
\begin{split}
R_y^2 \=   \frac{M_1^2\,e^{3 b}}{2\sinh^3b\,k_1^2}\,  \left(\frac{4 u +M_1+M_2}{4 u +M_1-M_2}\right)^{2} \=\frac{M_2^2\,e^{3 b}}{2\sinh^3b\,k_2^2}\, \left(\frac{4 u +M_1+M_2}{4 u +M_2-M_1}\right)^{2}\,.
\end{split}
\label{eq:RyconstraintEx}
\end{equation}
We can use $u$ to solve the constraint from the second equality and get
\begin{equation}
u \= \frac{(M_1-M_2)(k_2 M_1+k_1 M_2)}{4(k_2 M_1-k_1 M_2)}\,.
\end{equation}
The physicality condition \eqref{eq:distinctrodcond} requires $\frac{M_2}{M_1} \,k_1 \,<\, k_2 \,<\, k_1$ and we have defined a family of two-bubble solutions supported by a strut parametrized by 6 parameters $(M_1,M_2,b,k_1,k_2,q)$.

For simplicity, we can also solve \eqref{eq:RyconstraintEx} by considering that the two bubbles are identical $M_1 = M_2 = M$ and $k_1 = k_2 =k$. We then have a class of two interacting topological stars very similar to the single topological star studied in section \ref{sec:TS4d/5d} but with one extra parameter that we can choose to be the separation between the two bodies, $\delta \= u - \frac{M}{2}$. We can take a similar parametrization by applying \eqref{eq:maptosinglecenter}, and each bubble is described by the same pair $(r_\text{B},r_\text{S})$ with $r_\text{B} > r_\text{S}$. Therefore, the strut that separates the bubbles is given by the conical excess 
\begin{equation}
d_2 \= \frac{4\,\delta \,(r_\text{B} - r_\text{S}+\delta)}{(r_\text{B} - r_\text{S}+2\,\delta)^2}\,.
\end{equation}
The ADM mass, the electric and the magnetic charge of the system are naturally twice the ones computed for the single-bubble solutions \eqref{eq:ADMmass&charges4d}. The radius of the $y$-circle however is given by
\begin{equation}
R_y^2 \= \frac{16\,r_\text{B}^3 \,\left(r_\text{B}-r_\text{S}+\delta \right)^2}{k^2 \,(r_\text{B}-r_\text{S})\,(r_\text{B}-r_\text{S}+2\delta)^2}\,.
\end{equation}

There are many more configurations that we can think about by adding bubble or black string rods. However, the take-away message is that with the current constructions we cannot use the magnetic and electric charges to get rid of the struts in between the different objects. In the next section, we briefly discuss some extrapolations that can be used to get rid of the struts. 

\subsection{Going further}
We would like to explore the other classes of solutions one can obtain by considering the other branches of solutions for $W_\text{GF}$ given by the five possible functions $F_I$ \eqref{eq:DefFi}. We can be especially interested in the $\cosh$ branch, $F_2$, for which succession of rods with different sign of charges can be constructed.

Second, we can add an extra circle and consider geometries that are asymptotically $\IR^{1,3} \times$T$^2$. This does not change the underlying structure of the equations of motion. The ansatz can be written as follows
\begin{align}
&ds^2 = W_\text{GF}^{-1} \left[ -W_0 W_1 \,dt^2 +W_0^{-1} \, dy_0^2 + W_1^{-1} \, dy_1^2 \right]+ W_\text{GF}^3 \left[ e^{2(\nu_0 +\nu_1 + \nu_\text{GF})} \left(d\rho^2 + dz^2 \right) + \rho^2 d\phi^2 \right],\nonumber \\
&F^{(m)} \= d H(\rho,z) \wedge d\phi\,,\qquad F^{(e)} \= \frac{1}{\rho\,W_\text{GF}^3} \,\star_2 dH\,\wedge dt\wedge dy_0 \wedge dy_1 \,,
\label{eq:WeylSol2circle}
\end{align}
and the equations of motion for the new pair $(W_1,\nu_1)$ is identical to the ones for $(W_0,\nu_0)$ that we have already studied \eqref{eq:EOMW0} and \eqref{eq:Eqfornu1}. The only change is in the Maxwell and base layer for $(H,W_\text{GF},\nu_\text{GF})$ for which the source parts have different coefficients. Within this ansatz, it might be possible to alternate bubbles where the $y_0$-circle shrinks with bubbles where the $y_1$-circle shrinks without having connecting segments where the $\phi$-circle shrinks inducing struts. 

Finally, we will be interested in using Gibbons-Hawking type of sources. This will require to add a possible NUT charge in the metric ansatz \eqref{eq:WeylSolfin} by replacing $dy^2 \rightarrow (dy+A(\rho,z) \,d\phi)^2$. From the resolution in four dimensions by Papapetrou \cite{Papapetrou:1953zz}, we hope that the structure of the equations of motion is also not changed and their solvability remains. By doing so, we will be able to source the solutions by Gibbons-Hawking centers. This will allow first to resolve the conical defect of the single-bubble solutions as discussed in section \ref{sec:ResolutionConical} and possibly remove the need for the struts. 

\section{Generalization to $D+1$ dimensions and type IIB embedding}
\label{sec:ExtensionDdim}

Our strategy so far has been a ``bottom-up'' approach to building ultra-compact smooth objects that can mimic astrophysical black holes. However, unlike other bottom-up toy models such as gravastars \cite{Mazur:2001fv} or boson stars \cite{Schunck:2003kk}, our class of solutions can be easily embedded into string theory and therefore can be motivated from a UV theory. In this section, we use a ``top-down'' approach and discuss how to embed our solutions in type IIB string theory on S$^1\times$T$^4$. We will perform the uplift of the five-dimensional solutions and compare them to known type IIB solutions. We will show that our class of spherically symmetric solutions that describe topological stars can be obtained from an analytic continuation on the parameter space of specific black hole solutions in string theory. Moreover, the embedding of the Weyl solutions constructed in the previous section will give a new class of non-supersymmetric non-extremal type IIB solutions consisting of a stack of D1-D5-KKm black holes and D1-D5-KKm smooth bubbles.

First, we consider generalizations of the spherically symmetric solutions in five dimensions to solutions in arbitrary dimensions. Indeed, these solutions have been constructed as a superposition of Schwarzschild solutions and bubble of nothing supported by fluxes. A similar strategy can be applied to construct solutions that are asymptotic to $D$-dimensional Minkowski times an extra S$^1$ using the Schwarzschild-Tangherlini solution \cite{Tangherlini:1963bw}. We will show that the class of $(D+1)$-dimensional solutions will have the same properties as the one studied in five dimensions. Moreover, we will study the type IIB embedding of the solutions for $D=5$ as D1-D5 solutions on T$^4$. This framework is a common playground for the microstate geometry program to study smooth bubbling geometries in the same regime as five-dimensional black holes. We will compare our solutions to the smooth JMaRT solutions \cite{Jejjala:2005yu} and see how our solutions can be non-rotating while JMaRT is forced to live in the unphysical over-rotating regime of black hole. 

\subsection{The class of two-charge solutions in $D+1$ dimensions}
\label{sec:Ddimext}

In this section, we construct a class of spherically symmetric topological stars and black strings that are asymptotic to a S$^1$ fibration over $D$-dimensional Minkowki. 
Since the analysis and the properties of the solutions are very similar to the one performed for $D=4$ in section \ref{sec:bubble&BH5d}, we will be brief and we refer the reader to the Appendix \ref{app:D+1dimext} for more details.

We consider a $(D+1)$-dimensional Einstein-Maxwell theory with the following action
\begin{equation}
S_{D+1}=\int \mathrm{d}^{D+1} x \sqrt{-\det g}\left(\frac{1}{2 \kappa_{D+1}^{2}} R-\frac{1}{2} \left|F^{(m)}\right|^2-\frac{1}{2} \left|F^{(e)}\right|^2\right)\,,
\label{eq:ActionGenD+1}
\end{equation}
where $\kappa_{D+1} $ is the $(D+1)$-dimensional Einstein gravitational constant, $F^{(m)}$ and $F^{(e)}$ are magnetic $(D-2)$-form and electric three-form field strengths respectively. The equations of motion are 
\begin{equation}
d\star_{D+1} F^{(m)} = 0 \,,\qquad d\star_{D+1} F^{(e)} = 0 \,,\qquad R_{\mu \nu} = \kappa_{D+1}^2\left( T_{\mu \nu} - \frac{1}{D-1} \,g_{\mu \nu} \, {T_\alpha}^\alpha\right)\,,
\label{eq:EOMgenD+1}
\end{equation}
where $T_{\mu \nu}$ is the stress tensor
\begin{align}
T_{\mu \nu} ~=~ &\frac{1}{(D-3)!} \, \left[{F^{(m)}}_{\mu \alpha_2 \ldots \alpha_{D-2}}{{F^{(m)}}_{\nu}}^{ \alpha_2 \ldots \alpha_{D-2}} - \frac{1}{2(D-2)}\,g_{\mu \nu} {F^{(m)}}_{\alpha_1 \ldots \alpha_{D-2}}{F^{(m)}}^{\alpha_1 \ldots \alpha_{D-2}}\right] \nonumber \\
&+  \frac{1}{2} \, \left[{F^{(e)}}_{\mu \alpha \beta }{{F^{(e)}}_{\nu}}^{ \alpha\beta} - \frac{1}{6}\,g_{\mu \nu} {F^{(e)}}_{\alpha \beta \gamma}{F^{(e)}}^{\alpha \beta \gamma}\right]\,.
\end{align}
As in five dimensions, we use a spherically symmetric ansatz that satisfies a double-Wick-rotation symmetry $(t,y) \to (iy,it)$, 
\begin{equation}
\begin{split}
ds^2_{D+1} &\= - f_\text{S}(r) \,dt^2 + f_\text{B}(r)\, dy^2 + \frac{dr^2}{f_\text{S}(r)\, f_\text{B}(r)} + r^2 \,d\Omega^2_{D-2} \,, \\
F^{(e)} &\= \frac{Q}{r^{D-2}}\, dr \wedge dt \wedge dy\,,\qquad F^{(m)} \= P\,dV_{S^{D-2}}\,,
\end{split}
\label{eq:metricGenD+1Dim}
\end{equation}
where $d\Omega_{D-2}$ and $dV_{S^{D-2}}$ are the line element and the volume form of a round S$^{D-2}$ sphere and $P$ and $Q$ correspond to magnetic and electric charges respectively. We consider the superposition of a $(D+1)$-dimensional bubble of nothing and a S$^1$ fibration over $D$-dimensional Schwarzschild-Tangherlini solution \cite{Tangherlini:1963bw},
\begin{equation}
f_\text{B}(r) \= 1 - \left(\frac{r_\text{B}}{r}\right)^{D-3}\,,\qquad f_\text{S}(r) \= 1 - \left(\frac{r_\text{S}}{r}\right)^{D-3}\,.
\label{eq:SolD+1dim}
\end{equation}
The Einstein equations are solved if the fluxes satisfy
\begin{equation}
P^2+Q^2 = \frac{(D-3)(D-1) \, r_\text{S}^{D-3} r_\text{B}^{D-3}}{2\,\kappa_{D+1}^2}\,.
\label{eq:P&QD+1dim}
\end{equation}
The solutions have two coordinate singularities at $r^{D-3}=r_\text{S}^{D-3}$ and $r_\text{B}^{D-3}$. The former corresponds to a horizon while the latter corresponds to the degeneracy of the $y$-circle.

As detailed in the Appendix \ref{app:redtoD}, we have solutions of Einstein-Maxwell-dilaton in $D$ dimensions with ADM mass, $\cM$, electric and magnetic charges, $Q_e$ and $Q_m$, given by
\begin{equation}
\begin{split}
\cM &\= \frac{\pi^{\frac{D-1}{2}}}{\kappa_D^2\,\Gamma\left(\frac{D-1}{2} \right)}\,\left( (D-2)\, r_\text{S}^{D-3}  + r_\text{B}^{D-3} \right)\,,\\
 \mathcal{Q}^2 &\equiv Q_m^2 + Q_e^2 \=  \frac{(D-3)(D-1) \, r_\text{S}^{D-3} r_\text{B}^{D-3}}{2\,\kappa_{D}^2}\,,\qquad Q_m \= \frac{P}{e}\,,\qquad Q_e \= \frac{Q}{e}\,.
\end{split}
\label{eq:ADMmass&chargesDd}
\end{equation}

The phase space of solutions has the same properties as in five dimensions depicted in Fig. \ref{fig:phasespace} but with different delimitations between the regions (see Appendix \ref{app:phasespace}). For given mass $\cM$ and charge $\cQ$, we have two solutions that are, depending on $r_\text{S}^{D-3} \lessgtr r_\text{B}^{D-3}$, either topological stars or black strings:
\begin{itemize}
\item[•] If $r_\text{S}^{D-3} < r_\text{B}^{D-3}$, the outermost coordinate singularity, $r=r_\text{B}^{D-3}$, corresponds to the degeneracy of the $y$-circle providing an end to spacetime. We have a horizonless solution which ends as a smooth bolt with a potential conical defect, $\IR^2 / \mathbb{Z}_k \times S^{D-2}$. The parameters $(r_\text{S}^{D-3}, r_\text{B}^{D-3},k)$ are constrained according to the radius of the $y$-circle as
\begin{equation}
R_y^2 \= \frac{4 \,r_\text{B}^{D-1}}{k^2\,(D-3)^2\,(r_\text{B}^{D-3} - r_\text{S}^{D-3})}\,, \qquad k \in \mathbb{Z}_+\,.
\end{equation}
It would be interesting to study whether the conical defect can be resolved by blowing up smooth Gibbons-Hawking bubbles at the vicinity of the poles of the bolt as for five-dimensional topological stars.
\item[•] If $r_\text{S}^{D-3} \geq r_\text{B}^{D-3}$, the outermost coordinate singularity corresponds to a horizon at $r=r_\text{S}^{D-3}$. The horizon has a S$^{D-2}\times$S$^1$ topology corresponding to a black string. The Bekenstein-Hawking entropy and the temperature are given by
\begin{equation}
S \= \frac{4\pi^{\frac{D+1}{2}}}{\Gamma\left(\frac{D-1}{2} \right)\,\kappa_D^2}\,\left(r_\text{S}^{D-1}\left( r_\text{S}^{D-3} - r_\text{B}^{D-3}\right) \right)^{\frac{1}{2}}\,,\qquad T \= \frac{D-3}{4\pi\,r_\text{S}} \, \sqrt{1- \left(\frac{r_\text{B}}{r_\text{S}}\right)^{D-3}}\,.
\end{equation}
Moreover, the locus $r=r_\text{B}^{D-3}$ in the interior corresponds to a degeneracy of the spacetime as a Milne space as described in section \ref{sec:BS4d/5d}.
\end{itemize}

\subsection{Embedding in type IIB String Theory}
\label{sec:Embedding}

There are many ways to embed Einstein-Maxwell theories in string theory. We investigate the simplest embedding by considering torus compactification. We restrict the discussion to the five-dimensional and six-dimensional solutions by compactification of type IIB supergravity on S$^1\times$T$^4$ and T$^4$ respectively. It will be important to make the difference between the new S$^1$ and the previous $y$-circle since the former is supposed to be internal with a much smaller size. For that purpose, we will rename the S$^1$ that describes the $y$-circle as S$^1_y$ and use S$^1$ for the new internal circle.

We will start by the embedding of the six-dimensional solutions that we obtain from the previous section with $D=5$ since it is less involved than the five-dimensional solutions. They will correspond to D1-D5 solutions on S$^1_y\times$T$^4$ with equal charges. We pay particular attention to these solutions since they are well-studied in the microstate geometry program. Our construction gives the first construction of smooth non-supersymmetric D1-D5 solutions with the same mass and charges as non-extremal D1-D5 black holes.


The five-dimensional solutions correspond to type IIB solutions on S$^1_y\times$S$^1\times$T$^4$ with equal D1 and D5 charges, but one also needs to turn on a KK monopole on S$^1$ with the same charge. The magnetic field we observe in five dimensions corresponds to the sum of two magnetic fields that have different UV origins: one corresponds to a KKm charge while the other arises from a D5 charge.  The embedding of the Weyl solutions constructed in section \ref{sec:Multi} gives interesting and brand new configurations in type IIB made of a chain of D1-D5-KKm objects in the non-supersymmetric and non-extremal regime.


 Before going through the details, we first fix the conventions. We consider the action of type IIB supergravity in the string frame as
\begin{align}
S_{\mathrm{IIB}} \=& \frac{1}{2\kappa_{10}^2 } \int \sqrt{-\det g}\left[e^{-2 \Phi}\left(R+4(\partial \Phi)^{2}-\frac{H^{2}}{12}\right) -\frac{1}{2}\left(\left|F_{1}\right|^{2}+\left|F_{3}\right|^{2}+\frac{1}{2}\left|F_{5}\right|^{2}\right)\right] \nonumber \\ &-\frac{1}{4\kappa_{10}^2 } \int  C_{4} \wedge H \wedge d C_{2}\,,
\end{align}
where the R-R field strengths, $F_p$, in terms of the potentials are
\begin{equation}
F_{1}=d C_{0}, \qquad F_{3}=d C_{2}-C_{0} H, \qquad F_{5}=d C_{4}-H \wedge C_{2}\,.
\end{equation}
In this convention, the matter fields are renormalized with the gravitational couplings, $2\kappa_{10}^2$, unlike our convention for the Einstein-Maxwell action \eqref{eq:ActionGenD+1} and \eqref{eq:ActionGen}. 

The solutions of interest will have NS-NS fields turned off $(H=0,\Phi=0)$, the equations of motions that are relevant to us are
\begin{equation}
\begin{split}
&R_{\mu \nu} \=\frac{1}{2} \left[F_{1\,\mu}F_{1\,\nu} +\frac{1}{2}F_{3\,\mu ab}F_{3\,\nu}^{\,\,\,\,ab} +\frac{1}{48}F_{5\,\mu abcd}F_{5\,\nu}^{\,\,\,\,abcd} - \frac{1}{24} g_{\mu\nu}F_{3\, abc}F_{3}^{\,\,abc}\right]\,,\\
&d\star_{10} F_1 \= d\star_{10} F_3 \= d\star_{10} F_5 \= 0\,,\qquad F_5 \= \star_{10} F_5\,,\\
&R \= 0 \,,\qquad F_1 \wedge \star F_3 \+ F_3 \wedge F_5 \= 0\,.
\end{split}
\label{eq:EOMtypeIIB}
\end{equation}

\subsubsection{The embedding of the six-dimensional solutions}

The class of solutions, \eqref{eq:metricGenD+1Dim}, is special for $D=5$ since both electric and magnetic field strengths are three-forms. We can embed the electromagnetic fields in $F_3$ under a compactification on T$^4$:
\begin{equation}
\begin{split}
ds_{10}^2 \=& - \left(1- \left(\frac{r_\text{S}}{r}\right)^2 \right) \,dt^2 \+  \left(1- \left(\frac{r_\text{B}}{r}\right)^2\right)\,dy^2 \+ \frac{r^4 \,dr^2}{(r^2-r_\text{S}^2)(r^2-r_\text{B}^2)}\\
&\+ r^2 \,\left(d\vartheta^2 + \sin^2  \vartheta \,d\phi^2 + \cos^2 \vartheta\,d\psi^2 \right) \+ \sum_{i=1}^4 dz_i^2\,, \\
 C_2 \=& -\frac{Q_1}{r^2} \,dt \wedge dy \- Q_5 \,\cos^2 \vartheta \,d\psi \wedge d\phi\,,
\end{split}
\label{eq:met&GFtypeIIB5d}
\end{equation}
where $(z_i)_{i=1,\ldots,4}$ are the coordinates of the rigid T$^4$ of volume $V_{T^4}$, and $(\vartheta,\phi,\psi)$ are the Hopf coordinates of the S$^3$ with $0\leq \vartheta \leq \frac{\pi}{2}$ and $0\leq \phi,\psi\leq 2\pi$. The equations of motion \eqref{eq:EOMtypeIIB} are solved if
\begin{equation}
Q_1 \= Q_5 \= \pm\, r_\text{S} \, r_\text{B}\,.
\label{eq:chargetypeIIB5d}
\end{equation}
We have then a class of non-rotating type IIB solutions with $Q_1$ D1-brane charge, $Q_5$ D5-brane charges and with ADM mass
\begin{equation}
\cM \= \frac{\pi^2}{\kappa_5^2} \,\left(3 r_\text{S}^2 \+ r_\text{B}^2\right)\,,\qquad \kappa_5^2 \= \frac{\kappa_{10}^2}{2\pi R_y \, V_{T^4}}\,.
\end{equation}
After reduction on T$^4$, we retrieve the $D=5$ solutions \eqref{eq:metricGenD+1Dim} with \eqref{eq:SolD+1dim} with the identification $$F^{(e)} + F^{(m)} \= \frac{\sqrt{V_{T^4}}}{\sqrt{2}\,\kappa_{10}}\,dC_2 \= \frac{1}{\sqrt{2}\,\kappa_6} \,dC_2\,.$$
Note that the constraint on the charges \eqref{eq:chargetypeIIB5d} matches the one we obtained earlier \eqref{eq:P&QD+1dim} if we require the magnetic charge to be equal to the electric charge, $P=Q$.  The fact that the charges are identified is a consequence of imposing a rigid-T$^4$ compactification.

The black string obtained from \eqref{eq:met&GFtypeIIB5d} with $r_\text{S}^2 \geq r_\text{B}^2 $ is identical to the Cvetic-Youm non-rotating D1-D5 black string with equal charges \cite{Cvetic:1996xz,Cvetic:1997uw,Giusto:2004id}. The map can be done considering
\begin{equation}
r^2 \rightarrow r^2 + r_\text{B}^2 \,,\qquad r_\text{S}^2 \equi M \cosh^2 \delta\,,\qquad r_\text{B}^2 \equi M \sinh^2 \delta\,,
\end{equation}
where $\delta$ is the boost parameter giving rise to the D1 and D5 charges. From this point of view, taking $r_\text{B}^2 > r_\text{S}^2 $ requires an analytic continuation of the boost parameter that keeps the metric real valued.  This possibility is a consequence of the double Wick rotation symmetry between the time direction and the $y$-circle.

In \cite{Jejjala:2005yu}, a procedure has been applied to construct smooth solutions with the same topology as our topological stars using the class of Cvetic-Youm solutions. However, if $r_\text{S}^2$ is taken to be greater than $r_\text{B}^2$, large angular momenta are required to impose a circle degeneracy ``before'' the horizon. Therefore, the solutions have the same conserved quantities as unphysical over-rotating black holes in five dimensions. Our analytic continuation allows to bypass this constraint and our non-rotating smooth solutions are direct examples. We can apply the procedure of \cite{Jejjala:2005yu} with the analytic continuation of the parameters of Cvetic-Youm solutions and try to obtain a rotating smooth solution with charges in the physical regime of the corresponding black hole.  It is given as
\begin{align}
ds_{10}^2 \=& -\left(c_p dt-s_p dy\right)^2 + \frac{r_\text{S}^2}{\Sigma} \left(c_p dt-s_p dy + \omega_t\right)^2 + \left(c_p dy-s_p dt\right)^2 -\frac{r_\text{B}^2}{\Sigma} \left(c_p dy-s_p dt -\omega_y \right)^2 \nonumber \\
&+ \Sigma\, \left[ \frac{r^2dr^2}{g(r)}  + d\vartheta^2 \right] + \left(r^2+a_\phi^2\right)\,\sin^2\vartheta\,d\phi^2+ \left(r^2+a_\psi^2\right)\,\cos^2\vartheta\,d\psi^2 \+\sum_{i=1}^4 dz_i^2\,, \nonumber \\
C_2 \=& \pm \frac{r_\text{S} r_\text{B}}{\Sigma}\,\left(c_p dt-s_p dy+\omega_t\right)\wedge \left(c_p dy-s_p dt-\omega_y \right) \pm r_\text{S} r_\text{B} \,\cos^2 \vartheta \,d\psi\wedge d\phi \,, 
\label{eq:metricrotation}
\end{align}
where $(c_p,s_p)= (\cosh \delta_p, \sinh \delta_p)$ corresponds to the boost parameter giving rise to the P charge, $(a_\psi,a_\phi)$ are the angular-momentum parameters and we have defined
\begin{align}
\Sigma &\equi r^2 +a_\psi^2 \,\sin^2 \theta + a_\phi^2 \,\cos^2 \theta\,,\quad  g(r) \equiv (r^2-r_\text{B}^2+a_\phi^2)(r^2-r_\text{B}^2+a_\psi^2) -(r_\text{S}^2-r_\text{B}^2)(r^2-r_\text{B}^2)\,,\nonumber \\
\omega_t &\equi  a_\psi \cos ^{2} \theta\,d\psi+a_\phi \sin ^{2} \theta \,d\phi\,,\quad \omega_y \equi a_\phi \cos ^{2} \theta \,d \psi+ a_\psi\sin ^{2} \theta\, d \phi\,.
\label{eq:omegadef}
\end{align}
The double Wick rotation symmetry is more subtle but still present. The class of solutions is symmetric under $$(t,y)\rightarrow (iy,it) \qquad \mbox{with} \qquad (r^2,\vartheta,\phi,\psi)\rightarrow \left(r^2 +a_\phi^2 +a_\psi^2 , \vartheta+\frac{\pi}{2},\psi,\phi\right)\,,$$ and transform the parameters $(r_\text{S}, r_\text{B} , a_\psi ,a_\phi) \rightarrow (r_\text{B} , r_\text{S} , -i a_\phi, -i a_\psi)$. This shows, as for the non-rotating class, that each solution with a vanishing timelike Killing vector has a symmetric partner in the same class for which a spacelike Killing shrinks. However, imposing a smooth degeneracy requires more works. From a first study, it seems that the presence of angular momenta imposes smooth solutions to have either $r_\text{S}^2 > r_\text{B}^2$, that is to be JMaRT solutions, or to have a conical excess where the circle degenerates\footnote{We are grateful to David Turton for the discussion in that regards.} which corresponds to presence of struts.  We will study this issue in future work.

\subsubsection{The embedding of the five-dimensional solutions}

In this paper, we have constructed two classes of five-dimensional solutions: the spherically symmetric solutions in section \ref{sec:bubble&BH5d} and their axisymmetric Weyl generalizations in section \ref{sec:Multi}. We will first study the embedding of the former for which the formalism is slightly simpler.

\subsubsection*{The spherically symmetric solutions}

It is natural to embed the five-dimensional metric \eqref{eq:metric5d} in type IIB by considering a rigid T$^5$. However, there are many ways to embed the magnetic two-form field strength, $F^{(m)}$, as it can arise from $F_3$ or from a KK monopole charge along a circle of the T$^5$. Therefore, we take the following ansatz in type IIB by considering a S$^1\times$T$^4$
\begin{equation}
\begin{split}
ds_{10}^2 \=& - \left(1- \frac{r_\text{S}}{r}\right) \,dt^2 \+  \left(1-\frac{r_\text{B}}{r}\right)\,dy^2 \+ \frac{r^2 \,dr^2}{(r-r_\text{S})(r-r_\text{B})}\\
&\+ r^2 \,\left(d\theta^2 + \sin^2  \theta \,d\phi^2  \right) \+ \,\left(dz_5 + A \right)^2 \+ \sum_{i=1}^4 dz_i^2\,, \\
 C_2 \=& -\frac{Q_1}{r} \,dt \wedge dy \-  \,Q_5 \,\cos \theta \,dz_5 \wedge d\phi\,,\\
 dA \=&\, p \,\sin \theta \,d\theta \wedge d\phi\,,
\end{split}
\label{eq:met&GFtypeIIB4d}
\end{equation}
where $(z_i)_{i=1,\ldots,4}$ are the coordinates of the rigid T$^4$ of volume $V_{T^4}$, and $z_5$ is the coordinate of the extra S$^1$ of radius $L$. The equations of motion are solved providing
\begin{equation}
Q_1 \= Q_5 \= p \= \pm\, \sqrt{r_\text{S} \, r_\text{B}}\,.
\label{eq:chargetypeIIB4d}
\end{equation}
We have then a class of non-rotating type IIB solutions with equal D1-brane charge, D5-brane charge, and KKm charge. The ADM mass is
\begin{equation}
\cM \= \frac{2\pi}{\kappa_4^2} \,\left(2\, r_\text{S}^2 \+ r_\text{B}^2\right)\,,\qquad \kappa_4^2 \= \frac{\kappa_{10}^2}{4\pi^2 L\,R_y \, V_{T^4}}\,.
\end{equation}
After compactification along S$^1\times$T$^4$, the metric matches the five-dimensional metric \eqref{eq:metric5d}, but we obtain three gauge fields, two magnetic and one electric that we appropriately renormalize to have the same convention as the Einstein-Maxwell action \eqref{eq:ActionGen}
\begin{equation}
\begin{split}
F^{(e)} &\= \frac{\sqrt{\pi L V_{T^4}}}{\kappa_{10}}\,\frac{Q_1}{r^2}\,dr \wedge dt \wedge dy \= \frac{1}{\sqrt{2}\,\kappa_5}\,\frac{Q_1}{r^2}\,dr \wedge dt \wedge dy\,,\\
F^{(m1)} &\= \frac{1}{\sqrt{2}\,\kappa_5}\,Q_5 \sin \theta \,d\theta \wedge d\phi\,, \qquad 
F^{(m2)} \= \frac{1}{\sqrt{2}\,\kappa_5}\,p \sin \theta \,d\theta \wedge d\phi\,.
\end{split}
\label{eq:5dGFfromtypeIIB}
\end{equation}
In five dimensions, the magnetic gauge fields are indistinguishable, that is why it has been appropriate to recast into an unique gauge field. However, their UV origin is very different, one corresponds to a D5 charge while the other one corresponds to a KKm charge. Considering two magnetic gauge fields in section \ref{sec:class4d} will change the constraint on the charges \eqref{eq:Field5dGen} to $$Q^2+ P_1^2+P_2^2 \= \frac{3 r_\text{S} r_\text{B}}{2 \kappa_5^2} \,,$$
which is indeed compatible to the constraint we obtain in type IIB \eqref{eq:chargetypeIIB4d}. As in the previous section, the degree of freedom between the charges in Einstein-Maxwell theory has been frozen due to the compactification on a S$^1\times$T$^4$.

The black string obtained from \eqref{eq:met&GFtypeIIB4d} with $r_\text{S} \geq r_\text{B}$ is identical to the four-dimensional non-rotating three-charge STU black hole \cite{Cvetic:1995kv,Cvetic:2011dn} which can be embedded in type IIB following \cite{Chow:2014cca}. The map can be done as in the previous section,
\begin{equation}
r \rightarrow r + r_\text{B} \,,\qquad r_\text{S} \equi M \cosh^2 \delta\,,\qquad r_\text{B} \equi M \sinh^2 \delta\,,
\end{equation}
where $\delta$ is the boost parameter giving rise to the D1, D5 and KKm charges. Our construction offers an enlargement of the class of STU single-center solutions exploiting a double-Wick rotation symmetry to replace the $\cosh$ and $\sinh$ to arbitrary values. As it has been done with the five-dimensional Cvetic-Youm solutions in the previous section, we can use the embedding of the more general class of rotating four-charge STU solutions to generalize our present class to rotating solutions. This will be a subject of future studies.

\subsubsection*{The axisymmetric Weyl solutions}

We aim to generalize the embedding to the class of two-charge Weyl solutions \eqref{eq:WeylSolfin}. We will use a similar ansatz as before by considering an extra S$^1\times$T$^4$ and the magnetic field in five dimensions will arise from the connection along the S$^1$ and from the R-R two-form field:
\begin{equation}
\begin{split}
ds_{10}^2 \=& W_\text{GF}^{-1} \left[ -W_0 \,dt^2 +W_0^{-1} \, dy^2 \right]+ W_\text{GF}^2 \left[ e^{2(\nu_0 + \nu_\text{GF})} \left(d\rho^2 + dz^2 \right) + \rho^2 d\phi^2 \right] \\
& \+ \,\left(dz_5 + A \right)^2 \+ \sum_{i=1}^4 dz_i^2\,, \\
F_3 \=& dC_2 \=  \frac{q_1}{\rho\,W_\text{GF}^2} \,\star_2 dH\,\wedge dt\wedge dy \+ q_5 \,  d H  \wedge d\phi \wedge dz_5\,,\\
 dA \=&\, q_\text{KKm}\,d H \wedge d\phi\,.
\end{split}
\label{eq:met&GFtypeIIB4drod}
\end{equation}
Once again, the Einstein equations along the rigid T$^4$ and S$^1$ requires to take
\begin{equation}
q_1 \= q_5 \= q_\text{KKm}\,,
\end{equation}
which we can be fix to $1$ by reabsorbing into $H$. Therefore, we also end with D1-D5-KKm configurations with equal charges. Otherwise, the solutions work the same way as in five dimensions with a different normalization for the gauge-field scalar $H$. The solutions are given by two arbitrary functions that solve a Laplace equation on the three-dimensional base
\begin{equation}
\cL \left(\log W_0 \right) \=0 \,,\qquad \cL \left(\frac{1}{\rho} \,\partial_\rho K \right) \= 0\,, \qquad \text{with} \quad \cL \equi \frac{1}{\rho} \,\partial_\rho \left( \rho \partial_\rho \right) \+ \partial_z^2\,.
\end{equation}
The scalars $(W_\text{GF}, H)$ are given by
\begin{equation}
W_\text{GF}^2 \= F_I\left(\frac{1}{\rho} \partial_\rho K \right) \,,\qquad H \= \partial_z K\,,
\end{equation}
where $F_I$ is one of the five generating functions of one variable given by two real parameters \eqref{eq:DefFi}. The base scalars $(\nu_0,\nu_\text{GF})$ are obtained by integrating \eqref{eq:Eqfornu1} and \eqref{eq:EqfornuGF}. We retrieve the class of five-dimensional solutions with an electric and magnetic gauge fields after compactification on S$^1\times$T$^4$ by appropriately reshuffling the gauge fields as in \eqref{eq:5dGFfromtypeIIB}.

We can uplift all the multi-rod solutions constructed by sourcing $(W_0,\rho^{-1} \partial_\rho K)$ with rod sources in section \ref{sec:Rodsources}. In type IIB, they correspond to D1-D5-KKm non-extremal black strings and D1-D5-KKm non-BPS bubbles stacked on a line and prevented from collapse by struts.

\section{Discussion}
\label{sec:concl}

In this paper, we have shown from a bottom-up approach that smooth ultra-compact structure à la microstate geometries can be constructed with minimum of ingredients: electromagnetic gauge fields, an extra dimension and allowing non-trivial topology wrapped by fluxes. The topological stars have a minimal degree of complexity as they are static and spherically symmetric, but are good prototypes for testing the features of microstate geometries in an astrophysical regime. We argue that their size can range from microscopic to the macroscopic scales compared to the size of the extra dimension.   

For macroscopic topological stars, the solutions have the same malleability as bottom-up  ECO models, which will allow to estimate many observable deviations with respect to expectations for GR black holes. However, their physical viability and the scope of the outcomes will be much more robust since their UV origin in a quantum gravity theory as D1-D5-KKm solutions of type IIB string theory has been established. For this aspect, it will be very interesting to describe the physical characteristics of topological stars as seen by an asymptotic observer. As a non-exhaustive list of interesting computations, this will consist in studying the geodesics and the photon shell in such backgrounds \cite{Bah-Heidmann-White}, quasi-normal modes and information recovery, gravitational radiation and tidal Love number. Because of their simple structure compared to known microstate geometries, this will allow a more qualitative understanding of bubbles as microstate geometries.

For microscopic topological stars, which should rather be called topological particles, we have noted in this paper that nothing a priori prohibits the nucleation of microscopic objects of the size of the extra dimension and of mass $\cM \sim \frac{R_y}{\kappa_4^2}$. Such an observation could also have been made with known microstate geometries. If the size of the $y$-circle is slightly larger than the string scale\footnote{It must be larger than the string scale to avoid quantum instabilities \cite{Adams_2005}}, these objects have mass of order slightly larger than $10^2 M_P $ where $M_P$ is the Planck mass in four dimensions. They are generated by hidden electromagnetic fluxes that can be weakly coupled with the fields of the standard model. It is interesting to ask whether early universe processes could create stable configurations of massive bubbles that are long-lived as possible new candidates for dark matter.

The physical characteristics of the non-extremal two-charge black strings is also interesting. As noted in section \ref{sec:nonextremalBS4d/5d}, their curvature singularity is hidden by a curvature-free origin of Milne space. This could give interesting prototypes of traversable wormholes.

Furthermore, an important question to address is about stability. It is well-known that gravity with extra dimensions can lead to instabilities. Neutral black strings have a Gregory-Laflamme instability that forces them to decay to stable black holes \cite{Gregory:1993vy}, while static vacuum bubbles of nothing are semi-classically unstable, but the presence of gauge fields can drastically change this feature. The classical stability of similar black strings as ours, for which only the magnetic flux has been turned on, has been studied in \cite{Miyamoto:2006nd}. It has been shown that they are free from classical linear instability for $\frac{1}{2}r_\text{S} \leq r_\text{B} \leq r_\text{S}$. Extending to $r_\text{B} > r_\text{S}$ shows that the topological stars are classically stable for the full range of parameters\footnote{This follows from a stability analysis by Anindya Dey.}. 

In parallel, we have extended the construction of generalized Weyl solutions that has been derived in five-dimensional Einstein theory in \cite{Costa:2000kf,Emparan:2001wk,Elvang:2002br,Emparan:2008eg,Charmousis:2003wm}. Generic solutions consist in neutral black strings and bubble of nothing on a line and separated by struts. Adding gauge fields undermines the linearity of the Weyl equations of motion but highlights a non-trivial backreaction nature. We have been able to solve the equations of motion and find closed-form solutions by defining five types of gauge-field backreactions. By studying one in particular, we constructed the generalized charged Weyl solutions that consist of two-charge black strings and topological stars on a line. Unfortunately, regularity did not allow to have different orientations between the fluxes and then the objects are still separated by struts. From this result, we can wonder if the struts are ``quantum'' ingredients that must been taken into account to support structure at the vicinity of non-extremal black holes or if we did not turn on enough classical degrees of freedom to get rid of them. It would be interesting if the need of struts can be made manifest with orientifold planes in string theory. However, we still believe that the second option is possible by allowing NUT charges, angular momentum along $\phi$ and momentum along $y$. We motivate this argument by the well-known two Kissing Kerr solutions in four dimensions that has resolved the struts between two Schwarzschild black hole by imposing opposite angular momenta \cite{KRAMER1980259}. 

An interesting aspect of the generalized charged Weyl solutions is that they can be also embedded in type IIB string theory as multiple D1-D5-KKm static black strings and smooth bubbles stacked on a line. They give the first non-trivial examples of multi-center three-charge solutions in such framework far within the non-supersymmetric and non-extremal regime. With that regards, studying the interactions between centers and their dynamics would be interesting for further studies.

\section*{Acknowledgments}
We are grateful to Iosif Bena, Kim Berghaus, Frederico Bonetti, Anindya Dey, David Kaplan, Daniel R. Mayerson, David Turton, Nick Warner and Zackary White for interesting conversations and correspondence. The work of IB and PH is supported in part by NSF grant PHY-1820784.

\vspace{2cm}

\appendix
\leftline{\LARGE \bf Appendices}

\section{Charged Weyl solutions in five dimensions}
\label{app:WeylSol}
In this section we will give all the details of the construction of axisymmetric solutions of the Einstein-Maxwell theory \eqref{eq:ActionGen} in five dimensions using the Weyl formalism. We want to construct solutions that are asymptotic to a S$^1$ fiber over a flat four-dimensional spacetime. We start with a general axisymmetric ansatz for the metric 
\begin{equation}
ds^2 \= -f_\text{S}(\rho,z) \,dt^2 + f_\text{B}(\rho,z) \, dy^2 +h(\rho,z)  \left[ e^{2\nu(\rho,z)} \left(d\rho^2 + dz^2 \right) + \rho^2 d\phi^2 \right]\,,
\label{eq:metricaxisymApp}
\end{equation}
where $t$ is the time direction, $y$ is the coordinate parametrizing the extra S$^1$ with periodicity $2\pi R_y$, $(\rho,z,\phi)$ defines the cylindrical coordinates of the three-dimensional base. Moreover, the assumption of axisymmetry also constrains the two field strengths to be
\begin{equation}
F^{(m)} \= d H(\rho,z) \wedge d\phi\,,\qquad F^{(e)} \= dZ(\rho,z) \wedge dt\wedge dy\,.
\end{equation}
We will first detail the computation of the Ricci and the stress energy tensors. Then, we will appropriately order the equations of motion to obtain the different layers written down in section \ref{sec:EOMaxisym}. Finally we will solve the equations, find closed-form solutions. We will explicitely derive solutions for rod sources and analyze carefully their regularity.

\subsection{Ricci tensor}
We will label the coordinates of the two-dimensional base as $(x_1,x_2)=(\rho,z)$ with the latin letter ``$a,b,c...$''. We will use the tetrad formalism for which the indices are raised and lowered by the Minkowski metric $\eta_{MN} =\text{Diag}(-1,1,1,1,1)$. The tetrad one-forms obtained from the metric \eqref{eq:metricaxisymApp} are 
\begin{equation}
E^t = \sqrt{f_\text{S}}\, dt\,, \quad E^y =  \sqrt{f_\text{B}}\, dy\,, \quad E^a =  \sqrt{h} e^{\nu}\, dx^a\,, \quad E^\phi =  \rho  \sqrt{h}\,d\phi\,.
\end{equation} 
The spin connections, given by $dE^M = E^N\wedge \omega^M_{\,\,\,N} $, are 
\begin{equation}
\begin{split}
\omega^t_{\;\; a} &\=  \frac{1}{2} \,\sqrt{\frac{f_\text{S}}{h}}\,e^{-\nu}\, \partial_a \log(f_\text{S}) \,dt \,,\qquad \omega^y_{\;\; a} \=\frac{1}{2} \,\sqrt{\frac{f_\text{B}}{h}}\,e^{-\nu}\, \partial_a \log(f_\text{B}) \,dy\,, \\
\omega^\phi_{\;\; a} &\= \frac{1}2 \,\rho e^{-\nu}\,\partial_a \log (\rho^2 h) \,d\phi\,,\qquad \omega^a_{\;\; b}  \= \frac{1}2 \left( \partial_b\log(h e^{2\nu}) \,dx^a - \partial^a\log(h e^{2\nu}) \,dx_b \right) .
\end{split}
\end{equation} 
The curvature components, given by $\cR^M_{\;\;N}=d\omega^M_{\;\;N}+ \omega^M_{\;\;O}\wedge \omega^O_{\;\;N}$, lead to
\begin{equation}
\begin{split}
\mathcal{R}^t_{\;\; a} \=&- \frac{1}{2} \left[ \partial_b \left(\sqrt{\frac{f_\text{S}}{h}}\,e^{-\nu}\partial_a\,\log(f_\text{S})\right) - \frac{1}2 \,\sqrt{\frac{f_\text{S}}{h}}\,e^{-\nu}\,\partial_a \log(h e^{2\nu} )\partial_b \log(f_\text{S})  \right] dt \wedge dx^b \\
&- \frac{1}{4}\,\sqrt{\frac{f_\text{S}}{h}}\,e^{-\nu}\,\partial^b \log(h e^{2\nu} )\partial_b \log(f_\text{S})  \, dt \wedge dx_a \,, \\
\mathcal{R}^t_{\;\; y} \=& \frac{1}4 \,\frac{\sqrt{f_\text{B} f_\text{S}}}{h}\,e^{-2\nu}\, \partial^a \log (f_\text{B})\, \partial_a \log (f_\text{S})\,dy\wedge dt \,,\\
\mathcal{R}^t_{\;\; \phi} \=& -\frac{1}4\,\frac{\rho \,\sqrt{f_\text{S}}}{\sqrt{h}}\,e^{-2\nu}\, \partial^a \log(\rho^2 h)\, \partial_a \log (f_\text{S}) \, dt \wedge d\phi \,,\\
\mathcal{R}^y_{\;\; a} \=&- \frac{1}{2} \left[ \partial_b \left(\sqrt{\frac{f_\text{B}}{h}}\,e^{-\nu}\partial_a\,\log(f_\text{B})\right) - \frac{1}2 \,\sqrt{\frac{f_\text{B}}{h}}\,e^{-\nu}\,\partial_a \log(h e^{2\nu} )\partial_b \log(f_\text{B})  \right] dy\wedge dx^b \\
&- \frac{1}{4}\,\sqrt{\frac{f_\text{B}}{h}}\,e^{-\nu}\,\partial^b \log(h e^{2\nu} )\partial_b \log(f_\text{B})  \, dy \wedge dx_a \,, \\
\mathcal{R}^y_{\;\; \phi} \=& -\frac{1}4\,\frac{\rho \,\sqrt{f_\text{B}}}{\sqrt{h}}\,e^{-2\nu}\, \partial^a \log(\rho^2 h)\, \partial_a \log (f_\text{B}) \, dy \wedge d\phi \,,\\
\mathcal{R}^\phi_{\;\; a} \=& \frac{1}{2} \left[ \partial_b \left(\rho e^{-\nu} \partial_a \log (\rho^2 h) \right) - \frac{1}2 \partial_a \log(h e^{2\nu} ) ~ \partial_b\log (\rho^2 h) \rho e^{-\nu}   \right] dx^b \wedge d\phi \\
&+ \frac{1}{4} \rho e^{-\nu}  ~\partial^b \log(h e^{2\nu} ) ~\partial_b \log (\rho^2 h)~   dx_a \wedge d\phi \,, \\
\mathcal{R}^a_{\;\; b} \=& - \frac{1}{2}\,\left(\partial_1^2 + \partial_2^2 \right) \left[\log (h e^{2\nu})\right] dx^a \wedge dx_b\,.
\end{split}
\end{equation}
Therefore, the non-vanishing components of the Ricci tensor, $R_{MN}=\cR^O_{\;\;MON}$, will be given as
\begin{equation}
\begin{split}
2he^{2\nu}\,\,R_{tt} \=& \frac{1}{\rho\sqrt{h f_\text{S} f_\text{B}}}\,\, \partial^a \left[\rho\sqrt{h f_\text{S} f_\text{B}}\,\,\partial_a \log (f_\text{S})\right]\,, \\
2he^{2\nu}\,\,R_{yy} \=&- \frac{1}{\rho\sqrt{h f_\text{S} f_\text{B}}}\,\, \partial^a \left[\rho\sqrt{h f_\text{S} f_\text{B}}\,\,\partial_a \log (f_\text{B})\right] \,,\\
2he^{2\nu}\,\,R_{\phi \phi} \=&- \frac{1}{\rho\sqrt{h f_\text{S} f_\text{B}}}\,\,\partial^a \left[\rho\sqrt{h f_\text{S} f_\text{B}}\,\,\partial_a \log (\rho^2 h)\right]\,,\\
2he^{2\nu}\,\, R_{ab} \=& -\partial_a \partial_b \log\left(\rho^2 h f_\text{S} f_\text{B} \right)- \frac{1}{2} \,\partial_a \log f_\text{S} ~ \partial_b\log f_\text{S}- \frac{1}{2} \,\partial_a \log f_\text{B} ~ \partial_b\log f_\text{B}  \\
& -\frac{1}{2}\, \partial_a \log (\rho^2 h) ~ \partial_b \log (\rho^2 h)   +\frac{1}2 \,\partial_{(a} \log(h e^{2\nu}) \partial_{b)} \log \left( \rho^2 h f_\text{S} f_\text{B}\right) \\
& - \frac{1}2 \left[ \partial^c \log(h e^{2\nu}) \partial_c \log \left(\rho^2 h f_\text{S} f_\text{B} \right) +2\partial^c \partial_c (\log (h e^{2\nu})) \right] \delta_{ab}\,.
\end{split}
\end{equation}

\subsection{Stress energy tensor}

We rewrite the field strengths in the tetrad basis
\begin{align}
F^{(m)} &\= \frac{e^{-\nu}}{\rho\, h} \,\partial_a H E^a \wedge E^\phi \,,\qquad F^{(e)} \= \frac{e^{-\nu}}{\sqrt{hf_\text{S}f_\text{B}}}\,\partial_a Z \,E^a\wedge E^t\wedge E^y\,.
\label{eq:GFApp}
\end{align}
The stress energy tensor in the tetrad basis is given by
\begin{equation}
\begin{split}
T_{MN} ~=~ & {F^{(m)}}_{MO}{{F^{(m)}}_{N}}^{O} - \frac{1}{4}\,\eta_{MN} {F^{(m)}}_{OP}{F^{(m)}}^{OP}\\
&+  \frac{1}{2} \, \left[{F^{(e)}}_{\mu OP }{{F^{(e)}}_{\nu}}^{OP} - \frac{1}{6}\,\eta_{MN} {F^{(e)}}_{OPQ}{F^{(e)}}^{OPQ}\right]\,.
\end{split}
\end{equation}
We find
\begin{equation}
\begin{split}
T_{tt} &\= - T_{yy} = T_{\phi \phi} = -T_{MN} \eta^{MN}= \frac{1}{2} \frac{e^{-2\nu}}{\rho^2 h^2}\, \left(  \partial_a H \partial^a H + \frac{\rho^2 h}{f_\text{S}f_\text{B}}\, \partial_a Z \partial^a Z \right)\,, \\
T_{ab} &\= \frac{e^{-2\nu}}{\rho^2 h^2} \left(\partial_a H \partial_b H - \frac{\rho^2 h}{f_\text{S}f_\text{B}} \, \partial_a Z \partial_b Z - \frac{\delta_{ab}}{2} \left(  \partial_c H \partial^c H - \frac{\rho^2 h}{f_\text{S}f_\text{B}} \, \partial_c Z \partial^c Z \right) \right)\,.
\end{split}
\end{equation}

\subsection{The equations of motion}

The Einstein equation in five dimensions, 
\begin{equation}
 R_{MN} = \kappa_{5}^2\left( T_{MN} - \frac{1}{3} \,\eta_{MN} \, {T_O}^O \right)\,,
\end{equation}
gives along $(tt)$, $(yy)$ and $(\phi\phi)$
\begin{equation}
\begin{split}
&\sqrt{ \frac{\rho^2 h}{f_\text{S} f_\text{B}}}\,\, \partial^a \left[\rho\sqrt{h f_\text{S} f_\text{B}}\,\,\partial_a \log (f_\text{S})\right] \= \frac{2 \kappa_5^2}{3}\,\left(  \partial_a H \partial^a H + \frac{\rho^2 h}{f_\text{S}f_\text{B}}\, \partial_a Z \partial^a Z \right)\,, \\
&\sqrt{ \frac{\rho^2 h}{f_\text{S} f_\text{B}}}\,\, \partial^a \left[\rho\sqrt{h f_\text{S} f_\text{B}}\,\,\partial_a \log (f_\text{B})\right]  \= \frac{2 \kappa_5^2}{3}\,\left(  \partial_a H \partial^a H + \frac{\rho^2 h}{f_\text{S}f_\text{B}}\, \partial_a Z \partial^a Z \right)\,,\\
&\sqrt{ \frac{\rho^2 h}{f_\text{S} f_\text{B}}}\,\,\partial^a \left[\rho\sqrt{h f_\text{S} f_\text{B}}\,\,\partial_a \log (\rho^2 h)\right] \= -\frac{4 \kappa_5^2}{3}\,\left(  \partial_a H \partial^a H + \frac{\rho^2 h}{f_\text{S}f_\text{B}}\, \partial_a Z \partial^a Z \right)\,.
\end{split}
\end{equation}
The sum of the three equations and the difference of the two first are clearly independent of the gauge fields and we can use the sum of the two first as the last equation. Therefore, we introduce new warp factors that are more appropriate for the equations,
\begin{equation}
V \equi \rho \sqrt{h f_\text{S} f_\text{B}} \,,\qquad W_0 \equi \sqrt{\frac{f_\text{S}}{f_\text{B}}}\,,\qquad W_\text{GF} \equi \frac{1}{\sqrt{f_\text{S}f_\text{B}}}\,.
\label{eq:newwarpApp}
\end{equation}
The three equations above transform to
\begin{equation}
\begin{split}
&\partial^a \partial_a V \= 0 \,,\qquad \partial^a \left( V \partial_a \log W_0 \right) \= 0\,,\\
& V W_\text{GF}^2 \, \partial^a \left( V \partial_a \log W_\text{GF} \right) \= -\frac{2 \kappa_5^2}{3}\,\left(  \partial_a H \partial^a H + V^2 W_\text{GF}^4\, \partial_a Z \partial^a Z \right)
\end{split}
\label{eq:firstsetEqApp}
\end{equation}
The metric according to the new warp factors, \eqref{eq:newwarpApp}, is given in \eqref{eq:metricAxisym}. The two equations in the first line are independent of the gauge fields and then are the same equations that one obtains for vacuum solutions. Such equations and their solutions are well-known and well-studied \cite{Costa:2000kf,Emparan:2001wk,Elvang:2002br,Emparan:2008eg,Charmousis:2003wm}. Therefore, the warp factor $W_0$ is the ``purely massive'' warp factor while $W_\text{GF}$ is the warp factor cooresponding to the gauge-field backreaction.

The remaining Einstein equations along the two-dimensional flat base $(ab)$ give a priori three equations, $(\rho\rho)$, $(zz)$ and $(\rho z)$, that constrain $\nu$. The equations are not independent and we write a set of two equations, obtained from $(\rho z)$ and $(\rho \rho)-(zz)$, for $\partial_\rho \nu$ and $\partial_z \nu$ for which the integrability condition is guaranteed by the equations above \eqref{eq:firstsetEqApp}:
\begin{equation}
\begin{split}
\partial_\rho \log V\,\partial_z \nu \+ \partial_z \log V \,\partial_\rho \nu \= & \frac{1}{2} \partial_z \log W_0 \,\partial_\rho \log W_0 \+ \frac{3}{2}\partial_z \log W_\text{GF} \,\partial_\rho \log W_\text{GF} \\
& + \frac{\kappa_5^2}{V^2 W_\text{GF}^2} \left(\partial_\rho H\,\partial_z H - V^2W_\text{GF}^4 \partial_\rho Z \partial_z Z  \right) \\
& + \partial_\rho \partial_z \log V - \partial_z \log V \, \partial_\rho \log V + \frac{\partial_z \log V}{\rho}\,,\\
\partial_\rho \log V\,\partial_\rho\nu - \partial_z \log V \,\partial_z \nu \= & \frac{1}{4} \left( (\partial_\rho \log W_0)^2 -(\partial_z \log W_0)^2 \right) + \frac{3}{4} \left( (\partial_\rho \log W_\text{GF})^2 -(\partial_z \log W_\text{GF})^2 \right) \\
& +  \frac{\kappa_5^2}{2V^2 W_\text{GF}^2} \left[ (\partial_\rho H)^2 -(\partial_z H)^2 + V^2W_\text{GF}^4  \left( (\partial_z Z)^2 -(\partial_\rho Z)^2\right) \right] \\
& + \frac{1}{2}\left[(\partial_\rho^2 -\partial_z^2) \log V + (\partial_z \log V)^2 -(\partial_\rho \log V)^2 + 2 \frac{\partial_\rho \log V}{\rho}\right]\,.
\end{split}
\label{eq:secsetEqApp}
\end{equation}
It is natural to split $\nu$ in two pieces, \be \nu \equi \nu_0 + \nu_\text{GF}\,\ee where $\nu_0$ is only sourced by the ``purely massive'' warp factors and $\nu_\text{GF}$ is sourced by the gauge-field scalars. We will give these equations in a moment, but before that we write down the Maxwell equations, $d\star F^{(m)} = 0$ and $d\star F^{(e)} = 0$, and obtain
\begin{equation}
\partial^a \left[ \frac{1}{V W_\text{GF}^2} \,\partial_a H\right] \= 0\,,\qquad \partial^a \left[ V W_\text{GF}^2 \,\partial_a Z\right] \= 0\,.
\label{eq:thirdsetEqApp}
\end{equation}
We will restrict to gauge fields that are electromagnetic duals, that is we will assume that
\begin{equation}
F^{(e)} \= q \,\star F^{(m)}\,,
\end{equation}
where $q$ is the charge ratio between the electric and magnetic charges. From the expressions of the field strengths \eqref{eq:GFApp}, it implies
\begin{equation}
\partial_a Z \= \frac{q}{V W_\text{GF}^2}\,\epsilon^b_{\,\,a} \,\partial_b H\,,
\end{equation}
where $\epsilon_{ab}$ is the two-dimensional Levi-Civita tensor and the Maxwell equation for $Z$ is straightforwardly satisfied. 

We can now have a final version for the equations of motion \eqref{eq:firstsetEqApp}, \eqref{eq:secsetEqApp} and \eqref{eq:thirdsetEqApp}. We divide them in layers that will facilitate the construction of solutions. 
\begin{itemize}
\item[•] \underline{The zeroth layer:}
\begin{equation}
\partial_\rho^2 V \+ \partial_z^2 V \= 0\,.
\label{eq:zerothlayerApp}
\end{equation}
\item[•] \underline{The mass layer:}
\begin{equation}
\partial_\rho \left( V \,\partial_\rho \log W_0 \right) \+ \partial_z \left( V \,\partial_z \log W_0 \right) \= 0\,.
\label{eq:masslayerApp}
\end{equation}
\item[•] \underline{The Maxwell layer:}
\begin{equation}
\begin{split}
&\partial_\rho \left(\frac{1}{V\,W_\text{GF}^2} \partial_\rho H \right) \+ \partial_z \left(\frac{1}{V\,W_\text{GF}^2} \partial_z H \right) \= 0\,,\\
&V\,W_\text{GF}^2 \,\left[\partial_\rho \left( V \,\partial_\rho \log W_\text{GF} \right) \+ \partial_z \left( V \,\partial_z \log W_\text{GF} \right)  \right] \= - \frac{2(1+q^2)\kappa_5^2}{3} \,\left[ \left(\partial_\rho H\right)^2 +\left(\partial_z H\right)^2 \right]\,.
\end{split}
\label{eq:MaxwelllayerApp}
\end{equation}
\item[•] \underline{The  base layer:}
\end{itemize}
\begin{equation}
\begin{split}
\partial_\rho \log V\,\partial_z \nu_0 + \partial_z \log V \,\partial_\rho \nu_0 \= & \frac{1}{2} \partial_z \log W_0 \,\partial_\rho \log W_0  + \partial_\rho \partial_z \log V - \partial_z \log V \, \partial_\rho \log V + \frac{\partial_z \log V}{\rho}\,,\\
\partial_\rho \log V\,\partial_\rho\nu_0 - \partial_z \log V \,\partial_z \nu_0 \= & \frac{1}{4} \left( (\partial_\rho \log W_0)^2 -(\partial_z \log W_0)^2 \right) \\
& +\frac{1}{2} \left[(\partial_\rho^2 -\partial_z^2) \log V + (\partial_z \log V)^2 -(\partial_\rho \log V)^2 + 2 \frac{\partial_\rho \log V}{\rho}\right]\,, \\
\partial_\rho \log V\,\partial_z \nu_\text{GF}+ \partial_z \log V \,\partial_\rho \nu_\text{GF} \= & \frac{3}{2}\partial_z \log W_\text{GF} \,\partial_\rho \log W_\text{GF} + \frac{(1+q^2)\kappa_5^2}{V^2 W_\text{GF}^2}\,\partial_\rho H \partial_z H \,,\\
\partial_\rho \log V\,\partial_\rho\nu_\text{GF} - \partial_z \log V \,\partial_z \nu_\text{GF} \= & \frac{3}{4} \left( (\partial_\rho \log W_\text{GF})^2 -(\partial_z \log W_\text{GF})^2 \right) \\
& +  \frac{(1+q^2)\kappa_5^2}{2V^2 W_\text{GF}^2} \left[ (\partial_\rho H)^2 -(\partial_z H)^2 \right] \,.
\end{split}
\label{eq:EqfornuApp}
\end{equation}

\subsection{Closed-form solutions}
\label{app:closedformSol}
In this section, we detail the derivation of the solutions. Apart from the Maxwell layer, all the equations can be treated as linear equations with potential quadratic sources. Surprisingly, we can find closed-form solutions that rely only on one extra assumption to solve the Maxwell layer. As for vacuum solutions \cite{Costa:2000kf,Emparan:2001wk,Elvang:2002br,Emparan:2008eg,Charmousis:2003wm}, the solutions will be entirely given by two functions that solve a Laplace equation on the three-dimensional base. The main difference is that those functions will intervene in the warp factors and gauge fields in a much richer manner than in vacuum. 

\subsubsection{The zeroth layer}

Solutions of the zeroth layer \eqref{eq:zerothlayerApp} are a priori given by 
\begin{equation}
V \= f_1 (\rho + i z) + f_2(\rho - i z) \,,
\end{equation}
where $f_1$ and $f_2$ are arbitrary functions of one variable. As it has been showed in details in \cite{Emparan:2001wk}, one can pick a gauge by changing the coordinates of the two-dimensional base where 
\begin{equation}
V \= \rho\,.
\end{equation}
This coordinate system is commonly refered as the ``Weyl's canonical coordinates''. Very briefly, if we use the complex conjugate coordinates $w \equiv \rho + i z$ and $\bar{w} \equiv \rho - i z$, the two-dimensional metric is given by 
\begin{equation}
ds_{(\rho,z)}^2 \= dw \,d\bar{w}\,.
\end{equation}
The change of coordinates $(w,\bar{w}) \rightarrow ( f_1 (w),f_2 (\bar{w}))$, induces a conformal factor in the two-dimensional metric that can be absorbed in $\nu$. Moreover, the new $\rho$ and $z$ coordinates, given by $\rho/z = f_1 (w) \pm f_2 (\bar{w})$, imply that $V = \rho$.

\noindent From now on, we consider without loss of generality that 
\begin{equation}
V \= \rho\,.
\end{equation}
The benefit of choosing such a coordinate system is that two equations of motion transform to Laplace equation in the three-dimensional $(\rho,z,\phi)$  base space. Indeed, the Laplacian\footnote{The Laplacian is more rigorously $\partial_\rho \left( \rho \partial_\rho \right) \+ \rho\, \partial_z^2$, but we renormalize it for convenience.} in the base gives
\begin{equation}
\cL \equi \frac{1}{\rho} \,\partial_\rho \left( \rho \partial_\rho \right) \+ \partial_z^2\,,
\end{equation}
and the equations for $\log W_0$ and $\log W_\text{GF}$ are both Laplace equations with extra coupling terms for $\log W_\text{GF}$.

\subsubsection{The mass layer}

The mass layer is now given by the equation 
\begin{equation}
\cL (\log W_0 ) \= 0\,.
\end{equation}
This equation is well-known and well-studied. It is the equation one obtains for the warp factors in vacuum Einstein solutions \cite{Bonnor,Israel1964,Costa:2000kf,Emparan:2001wk}. We can source $\log W_0$ by massive rods or massive point particles. Rod sources are usually preferred by the fact that they give Schwarzschild-types of warp factors. We will study the solutions obtained by such sources in the section \ref{app:MultiRod}. 

\subsubsection{The Maxwell layer}

We aim to find closed-form solutions of the equations that govern the pair $(H,W_{GF})$:
\begin{equation}
(M):\,\,\partial^a \left[\frac{1}{\rho W^2_{GF}}\,\partial_a H \right]  =0\quad \text{and}\quad (E):\,\,\rho W_{GF}^2\,\partial^a \left[\rho \,\partial_a \log W_{GF}\right]= -\frac{2 \,(1+q^2)\,\kappa_5^2}{3}\, \partial_a H \partial^a H.
\end{equation}
First let's have a clean set of variables. We define
\begin{equation}
\gamma^2 \equi \frac{4 \,(1+q^2)\,\kappa_5^2}{3}\,,\qquad U \equi W^{-2}_{GF}\,,
\end{equation}
and we have
\begin{equation}
(M):\,\,\,\partial^a \left[\frac{U}{\rho}\,\partial_a H \right]  =0\quad \text{and}\quad (E):\,\,\,\partial^a \left[\rho \,\partial_a \log U\right]= \frac{\gamma^2}{\rho}\,U\, \partial_a H \partial^a H\,.
\end{equation}
The only assumption we will make is that we will decompose $(M)$ into two parts that will cancel rather than solving in full generality
\begin{equation}
(M1):\,\,\,\partial_a \left(\frac{1}{\rho}\partial^a H \right) =0 \quad \text{and}\quad (M2):\,\,\, \partial_a U \partial^a H \= 0\,.
\end{equation}
This is motivated by the fact that our single-center solutions satisfy this relation and that those equations are known to contain solutions for $H$ that can be sourced by magnetic charges \cite{Papapetrou:1953zz}. We now expand the equations:
\begin{equation}
\begin{split}
&(M1):\,\,\, \partial_\rho^2 H \+ \partial_z^2 H - \frac{1}{\rho}\,\partial_\rho H \= 0\,,\\
&(M2):\,\,\, \partial_\rho U \partial_\rho H \+ \partial_z U \partial_z H \=0\,,\\
&(E):\,\,\,  \partial_\rho^2 U \+ \partial_z^2 U + \frac{1}{\rho}\,\partial_\rho U  - \frac{1}{U}\left((\partial_\rho U)^2 +(\partial_z U)^2 \right) \= \frac{\gamma^2 \,U^2}{\rho^2} \,\left((\partial_\rho H)^2 +(\partial_z H)^2 \right)\,.
\end{split}
\end{equation}
The proof will be based on the following observation:

\noindent \emph{Consider an arbitrary solution of Laplace equation $\bar{K}(\rho,z)$, that is $$ \cL(\bar{K}) \= \partial_\rho^2 \bar{K} \+ \partial_z^2 \bar{K} + \frac{1}{\rho}\,\partial_\rho \bar{K}  \=0\,.$$ Therefore, if we write $\bar{K}$ as a function of $U$ only, $\bar{K}(U)$, then we have 
\begin{equation}
\cL(\bar{K}) \= \bar{K}'(U) \,\cL(U) \+ \bar{K}''(U)\, \left((\partial_\rho U)^2 +(\partial_z U)^2 \right) \=0\,.
\label{eq:KasU}
\end{equation}}
Note how close this equation is to $(E)$. From $(M2)$, it is very likely that $(\partial_\rho H)^2 +\partial_z H)^2 \propto (\partial_\rho U)^2 +(\partial_z U)^2$, then $(E)$ will be identical to the above equation. From the coefficients, we will be able to find the explicit $\bar{K}(U)$ that we will invert into $U(\bar{K})$. Therefore, the construction scheme consists in appropriately using $(M1)$ and $(M2)$ to obtain an equation $(E)$ that depends only on $U$.

The equation $(M2)$ implies that
\begin{equation}
\partial_\rho U \= \Gamma(\rho,z)\,\partial_z H\,,\qquad \partial_z U \= - \Gamma(\rho,z)\,\partial_\rho H\,,
\label{eq:resolutionM2}
\end{equation}
for which the integrability gives
\begin{equation}
\partial_\rho^2 H \+ \partial_z^2 H \+ \partial_\rho \log \Gamma \,\partial_\rho H \+ \partial_z \log \Gamma \,\partial_z H\= 0\,.
\end{equation}
Using now $(M1)$ to replace $\partial_\rho^2 H \+ \partial_z^2 H$ and \eqref{eq:resolutionM2} to replace $\partial_a H$ we get
\begin{equation}
\partial_\rho \log(\rho \Gamma) \,\partial_z U - \partial_z \log(\rho \Gamma) \,\partial_\rho U \= 0\,.
\end{equation}
This equation is trivial to integrate and we have
\begin{equation}
\Gamma \= \frac{1}{\rho}\,G(U)\,,
\end{equation}
where $G$ is an arbitrary differentiable function. Now let us pack everything to get an expression for $\cL(U)$, we obtain
\begin{equation}
\partial_\rho^2 U \+ \partial_z^2 U \= -\frac{G}{\rho^2}  \partial_z H \+ \frac{G'}{\rho} \partial_\rho U\, \partial_z H - \frac{G'}{\rho} \partial_z U\, \partial_\rho H \,,
\label{eq:GasU}
\end{equation}
which leads to
\begin{equation}
\cL(U) \= \frac{G'}{G} \, \left((\partial_\rho U)^2 +(\partial_z U)^2 \right) \,.
\end{equation}
Moreover, we have
\begin{equation}
(\partial^2_\rho H)^2 +(\partial_z H)^2 \= \frac{\rho^2}{G^2}\left((\partial_\rho U)^2 +(\partial_z U)^2 \right) \,.
\end{equation}
Replacing the two last expressions into $(E)$ we get\footnote{Note that to write down this equation we have divided by $(\partial^2_\rho H)^2 +(\partial^2_z H)^2$. This restricts the discussion to solution with non-constant $H$, that is to solutions with the gauge fields turned on. Therefore, we do not expect to retrieve in our closed-form solutions the expressions one would obtain in vacuum.}
\begin{equation}
\frac{G'(U)}{G(U)} \= \frac{1}{U} \+ \frac{\gamma^2 U^2}{G(U)^2}\quad \Leftrightarrow \quad (G^2)' -\frac{2}{U}\,G^2 \= 2 \gamma^2 U^2\,.
\label{eq:eqproofApp}
\end{equation}
This equation is integrable and gives
\begin{equation}
G(U)^2 \= U^2 \left(c \+ 2 \gamma^2 \,U \right)\,,  \qquad c\in \mathbb{C}\,.
\end{equation}
We can now use \eqref{eq:KasU} with \eqref{eq:GasU}. We can consider an arbitrary function $\bar{K}$ of Laplace equation such as
\begin{equation}
\frac{\bar{K}''(U)}{\bar{K}'(U)} \= -\frac{G'(U)}{G(U)} \quad \Rightarrow \quad \bar{K}'(U) \= \frac{a}{G(U)}\,, \qquad a\in \mathbb{C}\,.
\end{equation}
We finally obtain
\begin{equation}
\bar{K}(U) \=b \- \frac{2 a}{\sqrt{c}} \,\text{arctanh}\left[ \sqrt{1 + \frac{2\gamma^2 U}{c}} \right]\,,\qquad b\in \mathbb{C}\,.
\end{equation}
This is simple to invert
\begin{equation}
U \= -\frac{c}{2\gamma^2} \,\frac{1}{ \cosh^2\left[\frac{\sqrt{c}}{2\,a} \left(\bar{K}-b \right) \right]}\,,\qquad (a,b,c)\in \mathbb{C}\,.
\end{equation}
To finish the resolution, we now have to find $H$ from
\begin{equation}
\partial_z H \=\frac{\rho}{G(U)} \,\partial_\rho U \= \frac{\rho}{a} \partial_\rho \bar{K}\,,\qquad \partial_\rho H \= - \frac{\rho}{G(U)} \,\partial_z U \= -\frac{\rho}{a} \partial_z \bar{K}\,.
\end{equation}
The best then is to define a function $K$ such that $\rho^{-1} \partial_\rho K \= \bar{K}$ and we immediatly obtain that
\begin{equation}
H \= - \frac{1}{a}\, \partial_z K\,.
\end{equation}

We can appropriately redefine the three constants to get
\begin{equation}
W^2_{GF} \= U^{-1} \=- \frac{\cosh^2 \left( a \,\rho^{-1} \partial_\rho K +b\right)}{c^2}\,,\qquad H \= \sqrt{\frac{3}{2(1+q^2)\kappa_5^2}}\,\frac{a}{c}\,\partial_z K\,,\qquad (a,b,c)\in \mathbb{C}\,,
\end{equation}
and $K$ is a function satisfying
\begin{equation}
\cL \left(\frac{1}{\rho} \,\partial_\rho K \right)\=0\,.
\label{eq:LapKApp}
\end{equation}
The third parameter is irrelevant since we can rescale $\frac{a}{c} K \rightarrow K$, $c \rightarrow a$ and we obtain
\begin{equation}
W^2_{GF} \= U^{-1} \=- \frac{\cosh^2 \left( a \,\rho^{-1} \partial_\rho K +b\right)}{a^2}\,,\qquad H \= \sqrt{\frac{3}{2(1+q^2)\kappa_5^2}}\,\partial_z K\,,\qquad (a,b)\in \mathbb{C}\,.
\end{equation}
To finish the proof one needs to check that those solutions indeed satisfy the equations. We find that this is the case if in addition 
\begin{equation}
\rho \partial_\rho \left( \frac{1}{\rho} \partial_\rho K \right) + \partial_z^2 K \= 0\,.
\end{equation}
If it looks a stronger constraint compared to \eqref{eq:LapKApp}, it is not. Indeed we have
\begin{equation}
\cL \left(\frac{1}{\rho} \,\partial_\rho K \right)\=0 \quad \Leftrightarrow \quad \rho \partial_\rho \left( \frac{1}{\rho} \partial_\rho K \right) + \partial_z^2 K \=  f(z)\,,
\end{equation}
where $f$ is an arbitrary function of $z$. Moreover, by considering $\cL \left(\frac{1}{\rho} \,\partial_\rho K \right) =0$ we have an integration freedom $K \rightarrow K + g(z)$. The new constraint is just fixing correctly this integration freedom to the unique solution satisfying $f(z) =0$. \\

The solutions being mostly complex, we can define five branches of real solutions by appropriately playing with $(a,b)$. The Maxwell layer is then solved by considering a generating function $K$ that solves the following Laplace equation
\begin{equation}
\cL \left(\frac{1}{\rho} \,\partial_\rho K \right) \= 0\,,
\label{eq:KeqApp}
\end{equation}
and the pair $(H,W_\text{GF})$ is given by 
\begin{equation}
W^2_{\text{GF}} \=F_I \left(\frac{1}{\rho} \,\partial_\rho K \right)\,,\qquad H \= \sqrt{\frac{3}{2(1+q^2)\kappa_5^2}}\,\partial_z K\,,
\label{eq:WGF&HApp}
\end{equation}
where $F_I$ is taken from one of the five following choices
\begin{equation}
\begin{split}
F_1 (x) &\= \left(\frac{\sinh(a x+b)}{a} \right)^2\,,\qquad F_2(x) \= -\left(\frac{\cosh(a x+b)}{a} \right)^2\,,\qquad F_5 (x) \= (x +b)^2\,,\\
F_3 (x) &\= \left(\frac{\sin(a x+b)}{a} \right)^2\,,\qquad F_4(x) \= \left(\frac{\cos(a x+b)}{a} \right)^2\,,\qquad (a,b) \in \mathbb{R}\,.\\
\end{split}
\label{eq:FbranchApp}
\end{equation}
Note that we have necessarily assumed that the gauge fields have been turned on to write down \eqref{eq:eqproofApp}. If we want to retrieve the vacuum solutions we can simply take a function $K$ that satisfies \eqref{eq:KeqApp}, but we the pair $(H,W_\text{GF})$ is given by
\begin{equation}
W^2_{\text{GF}} \=\exp \left(\frac{1}{\rho} \,\partial_\rho K \right)\,,\qquad H \=0\,.
\end{equation}
Therefore, the presence of the gauge fields did not change the nature of the sources since we can source $\rho^{-1} \partial_\rho K$ by rods or point particles similarly as in vacuum. However, the way the sources for the gauge fields backreact in the metric warp factors have a richer form as we have five possible branches of backreaction and two arbitrary parameters $(a,b)$.

\subsubsection{The base layer}

The equations of the base layer drastically simplify with $V=\rho$,
\begin{equation}
\begin{split}
\partial_z \nu_0 \= & \frac{\rho}{2} \partial_z \log W_0 \,\partial_\rho \log W_0 \,,\qquad \partial_\rho\nu_0  \=  \frac{\rho}{4} \left( (\partial_\rho \log W_0)^2 -(\partial_z \log W_0)^2 \right) \,, \\
\partial_z \nu_\text{GF} \= & \frac{3\rho}{2}\partial_z \log W_\text{GF} \,\partial_\rho \log W_\text{GF} + \frac{(1+q^2)\kappa_5^2}{\rho W_\text{GF}^2}\,\partial_\rho H \partial_z H\,, \\
\partial_\rho\nu_\text{GF} \= & \frac{3\rho}{4} \left( (\partial_\rho \log W_\text{GF})^2 -(\partial_z \log W_\text{GF})^2 \right) +  \frac{(1+q^2)\kappa_5^2}{2\rho W_\text{GF}^2} \left[ (\partial_\rho H)^2 -(\partial_z H)^2 \right] \,.
\end{split}
\end{equation}
These equations are simple integral equations for which the integrability condition is guaranteed by the previous layers. The equation for $\nu_0$ is well-known as it is the same equation for vacuum solutions \cite{Bonnor,Israel1964,Costa:2000kf,Emparan:2001wk}. However, it needs to be integrated in a case-by-case manner depending on the type of sources for $\log W_0$. We will give the solutions for rod sources in section \ref{app:MultiRod}. 

As for $\nu_\text{GF}$, let us replace $(W_\text{GF},H)$ by their expressions to have an equation that depends on $K$ only. We find
\begin{equation}
\begin{split}
\partial_z \nu_{GF} &\= \frac{3\rho}{2} \,\partial_\rho \left(\frac{1}{\rho} \partial_\rho K \right)\partial_z \left(\frac{1}{\rho} \partial_\rho K \right) \,\frac{\frac{1}{4}\,F_I'^2-F_I}{F_I^2}\,,\\
 \partial_\rho \nu_{GF} &\= \frac{3\rho}{4} \left[\left(\partial_\rho \left(\frac{1}{\rho} \partial_\rho K \right)\right)^2-\left(\partial_z \left(\frac{1}{\rho} \partial_\rho K \right)\right)^2 \right] \,\frac{\frac{1}{4}\,F_I'^2-F_I}{F_i^2}\,,
\end{split}
\end{equation}
and we have
\begin{equation}
\frac{\frac{1}{4}\,F_I'^2-F_I}{F_I^2} \= \begin{cases} a^2 \quad \text{if} \quad I=1,2\,, \\
-a^2 \quad \text{if} \quad I=3,4 \,, \\
0  \quad \text{if} \quad I=5 \,,
\end{cases}\,.
\end{equation}
Therefore, 
\begin{equation}
\begin{split}
\partial_z \nu_{\text{GF}} &\= \begin{cases} 
\frac{3 a^2 \,\rho}{2} \,\partial_\rho \left(\frac{1}{\rho} \partial_\rho K \right)\,\partial_z \left(\frac{1}{\rho} \partial_\rho K \right) \quad \text{if} \quad I=1,2 \,,\\
-\frac{3 a^2 \,\rho}{2} \,\partial_\rho \left(\frac{1}{\rho} \partial_\rho K \right)\,\partial_z \left(\frac{1}{\rho} \partial_\rho K \right) \quad \text{if} \quad I=3,4 \,, \\
~~ 0 \qquad \text{if} \quad I=5\,,
\end{cases}\,,\\
 \partial_\rho \nu_{\text{GF}} &\= \begin{cases} 
\frac{3 a^2 \,\rho}{4} \,\left[\left(\partial_\rho \left(\frac{1}{\rho} \partial_\rho K \right)\right)^2-\left(\partial_z \left(\frac{1}{\rho} \partial_\rho K \right)\right)^2 \right]\quad \text{if} \quad I=1,2 \,,\\
-\frac{3 a^2 \,\rho}{4} \,\left[\left(\partial_\rho \left(\frac{1}{\rho} \partial_\rho K \right)\right)^2-\left(\partial_z \left(\frac{1}{\rho} \partial_\rho K \right)\right)^2 \right] \quad \text{if} \quad I=3,4 \,, \\
~~0 \qquad \text{if} \quad I=5\,,
\end{cases}\,.
\end{split}
\label{eq:EqfornuGFApp}
\end{equation}
The equations have exactly the same form as the equations for $\nu_0$ but with $\log W_0$ replaced by $\rho^{-1} \partial_\rho K$ and with different weights that depend on $a^2$. Thus, the base layer gives simple integral equations that must be integrated in a case-by-case manner depending on the kind of sources for $(\log W_0,\,\rho^{-1} \partial_\rho K)$.

To conclude, we have found closed-form axisymmetric two-charge solutions of the Einstein-Maxwell theory \eqref{eq:ActionGen} in five dimensions. We refer to \eqref{eq:WeylSolfin} and the paragraph that follows for a complete summary of the solutions.

\subsection{Rod solutions}
\label{app:MultiRod}

We consider that the sources are obtained from $n$ distinct rods on the $z$-axis centered around $z=a_i$ and with length $M_i$. We refer to the section \ref{sec:Rodsources} for the derivation of the main functions $(H,W_0,W_\text{GF},\nu_0,\nu_\text{GF})$. We will discuss in more details the constraints on the asymptotics and the regularity of the metric and gauge fields.

\subsubsection{Asymptotics}

Far away from the rods, $\rho \gg z^\pm_i$ or/and $z \gg z^\pm_i$,  we have
\begin{equation}
r_\pm^{(i)} \sim \sqrt{\rho^2 +z^2}\,, \qquad R_\pm^{(i)} \sim 2 \sqrt{\rho^2 +z^2} \,,\qquad E_{\pm \pm}^{(i,j)} \sim 2 \left(\rho^2 +z^2 \right)\,.
\end{equation}
It is then appropriate to use spherical coordinates
\begin{equation}
\rho = r \sin \theta \,,\qquad z = r\cos \theta\,.
\end{equation}
The main scalars at large $r$ are given by
\begin{equation}
H \sim  \frac{\sqrt{3}}{\sqrt{2(1+q^2)}\kappa_5}\,\sum_{i=1}^n M_i P_i\,\cos \theta  \,,\qquad W_\text{GF} \sim \frac{\sinh b}{a} \,,\qquad W_0 \sim 1\,,\qquad e^{2(\nu_0+\nu_\text{GF})} \sim 1\,.
\end{equation}
The metric and field strengths \eqref{eq:WeylSolfin} are asymptotic to
\begin{equation}
\begin{split}
ds^2&\, \sim\, \frac{a}{\sinh b} \left( -dt^2 +dy^2\right) \+ \frac{\sinh^2 b}{a^2} \left( dr^2 +r^2 d\Omega_2^2\right)\,,\\
F^{(m)} &\,\sim\, -  \frac{\sqrt{3}}{\sqrt{2(1+q^2)}\kappa_5} \sum_{i=1}^n M_i P_i \,\sin \theta \,d\theta \wedge d\phi\,,\qquad F^{(e)} \,\sim\,   \frac{\sqrt{3}a^2 q\,\sum_{i=1}^n M_i P_i }{\sqrt{2(1+q^2)}\kappa_5\,\sinh^2 b}\,\frac{1}{r^2}\,dr \wedge dt \wedge dy\,.
\end{split}
\end{equation}
The solutions are asymptotic to $\IR^{1,3}\times$S$^1$ if
\begin{equation}
a = \sinh b\,.
\label{eq:asymptoticsCond}
\end{equation}
In the four-dimensional framework described in section \ref{eq:reduction4d} obtained after KK reduction along $y$, the solutions are massive solutions with magnetic and electric charges. The conserved quantities are given by
\begin{equation}
\cM \= \frac{2\pi}{\kappa_4^2}\sum_{i=1}^n M_i \left(3  P_i \cosh b - G_i\right)\,,\qquad {Q_e} \= q\,{Q_m} \= \frac{\sqrt{3}\,q}{\sqrt{2(1+q^2)}\,\kappa_4} \,\sum_{i=1}^n M_i P_i \,.
\end{equation}

\subsubsection{Regularity}
\label{app:regularity}

The ratio $R_+^{(i)}/R_-^{(i)}$ and $\nu_{ij}$ \eqref{eq:GeneratingNu} have potential zeroes or divergences on the $z$-axis only. Thus, $W_0$ and $e^{2(\nu_0+\nu_\text{GF}})$ are regular and positive out of the $z$-axis. However, $W_\text{GF}$ can have zeroes out of the $z$-axis. Indeed, due to the $\sinh$ expression of $W_\text{GF}$ \eqref{eq:WarpfactorsRods}, if two successive rods have different signs of charges  $P_i< 0$ and $P_{i+1}>0$, then $W_\text{GF} \rightarrow - \infty$ at the $i^\text{th}$ rod and $+\infty$ at the $(i+1)^\text{th}$ rod. By continuity, there is a closed curve in between the two rods and that is not only supported on the $z$-axis where $W_\text{GF}$ vanishes. On this curve, the metric \eqref{eq:WeylSolfin} is singular. We are then forced to require that all charges have the same sign,
\begin{equation}
P_i \,>\,0\,.
\end{equation}
This is a strong restriction since the rods will not be repulsed from each other using fluxes with opposite orientations. Taking the branches $F_2$ or even $F_4$ for $W_\text{GF}$ \eqref{eq:FbranchApp} would have allowed for charges with different sign. However, we have observed that those branches have other complications to deal with.

Assuming that $P_i$ is positive for each rod implies that $W_\text{GF}$ is positive everywhere. Therefore, the metric is regular out of the $z$-axis and free from closed timelike curves. We will now discuss the regularity on the $z$-axis which is more involved. We will divide the discussion in two parts: the regularity at the rods where the $y$-circle or the time direction shrink and the regularity out of the rods where the $\phi$-circle shrinks.

\begin{itemize}
\item[•] \underline{At the $i^\text{th}$ rod:}
\end{itemize}

The local spherical coordinates around the i$^{\text{th}}$ rod are given by $r_i \rightarrow 0$ for $0\leq \theta_i <\pi$ with
\begin{equation}
\rho = \sqrt{r_i(r_i+M_i)}\,\sin\theta \,,\qquad z = \left(r_i+\frac{M_i}{2} \right) \cos\theta_i +a_i\,.
\end{equation}
The two-dimensional base behaves as
\begin{equation}
d\rho^2 + dz^2 \sim \frac{M_i \sin^2 \theta_i}{4} \left(\frac{dr_i^2}{r_i}+M_i\, d\theta_i^2 \right)\,.
\end{equation}
Moreover,
\begin{equation}
\frac{R_+^{(i)}}{R_-^{(i)}} \sim \frac{M_i}{r_i} \,,\qquad \frac{R_+^{(j)}}{R_-^{(j)}} \sim \frac{\frac{M_j}{2} + \left|a_i-a_j+\frac{M_i}{2}\cos\theta_i \right|}{-\frac{M_j}{2} + \left|a_i-a_j+\frac{M_i}{2}\cos\theta_i \right|}\,,\quad j\neq i\,.
\label{eq:RpRmexp}
\end{equation}
Thus,
\begin{equation}
W_\text{GF} \sim \frac{e^b}{2\sinh b}\,\prod_{j\neq i} \left( \frac{R_+^{(j)}}{R_-^{(j)}} \right)^{ P_j\,\sinh b}\,\left(\frac{M_i}{r_i} \right)^{P_i\, \sinh b }\, \,,\qquad W_0 \sim\left( \frac{M_i}{r_i} \right)^{ G_i}\,\prod_{j\neq i} \left(  \frac{R_+^{(j)}}{R_-^{(j)}}\right)^{G_j}\,.
\end{equation}
The metric components along $(t,y,\phi)$ give
\begin{equation}
\begin{split}
g_{tt} &\sim -\frac{2\sinh b}{e^b}  \,\prod_{j\neq i} \left(  \frac{R_+^{(j)}}{R_-^{(j)}}\right)^{G_j- P_j\,\sinh b}\, \left(\frac{M_i}{r_i} \right)^{ G_i-P_i\,\sinh b}\,,\\
g_{yy} &\sim  \frac{2\sinh b}{e^b}\,\prod_{j\neq i} \left(  \frac{R_+^{(j)}}{R_-^{(j)}}\right)^{-G_j-P_j\,\sinh b} \,\left(\frac{M_i}{r_i} \right)^{- G_i-P_i\,\sinh b} \,,\\
g_{\phi\phi} &\sim \frac{M_i^2\,e^{2b}}{4\sinh^2 b}\,\prod_{j\neq i} \left( \frac{R_+^{(j)}}{R_-^{(j)}} \right)^{2 P_j\,\sinh b} \,\left(\frac{M_i}{r_i} \right)^{2P_i\,\sinh b-1}\, \sin^2 \theta_i  \,.\\
\end{split}
\label{eq:ytphibehaviorBubblerod}
\end{equation}
The $\phi$-circle must have a finite size at $r_i =0$. This fixes the magnetic charges to be
\begin{equation}
P_i = \frac{1}{2 \sinh b}\,.
\end{equation}
Moreover, the condition that $g_{tt}$ and $g_{yy}$ do not diverge requires that
\begin{equation}
|G_i| \leq  \frac{1}{2}\,.
\end{equation}
We have three interesting cases: $G_i = \frac{1}{2}$ and $g_{tt}$ is finite while $g_{yy} \propto r_i$, $G_i = -\frac{1}{2}$ and $g_{yy}$ is finite while $g_{tt} \propto r_i$ and $G_i = 0$ and $g_{tt} \propto  g_{yy} \propto \sqrt{r_i}$. To derive the metric components along $(r_i,\theta_i)$ we first need the limits for $\nu_{jk}$ \eqref{eq:GeneratingNu}. We have different situations:
\begin{equation}
\begin{split}
e^{\nu_{ii}}&\sim \frac{16 r_i^2}{M_i^2\sin^4 \theta_i}\,,\qquad \quad e^{\nu_{jk}}\ \sim \begin{cases}
1 \qquad &\text{if } i<j\leq k  \text{ or } j\leq k<i  \\
\dfrac{(z_k^- - z_j^+)^2(z_k^+ - z_j^-)^2}{(z_k^+ - z_j^+)^2(z_k^- - z_j^-)^2} \qquad &\text{if } j<i<k 
\end{cases}\,, \\
e^{\nu_{ij}} &\sim \begin{cases} 
\dfrac{(z_j^+ - z_i^-)^2\left(z_j^- - \left(a_i +\frac{M_i}{2} \cos\theta_i \right)\right)^2}{(z_j^- - z_i^-)^2\left(z_j^+ - \left(a_i +\frac{M_i}{2} \cos\theta_i \right)\right)^2} \qquad &\text{if } j>i,  \\
\dfrac{(z_j^- - z_i^+)^2\left(z_j^+ - \left(a_i +\frac{M_i}{2} \cos\theta_i \right)\right)^2}{(z_j^+ - z_i^+)^2\left(z_j^- - \left(a_i +\frac{M_i}{2} \cos\theta_i \right)\right)^2} \qquad &\text{if } j<i, 
\end{cases}\,,
\end{split}
\label{eq:EEEEatrod}
\end{equation}
where we remind that $z^\pm_j$ are the rod endpoints \eqref{eq:coordinatesEndpoints}. We gather everything to derive the behavior of $\nu_\text{GF}  + \nu_0 = \frac{1}{4} \sum_{j,k=1}^n (3a^2 P_j P_k + G_j G_k) \,\nu_{jk}= \frac{1}{4} \sum_{j,k=1}^n (\frac{3}{4} + G_j G_k)\,\nu_{jk}$ \eqref{eq:nufinal},
\begin{equation}
\begin{split}
e^{2 \nu} \sim & \left(\frac{4 r_i}{M_i \sin^2 \theta_i} \right)^{\alpha_{ii}} \,\prod_{j\neq i} \left( \frac{R_+^{(j)}}{R_-^{(j)}}\right)^{-2\alpha_{ij}} \, \\
&\times \prod_{j=1}^{i-1} \prod_{k=i}^n \left(\dfrac{(z_k^- - z_j^+)(z_k^+ - z_j^-)}{(z_k^+ - z_j^+)(z_k^- - z_j^-)}  \right)^{2\alpha_{jk}}  \prod_{j\neq i} \left(\frac{z_j^+ - z_i^-}{z_j^- -z_i^-} \right)^{2\,\text{sign}(j-i) \,\alpha_{ij}} \,,
\end{split}
\end{equation}
where we have defined the exponents $\alpha_{jk}$ as
\begin{equation}
\alpha_{jk} \equi \frac{3+4 G_j G_k}{4}\,,
\end{equation}
and the expansion of $\frac{R_+^{(j)}}{R_-^{(j)}}$ is $\theta_i$-dependent given in \eqref{eq:RpRmexp}. We have also considered that the product ``$\prod_{j=1}^{i-1}$'' is equal to $1$ for the first rod, $i=1$. We define the constants $d_i$ that depend on the geometry of the rods only and are independent of $r_i$ and $\theta_i$
\begin{equation}
d_1 \equi 1\,,\qquad d_i \equi  \prod_{j=1}^{i-1} \prod_{k=i}^n \left(\dfrac{(z_k^- - z_j^+)(z_k^+ - z_j^-)}{(z_k^+ - z_j^+)(z_k^- - z_j^-)}  \right)^{ \frac{3 + 4 G_j G_k}{4}}\quad \text{when } i=2,\ldots n\,.
\end{equation}
Thus, $g_{r_i r_i}$ and $g_{\theta_i \theta_i}$ behave around the i$^\text{th}$ rod as
\begin{equation}
\begin{split}
g_{\theta_i \theta_i} \sim &  \frac{M_i^2\,e^{2b}\,d_i^2}{4\sinh^2 b} \, \prod_{j\neq i} \left(\frac{z_j^+ - z_i^-}{z_j^- -z_i^-} \right)^{2\,\text{sign}(j-i) \,\alpha_{ij}}\,\prod_{j\neq i} \left( \frac{R_+^{(j)}}{R_-^{(j)}}\right)^{1-2\alpha_{ij}}\, \left(\frac{4r_i}{M_i \sin^2 \theta_i} \right)^{G_i^2- \frac{1}{4}} \,, \\
g_{r_i r_i} \sim &  \frac{M_i^2\,e^{2b}\,d_i^2}{4\sinh^2 b} \, \prod_{j\neq i} \left(\frac{z_j^+ - z_i^-}{z_j^- -z_i^-} \right)^{2\,\text{sign}(j-i) \,\alpha_{ij}}\,\prod_{j\neq i} \left( \frac{R_+^{(j)}}{R_-^{(j)}}\right)^{1-2\alpha_{ij}}\,\frac{1}{r_i} \left(\frac{4r_i}{M_i \sin^2 \theta_i} \right)^{G_i^2- \frac{1}{4}} \,.
\end{split}
\label{eq:rthetabehaviorrod}
\end{equation}
It is obvious from those expressions that we need $G_i = \pm \frac{1}{2}$ in order to have a well defined $g_{\theta_i \theta_i}$ so the choice $G_i=0$ mentioned above is singular. We now treat the two possible values separately:
\begin{itemize}
\item[•] If $\,G_i = P_i\,\sinh b = \frac{1}{2}$: 

We remind that this corresponds to a shrinking $y$-direction \eqref{eq:ytphibehaviorBubblerod}. From \eqref{eq:ytphibehaviorBubblerod} and \eqref{eq:rthetabehaviorrod}, we notice that the $\theta_i$-dependent factors in $g_{yy}$ and $g_{r_i r_i}$ are remarkably the same. We end with a local five-dimensional metric around the i$^\text{th}$ rod as
\begin{equation}
ds^2 \bigl|_{r_i = 0} \= g_{tt}(\theta_i) \,dt^2 + g_{\theta_i\theta_i}(\theta_i)  \,\left( d\theta_i^2 + \bar{g}_{\phi\phi}(\theta_i) \,\sin^2 \theta_i\, \,d\phi^2\right) + \bar{g}_{r_i r_i}(\theta_i) \left(d\rho_i^2 + \frac{\rho_i^2}{C_i} \,dy^2 \right)\,,
\label{eq:metricatrod}
\end{equation}
where $(g_{tt}(\theta_i),g_{\theta_i\theta_i}(\theta_i),\bar{g}_{\phi\phi}(\theta_i),\bar{g}_{r_i r_i}(\theta_i))$ can be obtained from \eqref{eq:ytphibehaviorBubblerod}  and \eqref{eq:rthetabehaviorrod} and are all finite and non-zero for $0\leq \theta_i <\pi$. Moreover, we have defined $\rho_i^2 \equiv 4 r_i$ and $C_i$ is given by
\begin{equation}
C_i \equiv \frac{M_i^2\,e^{3 b}}{2\sinh^3b}\,\,d_i^2\, \prod_{j\neq i} \left(\frac{z_j^+ - z_i^-}{z_j^- -z_i^-} \right)^{2\,\text{sign}(j-i) \,\alpha_{ij}}\,.
\end{equation}
The two dimensional subspace $(\rho_i,y)$ describes a smooth origin of $\IR^2$ or a smooth discrete quotient $\IR^2/\mathbb{Z}_{k_i}$ if the parameters are fixed according to the radius of the $y$-circle
\begin{equation}
\begin{split}
R_y^2 = \frac{C_i}{k_i^2}\,.
\end{split}
\label{eq:condRymultibubble}
\end{equation}
To conclude, the time slices of the five-dimensional space at the i$^\text{th}$ rod is a bolt described by a warped S$^2$ sphere times an origin of a $\IR^2/\mathbb{Z}_{k_i}$ space.
\item[•] If $\,G_i = -P_i \,\sinh b = -\frac{1}{2}$: 

It is now the time direction that shrinks \eqref{eq:ytphibehaviorBubblerod}. The analysis is identical to the one above and the metric at the rod is given by 
\begin{equation}
ds^2 \bigl|_{r_i = 0} \= g_{yy}(\theta_i) \,dy^2 + g_{\theta_i\theta_i}(\theta_i)  \,\left( d\theta_i^2 + \widetilde{g}_{\phi\phi}(\theta_i) \,\sin^2 \theta_i\, \,d\phi^2\right) + \widetilde{g}_{r_i r_i}(\theta_i) \left(d\rho_i^2 - \frac{\rho_i^2}{C_i} \,dt^2 \right)\,.
\label{eq:metricatrod2}
\end{equation}
The metric corresponds then to a horizon of a black string. The topology of its horizon is a warped S$^2\times$S$^1$. One can relate $C_i$ to the temperature of the solution by requiring smoothness of the Euclideanized solution. We find
\begin{equation}
T^{2} \= \frac{1}{4 \pi^2 C_i}\,.
\end{equation}
\end{itemize}

\begin{itemize}
\item[•] \underline{Out of the rods:}
\end{itemize}

We now study the behavior of the solutions on the $z$-axis out of the rods where the $\phi$-circle shrinks to zero size. On this segments, each $R^{(i)}_\pm$ is non-zero and finite,
\begin{equation}
R^{(i)}_\pm = 2 |z-a_i| \pm M_i\,.
\end{equation}
Thus, $W_\text{GF}$ and $W_0$ are also non-zero and finite there. The regularity reduces to the study of the three-dimensional subspace $(\rho,z,\phi)$,
\begin{equation}
ds_3^2 \= e^{2 (\nu_0 +\nu_\text{GF})} \left(d\rho^2 +dz^2 \right) +\rho^2 d\phi^2\,.
\end{equation}
At $\rho=0$ and out of the rods, we want this space to correspond at least to the origin of a $\IR^3$ with potential conical defects. First we have
\begin{equation}
e^{\nu_{jk}} \sim \begin{cases}
1 \qquad &\text{if } j\leq k  \text{ and } z \not\in [a_j + \frac{M_j}{2} , a_k - \frac{M_k}{2} ]  \\
\dfrac{(z_k^- - z_j^+)^2(z_k^+ - z_j^-)^2}{(z_k^+ - z_j^+)^2(z_k^- - z_j^-)^2} \qquad &\text{if } j<k   \text{ and } z \in [a_j + \frac{M_j}{2} , a_k - \frac{M_k}{2} ]
\end{cases}\,.
\end{equation}
Therefore, we get
\begin{equation}
e^{2 (\nu_\text{GF} +\nu_0)} \sim \begin{cases}
1  &\text{if } z < a_1 - \frac{M_1}{2}  \text{ and } z > a_n - \frac{M_n}{2} \\
 d_i^2=\prod_{j=1}^{i-1} \prod_{k=i}^n \left(\dfrac{(z_k^- - z_j^+)(z_k^+ - z_j^-)}{(z_k^+ - z_j^+)(z_k^- - z_j^-)}  \right)^{ 2\alpha_{jk}}~ &\text{if } z \in [a_{i-1} + \frac{M_{i-1}}{2} , a_i - \frac{M_i}{2} ]
\end{cases}
\end{equation}
First we notice that asymptotically, $z \sim \pm \infty$, the base space is directly flat $\IR^3$ without conical defect. Moreover, note that
\begin{equation}
\dfrac{(z_k^- - z_j^+)(z_k^+ - z_j^-)}{(z_k^+ - z_j^+)(z_k^- - z_j^-)}  = \frac{(a_k-a_j)^2 - \frac{1}{4} (M_k+M_j)^2}{(a_k-a_j)^2 - \frac{1}{4} (M_k-M_j)^2} < 1 \,,\qquad \alpha_{jk} \= \frac{3P_j P_k +4 G_j G_k}{2} >0\,.
\label{eq:simpleform}
\end{equation}
Then, we necessarily have $e^{2 (\nu_\text{GF} +\nu_0)} <1$ for $z \in [a_{i-1} + \frac{M_{i-1}}{2} , a_i - \frac{M_i}{2} ]$. Thus, the segment has a conical excess that manifests itself as a string with negative tension or a strut between the two rods. The string gives the necessary repulsion in order for the whole structure not to collapse.
The three-dimensional metric on the $z$-axis in between the $(i-1)^\text{th}$ and $i^\text{th}$ rods is given by
 \begin{equation}
ds_3^2 \= d_i^2 \left(d\rho^2 +dz^2 +\frac{\rho^2}{d_i^2} d\phi^2 \right) \,.
\end{equation}

To conclude, our class of solutions describes regular two-charge black strings (for rods with $G_i = - \frac{1}{2}$) and regular topological stars (for rods with $G_i = \frac{1}{2}$) that are stacked on a line and which are prevented from collapse by struts between them.

\section{Topological stars and black strings in $D+1$ dimensions}
\label{app:D+1dimext}

In section \ref{sec:Ddimext}, we have seen that one can construct solutions that are superposition supported by fluxes of $(D+1)$-dimensional bubble of nothings and S$^1$ fibration over $D$-dimensional Schwarzschild-Tangherlini solutions \cite{Tangherlini:1963bw},
\begin{equation}
\begin{split}
ds^2_{D+1} &\= -\left( 1 - \left(\frac{r_\text{S}}{r}\right)^{D-3} \right)\,dt^2 + \left( 1 - \left(\frac{r_\text{B}}{r}\right)^{D-3} \right)\, dy^2 + \frac{r^{2(D-3)}\,dr^2}{\left(r^{D-3} - r_\text{S}^{D-3}\right)\left(r^{D-3} - r_\text{B}^{D-3}\right)} \\
&~~~+ r^2 \,d\Omega^2_{D-2} \,, \\
F^{(e)} &\= \frac{Q}{r^{D-2}}\, dr \wedge dt \wedge dy\,,\qquad F^{(m)} \= P\,dV_{S^{D-2}}\,,
\end{split}
\label{eq:metricGenD+1DimApp}
\end{equation}
with
\begin{equation}
P^2+Q^2 = \frac{(D-3)(D-1) \, r_\text{S}^{D-3} r_\text{B}^{D-3}}{2\,\kappa_{D+1}^2}\,.
\label{eq:P&QD+1dimApp}
\end{equation}
The solutions correspond to either smooth topological stars for $r_\text{B}^{D-3} > r_\text{S}^{D-3}$ or two-charged black strings for $r_\text{B}^{D-3} \leq r_\text{S}^{D-3}$.

\subsection{Reduction to $D$ dimensions}
\label{app:redtoD}

We consider the minimum ingredients for the Kaluza-Klein reduction along the $y$-circle. For the matter fields, we assume that only $F^{(e)}$ has a component along $dy$ and define $F^{(e)}_y$ as in \eqref{eq:Fey}. The ansatz for the reduction of the metric is
\begin{equation}
ds_{D+1}^2 = e^{-2(D-2)\Phi}\, dy^2 \,+\, e^{2 \Phi}\,ds_{D}^2\,.
\end{equation}
The Einstein-Maxwell action \ref{eq:ActionGenD+1} reduced to an Einstein-Maxwell-dilaton theory in $D$ dimensions:
\begin{equation}
\begin{split}
S_{D}\=\int \mathrm{d}^{D} x \sqrt{-\text{det}\,g_{D}}\,& \left(\frac{1}{2 \kappa_D^{2}} R_{D-1} - \frac{(D-2)(D-1)}{2 \kappa_D^{2}}\, \partial_a \Phi\, \partial^a \Phi - \frac{e^{-2(D-3)\Phi}}{2e^2} \left|F^{(m)}\right|^2\right.\\
& ~~ \left. -\frac{e^{2(D-3)\Phi}}{2e^2}  \left|F^{(e)}_y\right|^2 \right)\,,
\end{split}
\end{equation}
where the gravitational and electric couplings are given by
\begin{equation}
\kappa_D^2 \equi \frac{\kappa_{D+1}^2}{2\pi R_y}\,,\qquad e^2 \equi \frac{1}{2\pi R_y}\,.
\end{equation}
Depending on taste, we can rescale $\Phi$ to get a canonical scalar Lagrangian with
\begin{equation}
\bar{\Phi} =\sqrt{2(D-2)(D-1)}\,\Phi\,.
\end{equation}
In this framework, the solutions are given by
\begin{equation}
\begin{split}
ds_{D}^2 & =  \left(1-\left(\frac{r_\text{B}}{r}\right)^{D-3} \right)^{\frac{1}{D-2}} \left[- \left(1-\left(\frac{r_\text{S}}{r}\right)^{D-3}\right) dt^2 + \frac{r^{2(D-3)} dr^2}{(r^{D-3}-r_\text{S}^{D-3})(r^{D-3}-r_\text{B}^{D-3})} + r^2 d\Omega_{D-2}^2 \right]\,,\\
e^{2 \Phi} &= \left(1-\left(\frac{r_\text{B}}{r}\right)^{D-3} \right)^{-\frac{1}{D-2}}\,,\\
F^{(e)} &\= \frac{Q}{r^{D-2}}\, dr \wedge dt \,,\qquad F^{(m)} = P\,dV_{S^{D-2}}\,,\qquad P^2+Q^2= \frac{(D-3)(D-1) e^2\, r_\text{S}^{D-3} r_\text{B}^{D-3}}{2\,\kappa_{D}^2}\,.
\end{split}
\label{eq:metric&GFDdim}
\end{equation}
From a $D$-dimensional perspective, the solutions are sourced by an electric charge and a magnetic charge. The conserved quantities in $D$ dimensions, as the ADM mass, $\cM$, the electric and the magnetic charges, $Q_e$ and $Q_m$, are given, following the conventions of \cite{Myers:1986un}, by
\begin{equation}
\begin{split}
\cM &\= \frac{\pi^{\frac{D-1}{2}}}{\kappa_D^2\,\Gamma\left(\frac{D-1}{2} \right)}\,\left( (D-2)\, r_\text{S}^{D-3}  + r_\text{B}^{D-3} \right)\,,\\
 \mathcal{Q}^2 &\equiv Q_m^2 + Q_e^2 \=  \frac{(D-3)(D-1) \, r_\text{S}^{D-3} r_\text{B}^{D-3}}{2\,\kappa_{D}^2}\,,\qquad Q_m \= \frac{P}{e}\,,\qquad Q_e \= \frac{Q}{e}\,.
\end{split}
\label{eq:ADMmass&chargesDdApp}
\end{equation}
As in five dimensions, one can invert those expressions and find two solutions $({r_\text{S}^{(i)}}^{D-3} ,{r_\text{B}^{(i)}}^{D-3})_{i=1,2}$ for a given $(\cM,\cQ)$. The expressions are not very useful and look like \eqref{eq:rs&rbgivenM&Q} with different coefficients. The important points are that the solutions exist only if
\begin{equation}
\Gamma\left(\frac{D+1}{2} \right)\,\sqrt{D-3}\,\kappa_D \,\cM \,\geq\, \sqrt{2(D-2)(D-1)} \,\pi^{\frac{D-1}{2}}\,\cQ\,,
\end{equation}
and we have the following relations
\begin{equation}
\begin{split}
{r_\text{B}^{(1)}}^{D-3}& \= (D-2) \,{r_\text{S}^{(2)}}^{D-3}\,,\\
{r_\text{B}^{(1)}}^{D-3}& \,>\, {r_\text{S}^{(1)}}^{D-3}\,,\\
{r_\text{S}^{(2)}}^{D-3}& \,\geq\, {r_\text{B}^{(2)}}^{D-3}\,,\qquad \text{when} \quad \Gamma\left(\frac{D+1}{2} \right)\,\sqrt{D-3}\,\kappa_D \,\cM \,\geq\,\frac{(D-1)^{\frac{3}{2}} }{\sqrt{2}}\,\pi^{\frac{D-1}{2}}\,\cQ\,.
\end{split}
\end{equation}

\subsection{Phase space}
\label{app:phasespace}

As in five dimensions, the class of spherically symmetric solutions defined above describes either topological stars or black strings:

\begin{itemize}
\item[•] \underline{Topological star:}

If $r_\text{B}^{D-3} > r_\text{S}^{D-3}$, the outermost singularity is where the $y$-circle degenerates thus corresponding to an end of spacetime. The topology at this locus is best described by the radial distance
\begin{equation}
\rho^2 \equi \frac{4}{(D-3)^2}\,\frac{r^{D-3}-r_\text{B}^{D-3}}{r_\text{B}^{D-3}-r_\text{S}^{D-3}}\,,
\end{equation}
and taking the limit $\rho \rightarrow 0$. The metric \eqref{eq:metricGenD+1DimApp} converges to
\begin{equation}
\begin{split}
ds^2_{D+1}  &\sim - \frac{r_\text{B}^{D-3} - r_\text{S}^{D-3}}{r_\text{B}^{D-3}}\,dt^2 \+ r_\text{B}^2 \, \left[ d\rho^2 + \frac{(D-3)^2\,(r_\text{B}^{D-3} - r_\text{S}^{D-3})}{4 \,r_\text{B}^{D-1}}\,\rho^2 \,dy^2 + d\Omega_{D-2}^2\right] \,.
\label{eq:localmetricD+1}
\end{split}
\end{equation}
The $(\rho,y)$ subspace corresponds to a smooth origin of $\IR^2$ providing that 
\begin{equation}
R_y^2 \= \frac{4 \,r_\text{B}^{D-1}}{(D-3)^2\,(r_\text{B}^{D-3} - r_\text{S}^{D-3})}\,.
\end{equation}
The spacetime at the coordinate singularity corresponds to a smooth S$^{D-2}$ bubble of radius $r_\text{B}$ sitting at an origin of a $\IR^2$. One can also show that the matter fields are regular there. Moreover, it is fairly straightforward that the solution has no closed timelike curves for $r^{D-3} \geq r_\text{B}^{D-3}$ from the metric \eqref{eq:metricGenD+1Dim}. Consequently, we have constructed a solution that caps off smoothly before the horizon as a round S$^{D-2}$ bubble wrapped by electric and magnetic fluxes.

We can also allow the solution to have a conical defect at the bubble locus. The local geometry given by \eqref{eq:localmetricD+1} will correspond to a smooth discrete quotient S$^{D-2} \times \IR^2/\mathbb{Z}_k$ providing that 
\begin{equation}
R_y^2 \= \frac{4 \,r_\text{B}^{D-1}}{k\,(D-3)^2\,(r_\text{B}^{D-3} - r_\text{S}^{D-3})}\,,\qquad k\in \mathbb{Z}_+\,.
\label{eq:RyorbifoldD+1dim}
\end{equation}
Having a large $k$ is the only way to have a macroscopic topological star compared to the size of the extra dimension.

\item[•] \underline{Black string:}

We start with the solutions where $r_\text{B}^{D-3} > r_\text{S}^{D-3}$. The outermost coordinate singularity corresponds to a horizon at $r^{D-3}=r_\text{S}^{D-3}$. The horizon has a S$^{D-2}\times$S$^1$ topology corresponding to a black string. At the horizon, the radius of the S$^1$ is $\left(\frac{r_\text{S}^{D-3} - r_\text{B}^{D-3}}{r_\text{S}^{D-3}}\right)^{1/2}R_y$ while the radius of the S$^{D-2}$ is $r_\text{S}$. The Bekenstein-Hawking entropy is then
\begin{equation}
S \= \frac{4\pi^{\frac{D+1}{2}}}{\Gamma\left(\frac{D-1}{2} \right)\,\kappa_D^2}\,\left(r_\text{S}^{D-1}\left( r_\text{S}^{D-3} - r_\text{B}^{D-3}\right) \right)^{\frac{1}{2}}\,.
\end{equation}
We read the temperature from the near-horizon metric and we get
\begin{equation}
T \= \frac{D-3}{4\pi\,r_\text{S}} \, \sqrt{1- \left(\frac{r_\text{B}}{r_\text{S}}\right)^{D-3}}\,.
\end{equation}
We have then defined a two-charge non-extremal black string that reduces to a two-charge non-extremal black hole in $D$ dimensions given by \eqref{eq:metric&GFDdim}, with mass and charges  \eqref{eq:ADMmass&chargesDdApp}. As in five dimensions, the locus $r^{D-3} = r_\text{B}^{D-3}$ in the interior is part of the spacetime and corresponds to a S$^{D-2}$ bubble on the origin of a two-dimensional Milne space. 

Finally, when $r_\text{S}^{D-3} = r_\text{B}^{D-3}=m$, the solution corresponds to an extremal two-charge black string given by
\begin{equation}
\begin{split}
ds^2_{D+1} &\=  \left(1+ \frac{m}{\rho^{D-3}} \right)^{-1}\,\left(-dt^2 + dy^2 \right)+ \left(1+ \frac{m}{\rho^{D-3}} \right)^\frac{2}{D-3}\left[d\rho^2 + \rho^2 \,d\Omega_{D-2}^2 \right]\,,\\
F^{(e)} &\= \frac{Q}{(D-3)\,m} d\left(\left(1 + \frac{m}{\rho^{D-3}} \right)^{-1}\right) \wedge dt \wedge dy\,,\qquad F^{(m)} = P\,\sin\theta\,d\theta \wedge d\phi\,,
\end{split}
\end{equation}
where we have defined $\rho^{D-3} \equiv r^{D-3} -m$ and $(P,Q)$ are still constrained by \eqref{eq:P&QD+1dim}.The near-horizon geometry corresponds to an AdS$_3\times$S$^{D-2}$ where the radius of AdS and the radius of the sphere are $\frac{2}{D-3}m$ and $m$ respectively.

\item[•] \underline{The phase space:}

The phase space of solutions for given mass, ``total'' charge and KK radius, $(\cM,\cQ,R_y)$ \eqref{eq:ADMmass&chargesDdApp}, has exactly the same properties as in five dimensions depicted in Fig.\ref{fig:phasespace}. The difference is in the delimitations between the different regions:
\begin{itemize}
\item[-] For $\Gamma\left(\frac{D+1}{2} \right)\,\sqrt{d-3}\,\kappa_D \,\cM \,<\, \sqrt{2(D-2)(D-1)} \,\pi^{\frac{D-1}{2}}\,\cQ$: no solutions exist.
\item[-] For $\sqrt{2(D-2)(D-1)} \,\pi^{\frac{D-1}{2}}\,\cQ\,\leq\,\Gamma\left(\frac{D+1}{2} \right)\,\sqrt{d-3}\,\kappa_D \,\cM \,<\, \frac{(D-1)^{\frac{3}{2}} }{\sqrt{2}}\,\pi^{\frac{D-1}{2}}\,\cQ$: two lattices of topological stars exist. Each node corresponds to an orbifold parameter following the quantization \eqref{eq:RyorbifoldD+1dim}.
\item[-] For $\frac{(D-1)^{\frac{3}{2}} }{\sqrt{2}}\,\pi^{\frac{D-1}{2}}\,\cQ\,\leq \, \Gamma\left(\frac{D+1}{2} \right)\,\sqrt{d-3}\,\kappa_D \,\cM$: One branch of solutions corresponds to the two-charge black string, while the other branch corresponds to a lattice of topological stars. In this region, topological stars and black strings live in the same regime of mass and charges.
\end{itemize}

\end{itemize}

\newpage



\bibliographystyle{utphys}      

\bibliography{microstates}       


\end{document}